\documentclass[traditabstract]{aa}  
\UseRawInputEncoding
\usepackage{graphicx}
\usepackage{txfonts}
\usepackage{bm}
\usepackage{threeparttable}
\usepackage{ltxtable}
\usepackage{ltablex}
\keepXColumns
\usepackage{lscape}
\usepackage{array}  
\usepackage{multicol}
\usepackage{placeins}
\usepackage[utf8]{inputenc}
\usepackage[symbol*]{footmisc}
\setfnsymbol{wiley}

\begin{document}

   \title{PENELLOPE. IX}

   \subtitle{Lithium, iron and barium elemental abundances  in eight nearby young clusters}
\titlerunning{Lithium, iron and barium abundances in YOCs}
   \author{R. Carini \inst{1},
          K. Biazzo \inst{1},
          A. Frasca \inst{2},
          C.F. Manara \inst{3},
          J.M. Alcal\'a \inst{4},
          P. \'Abr\'aham \inst{5,6,7},
          J. Campbell-White \inst{3},
          R. Claes \inst{3},
          M. Fang \inst{8},
          M. Gangi \inst{9,1},
          J.F. Gameiro \inst{10,11},
          \'A. K\'osp\'al \inst{5,6,12},
          K. Mauco  \inst{3,13},
          I. Mendigut\'ia \inst{14},
          B. Nisini \inst{1},
          M. Robberto \inst{15,16},
          C.E. Robinson \inst{17},
          C. Schneider \inst{18},
          M. Siwak \inst{5,19},
          T. Sperling \inst{20},
          L. Tychoniec \inst{21},
          \and
          L. Venuti \inst{22} 
          }

   \institute{
   INAF-Osservatorio Astronomico di Roma, via Frascati 33, 00078 Monte Porzio Catone (RM), Italy; \email{roberta.carini@inaf.it}
          \and
          INAF-Osservatorio Astrofisico di Catania, via S.Sofia 78, 95123 Catania, Italy   \and
             European Southern Observatory, Karl-Schwarzschild-Stra\ss{}e 2, 85748 Garching bei M\"unchen, Germany
          \and
           INAF-Osservatorio Astronomico di Capodimonte, via Moiariello 16, 80131 Napoli, Italy  
          \and
          Konkoly Observatory, HUN-REN Research Centre for Astronomy and Earth Sciences, MTA Centre of Excellence, Konkoly-Thege Mikl\'os \'ut 15-17, 1121 Budapest, Hungary
          \and 
          Institute of Physics and Astronomy, ELTE E\"{o}tv\"{o}s Lor\'and University, P\'azm\'any P\'eter s\'et\'any 1/A, 1117 Budapest, Hungary
          \and
          Institute for Astronomy (IfA), University of Vienna,T\"{u}rkenschanzstrasse 17, A-1180 Vienna, Austria
          \and
          Purple Mountain Observatory, Chinese Academy of Sciences, 10 Yuanhua Road, Nanjing 210023, People's Republic of China
          \and
          ASI, Italian Space Agency, Via del Politecnico snc, 00133, Rome, Italy
          \and
          Instituto de Astrof\'isica e Ci\^encias do Espa\c{c}o, Universidade do Porto, CAUP, Ruadas Estrelas, 4150-762 Porto, Portugal
          \and
          Departamento de F\'isica e Astronomia, Faculdade de Ci\^encias, Universidade do Porto, Ruado Campo Alegre 687, 4169-007 Porto, Portugal
          \and
          Max-Planck-Insitut f\"{u}r Astronomie, K\"{o}nigstuhl 17, 69117 Heidelberg, Germany
          \and
%          Independent Researcher, Porto, Portugal
%          \and
          Instituto de Astronom\'ia,Universidad Aut\'onoma de M\'exico Ensenada,B.C,M\'exico
          \and
          Centro de Astrobiolog\'ia (CAB), CSIC-INTA, Camino Bajo del Castillo s/n, 28692, Villanueva de la Ca\~nada, Madrid, Spain
          \and
          Space Telescope Science Institute, 3700 San Martin Dr., Baltimore, MD 21218, USA
          \and
          Johns Hopkins University, Bloomberg Center for Physics and Astronomy, 3400 N. Charles Street, Baltimore, MD 21218, USA
          \and
          Division of Physics and Astronomy, Alfred University, 1 Saxon Drive, Alfred, NY 14802, USA
          \and
          Institut f\"ur Theoretische Physik und Astrophysik Christian-Albrechts-Universit\"at zu Kiel, Leibnizstrasse 15 24118 Kiel, Germany
          \and
          Mt. Suhora Astronomical Observatory, University of  the National Education Commission, ul. Podchor\.zych 2, 30-084 Krak\'ow, Poland
          \and
          Th\"uringer Landessternwarte, Sternwarte 5, 07778 Tautenburg, Germany
          \and
          Leiden Observatory, Leiden University, PO Box 9513, 2300RA, Leiden, The Netherlands
          \and
          SETI Institute, 339 Bernardo Ave., Suite 200, Mountain View, CA 94043, USA}
        
\authorrunning{Carini et al.}          
   \date{}
% \abstract{}{}{}{}{} 
% 5 {} token are mandatory
\vspace{-2cm}
  \abstract
  {We conducted a homogeneous  chemical analysis of pre-main sequence stars  with effective temperatures ranging from $\sim$ 3000 K to $\sim$ 5500 K in eight  nearby star-forming regions (SFRs): Chamaeleon I,  $\eta$ Chamaeleonis, Lupus, Orion OB1a, Orion OB1b, $\sigma$\,Orionis, Taurus, and Corona-Australis.
   Our study  aims to: 1) derive the lithium abundance ($A{\rm (Li)}$) and highlight the impact of veiling correction on both  $A{\rm (Li)}$ and age determination; 2)  perform the iron (Fe) and barium (Ba) abundance analysis in regions with scarce previous measurements; 3)  investigate the possible Ba enhancement.
   
   The analyzed data  were obtained as part of the PENELLOPE Large Program using the ESPRESSO, UVES, and X-Shooter instruments. 
   We measured the equivalent width of the lithium line ($EW_{\rm Li}$) at $\lambda$ = 6707.8 \AA, from which $A{\rm (Li)}$ is derived   using  the curves of growth method. The Fe and Ba abundances have been measured through spectral synthesis analysis. Using the EAGLES code, we derived an upper limit on the age of the eight SFRs.
    
    Our findings underscore the necessity of veiling corrections on $EW_{\rm Li}$, which can shift  $A{\rm (Li)}$ and age estimates by up to $\sim$ 0.7 dex and $\sim$ 20 Myr, respectively.
   Accounting for veiling, the $A{\rm (Li)}$ distributions peak in a range between 3.3 and 3.8  dex for most clusters,  and  the upper age limit is approximately 5 Myr for all SFRs.
    We successfully  measured the mean iron and barium abundances in  Lupus, Taurus, Cha I, and $\eta$ Cha, showing  slightly sub-solar iron abundance, and a clear Ba overabundance, with  [Ba/H] values reaching up to 0.75 dex.}
              \maketitle
             
\section{Introduction}
The determination of the chemical composition of star-forming regions (SFRs) and young open clusters (YOCs) is  critically important for various  astrophysical issues, in both planetary and stellar contexts. These young regions  are key objects for tracing the current chemical pattern of the Galactic thin disk. Due to their recent formation, these regions have not undergone significant radial migration across the Galactic disk. Their chemical abundances are therefore expected to closely mirror the current composition of the local interstellar medium (ISM), showing minimal evidence of subsequent chemical enrichment (\citealt{spina14}).

%they are still close to their birthplace and contain  stellar populations that has not yet had time to spread out across the Galactic disk.  
One of the most important elements for studying young regions is  lithium ($^{7}\rm Li$). This element is indeed sensitive to stellar interior processes,  it serves as a tracer  of internal mixing processes, representing a  benchmark for stellar evolution models. In pre-main sequence (PMS) stars, the deviations of the observed lithium patterns from predictions by standard stellar models provide a crucial test for  theoretical models, highlighting the limitations in what concerns the treatment of % standard models lacking factors like convection, 
overshooting, and non-standard mixing mechanisms (e.g., rotation or magnetic fields; see e.g. \citealt{pins97}, \citealt{jeff06,somers15,baraffe17}). Furthermore, since lithium is easily destroyed at relatively low temperatures ($\sim$ 2.5 $\times$ $10^6$ K), its depletion in stellar atmosphere provides an age indicator for low-mass PMS stars in young (age < 200 Myr) populations \citep{bildsten}. The amount of depletion depend on mass, age, metallicity, and other mixing processes occurring during the early stellar evolution.
 Low-mass (<0.5 M$_\odot$) PMS stars reach these temperatures as they contract toward the main sequence \citep{boden65}. Low-mass stars require relatively long time to reach the critical temperature for lithium burning. Specifically, stars with mass lower than 0.2 M$_\odot$ initiate $^{7}\rm Li$ burning after $\sim$ 20-25 Myr, while those in the $\sim$ 0.2-0.5 M$_\odot$ begin  after $\sim$ 15-5 Myr  (\citealt{baraffe15} and reference within).
  Since these stars remain fully convective, they eventually deplete their entire $^{7}\rm Li$ content during the PMS phase. Stars more massive than 0.5 M$_\odot$ initiate lithium burning during the early stages of their PMS evolution. For example, a 0.6 M$_\odot$ star begins depletion at $\sim$ 3 Myr. The duration of this process is limited, concluding once a radiative core develops. This transition is mass-dependent; in more massive stars, the convective envelope gradually shrinks. Consequently, the temperature at the base of the envelope becomes too low, halting further $^{7}{\rm Li}$ destruction. As a result, these stars retain a portion of their initial lithium abundance (typically $A{\rm(Li)} \sim$ 3.3 dex). For instance, a 1.0 M$_\odot$ star is expected to deplete only 60\% of its initial $^{7}{\rm Li}$, while stars with masses greater than $\sim$ 1.2 M$_\odot$  are not expected to  destroy  lithium in their envelopes (\citealt{randich21} and reference within). Consequently, the depletion of $^{7}\rm Li$ serves as a robust chronometer for characterizing the ages of young stellar associations and open clusters (e.g, \citealt{song02,palla07,lim16,randich21}).
 
 %The amount of depletion  depends on mass, age, metallicity, and other mixing processes occurring during the  early stellar evolution.

%Moreover, $^{7}\rm Li$ is the heaviest element produced during the Big Bang.
%Unfortunately, the stellar Li measurements are inconsistent with the  measurement  of the cosmic microwave background (CMB), this  is referred to as the cosmological Li problem (see e.g. \citealt{cyburt16,mat20} and  the reference within).
% Observations of young stellar populations provide constraints on lithium destruction processes that contribute to the this problem (see e.g \citealt{fields11,prantzos17}).
% Finally, Lithium depletion occurs rapidly in low-mass PMS stars,  making it an effective chronometer for very young clusters (ages < 200 Myr). 

In addition to lithium, iron (Fe) is a fundamental tracer for investigating the origin and evolution of star-forming regions and the chemical evolution of the Galactic disk. As the primary proxy for overall metallicity, iron abundance [Fe/H]\footnote{Throughout the paper the abundance of the X element is given as [X/H]=$\log \frac{A{\rm (X)}}{A{\rm(H)}} + 12$, where $\log A(X)$ is the absolute abundance.} provides critical constraints on the star formation, possible chemical enrichment history, and chemical patterns of stellar populations. 
Furthermore, knowledge of the iron abundance in SFRs is essential for investigating the formation and evolution of exoplanets. A growing consensus supports the planet-metallicity correlation, wherein metal-rich environments facilitate the formation of planetary systems, particularly giant planets (e.g \citealt{mulders16A, swastik22} and references therein).
In SFRs and YOCs, the iron abundance is typically observed to be slightly sub-solar or near-solar (e.g \citealt{biazzo11a, biazzo11,spina14} and reference therein).
%Recent studies have shown that metallicity correlates with planet presence, in particular for giant planets.}
 
Finally, another important element used to investigate the chemical pattern of  young regions is barium (Ba).
Ba is produced by neutron capture reaction, mostly by the s-process occurring in low-mass AGB stars \citep{busso99,karakas14, koba20}.
%it represents an excellent tracer of the chemical enrichment mechanism in the Galaxy.
In the last decades, several studies have shown an overabundance  of  Ba content in young clusters;
in particular, \cite{dorazi09} discovered an anti-correlation between [Ba/Fe] and cluster age analyzing 20 open clusters (OCs) in the Galaxy. The old OCs (age $\gtrsim$ 4 Gyr) exhibited a solar Ba abundance, while the OCs  with ages $\sim$ 100-200 Myr showed an enhancement up to 0.2-0.3 dex, and the younger clusters ( $\lesssim$ 70 Myr) showed an higher Barium content up to 0.6-0.7 dex.
These results have been confirmed by other authors (e.g. \citealt{dorazi12,jacob11,mishe13,baratella21,spina21,magrini23}).
The contribution of the low-mass AGBs  to the Galactic chemical enrichment  can explain the 
enhancement  observed in intermediate-age OCs \citep{dorazi09}, but not in the younger ones.
Currently, it is not possible to reconcile this large Ba abundance ($\sim$ 0.7 dex)   with any standard nucleosynthesis  and galactic evolution model.
Moreover, it remains controversial whether all other s-process elements follow Ba's behavior. Specifically, elements formed in the second peak of the s-process (along with Ba), such as lanthanum (La) and cerium (Ce), would be expected to share the same patterns. However, some authors have found  a lack of significant trend with age \citep{dorazi12,jacob13,baratella21}, in contradiction with \cite{maiorca11}, adding further complexity to the mystery.

%The young regions are key objects for tracing the current chemical pattern of the Galactic thin disk, because they are still close to their birthplace and contain a homogeneous stellar population that hasn't had time to spread out across the Galactic disk.\\

In this work, we present a systematic and homogeneous analysis of lithium, iron and barium abundances of PMS stars in several star-forming regions: Chamaleon I (Cha I), $\eta$- Chamaleontis ($\eta$\,Cha), Lupus, Taurus, $\sigma$\,Orionis ($\sigma$\,Ori), Orion OB1a, Orion OB1b and Corona-Australis (CrA). We analyzed  spectra gathered as part of the PENELLOPE program. This ESO large program serves to complement the Hubble Space Telescope's (HST) UV Legacy Library of Young Stars (ULLYSES, \citealt{ulisse}). 
The goal of these two programs is to observe a large sample of young stars, probing a wide range of ages and masses to provide sufficient statistics for understanding the processes of accretion and ejection during the star formation.
For a comprehensive description of the PENELLOPE survey we refer to \cite{manara21}. The paper is organized as follows: in Sect. 2 we describe the data;  spectral analysis  of lithium  is in Sect. 3, the study of iron and  barium  is presented in Sect. 4, while we summarize our findings in Sect. 5.

 \section{Data}
 The data used in this work were acquired within the  PENELLOPE survey \citep{manara21}.
 The details on the observational strategy and data reduction of the PENELLOPE sample are reported in \cite{manara21}.
In brief,  the  spectra   were obtained using the instruments ESPRESSO  (Echelle SPectrograph for Rocky Exoplanets and Stable Spectroscopic Observations; \citealt{espresso}),  UVES (Ultraviolet and Visual Echelle Spectrograph; \citealt{uves}), and X-Shooter  \citep{xshooter}, all mounted on the ESO@VLT (Very Large Telescope). 

ESPRESSO spectra cover a wavelength range of 380-788 nm, with a resolution of $R$ $\sim$ 140,000.
UVES observations, conducted using the Red and Blue arms in dichroic mode, span the 330-450 nm and 480-680 nm with R  $\sim$  70,000.
X-Shooter provides broader coverage from $\sim$ 300 nm to $\sim$ 2500 nm, divided in three arms: UVB (300-500 nm), VIS (500-1000 nm) and NIR (1000-2500 nm) with a resolution $\sim$ 17500.

The analyzed sample comprises 75 PMS stars belonging to eight different associations: Lupus (30), Cha I (15), Orion OB1a (2), Orion OB1b (7), Taurus (8), $\eta$ Cha (7), $\sigma$\,Ori (3), and  Corana-Australis (2).
All  targets were observed with X-Shooter. The brightest stars (V < 16 mag) were observed with ESPRESSO, while UVES was employed for the fainter stars and those that could not otherwise be observed with ESPRESSO. These high resolution observations were carried out simultaneously, or quasi-simultaneously, with the X-shooter observations. The mean signal-to-noise ($S/N$) ratios measured around 6000 \AA\ are  about 50 for both ESPRESSO and X-Shooter spectra, and 40 for UVES spectra.

For both ESPRESSO and UVES, each target was observed at three distinct "epochs" (ep.) separated by intervals of a few days.  The specific dates of the observations are reported in Tables \ref{tabewli1}, \ref{tabewli2}, and \ref{tabewli3}.
Consequently, we were able to analyze multiple spectra for each target, ensuring a robust cross-instrument comparison and the monitoring of short-term variability.

%The brightest stars (V< 16 mag) were observed with ESPRESSO. These spectra have a wavelength range   between 380 and 788 nm, with a resolution of $R$ $\sim$ 140000.
%Each target is observed for three times, referred to as "epochs" (ep.).

%Spectra of fainter targets, as well as target that are not be observed whit ESPRESSO, were  taken with UVES with the Red and Blue arms in dichroic mode. This allowed us to cover the wavelength ranges 330-340 nm and 480-680 nm with a resolution of approximately  70000.
%UVES also observed the targets in three different epochs.

%Our sample consists of eight PMS stars belonging to Taurus, three to $\sigma$-Ori, 10 to Orion OB1,  30 to Lupus, 7 to $\eta$ Cha, 15 to Cha I associations and 2 to Corana-Australis.
%Most targets were observed with two instruments, ESPRESSO and X-Shooter or UVES and X-Shooter, so  we analyzed  multiple spectra for each target.

Estimates of the photospheric properties used in our analysis,  such as effective temperature ($T_{\rm eff}$) and surface gravity ($\log g$), radial  velocity (RV), projected rotational velocity ($v\sin i$), and veiling ($r$), were performed on both the medium-resolution and the high-resolution spectra using the ROTFIT \citep{frasca15}  code by the PENELLOPE's team (\citealt{manara21} and Antonio Frasca priv. comm.). Briefly, the code was developed for deriving $T_{\rm eff}$, $\log g$, $v\sin i$, and $r$ comparing the target spectrum with a grid of templates at the same resolution of the ESPRESSO, UVES, and X-shooter spectra. The code performs a $\chi^2$ minimization of the difference between the observed and template spectra parameters around selected spectral regions particularly suitable for the determination of the atmospheric stellar parameters. Veiling was therefore measured in five spectral regions (around 4500, 5400, 6200, 7100, 9700\,\AA) for X-shooter spectra and in four spectral regions (namely, $\sim$5000, 5500, 6000, 6500\,\AA) for UVES and ESPRESSO spectra. For a detailed explanation of the code, see \cite{frasca15, frasca17}, and \cite{manara21}.

These values are not available for two spectra observed with X-Shooter: RECX\,11  and SO\,1153 ep. 2.
For these stars we derived $T_{\rm eff}$ through the line-depth ratio (LDR, \citealt{grey94}), considering a typical surface gravity  for this kind of stars (4.5 dex) and a rotational velocity 0.0 km/s.
We used the relations from \cite{biazzo07} for non-rotating dwarf stars, analyzing LDR of two pairs of lines in the visible range: $\lambda$6199 V I-$\lambda$6200 Fe I and $\lambda$6252 V I-$\lambda$6253 Fe I.
These relations can be applied only for stars with a temperature between  4000 K and  6200 K;
temperatures outside this range are more uncertain because of the influence of molecular bands in the coolest stars and the very small depths of the low-excitation
lines in the hottest stars, respectively.
The list of  stars together with their $T_{\rm eff}$ and veiling values are reported in Table \ref{tabewli1}, \ref{tabewli2}, and  \ref{tabewli3}.

%--------------------------------------------------------------------

\section{Analysis of the lithium line }

\subsection{Equivalent Width Determination}

To determine  the  equivalent width of the lithium line ($EW_{\rm Li}$) at $\lambda$ = 6707.856 \AA\ \citep{campbel23}, we developed an \textit{IDL} code  that estimates the local continuum  through a linear fit  obtained in two narrow ranges ( $\sim$ 5 \AA\,) located near the wings of the $^{7}\rm Li$ absorbing line. This continuum is then used to normalize the spectrum, from which the $EW_{\rm Li}$ is derived by performing a Gaussian fit.
Errors in  $EW_{\rm Li}$ are evaluated from the fitting procedure, with typical values of   10-15 m\AA\ for ESPRESSO  and UVES data and 30 m\AA\ for X-Shooter data.
The results obtained from high-resolution (ESPRESSO, UVES) and medium-resolution (X-Shooter) data are consistent within the uncertainties. The mean difference in $EW_{\rm Li}$ is $\sim$ 8 m\AA\, with a standard deviation of $\sim$ 43 m\AA.

The spectra are affected by accretion veiling, which is  the (non-photospheric)  excess continuum  emission due to the accretion process (see \citealt{hart16} and references therein)  that can hide or "veil" the photospheric lines. To correct the $EW_{\rm Li}$ for this contribution, we applied the relationship $EW_{\rm Li}^{veil}$ = $EW_{\rm Li}^{raw}(1+r)$ where $EW_{\rm Li}^{raw}$ is the lithium equivalent width measured as explained above. 
Veiling estimates may be influenced by spectral resolution thus, we adopted the $r$ value  derived from the ROTFIT analysis closest to the $^{7}\rm Li$ line at 6707.8 \AA\, measured from the spectra acquired with the three instruments as follows: $r_{650}$ at 6500 \AA\ for ESPRESSO and UVES spectra, and $r_{710}$ at 7100 \AA\ for X-Shooter spectra.
%For the $^{7}\rm Li$  line at $6707.8$ \AA\,  we adopted the $r$ value measured closest to the $^{7}\rm Li$ line, that is,  at 6500 \AA\ ($r_{650}$) for ESPRESSO and UVES spectra, and $r$ at 7100 \AA\ ($r_{710}$) for X-Shooter spectra. 
Since the $^{7}\rm Li$ I line is blended  with the FeI $\lambda$ 6707.4 \AA\ line, we subtracted the iron contribution using the corrections by \cite{franciosini}, which are given as a function of effective temperature, gravity, and metallicity.

Determining lithium abundances in stars cooler than 4000 K (M-type stars)  is complicated because of the presence of molecular bands and additional spectral lines from other elements. These blends with the lithium feature significantly decrease the apparent continuum level (e.g., \citealt{zapatero}). This so-called pseudo-continuum obscures the actual intensity of the true continuum, making it impossible to measure the genuine equivalent width. As a result, only a pseudo-equivalent width (pEW) can be estimated and no iron corrections are available in the literature.

\begin{figure*}
      \centering
	   \begin{minipage}{0.45\linewidth}
		\includegraphics[width=\linewidth]{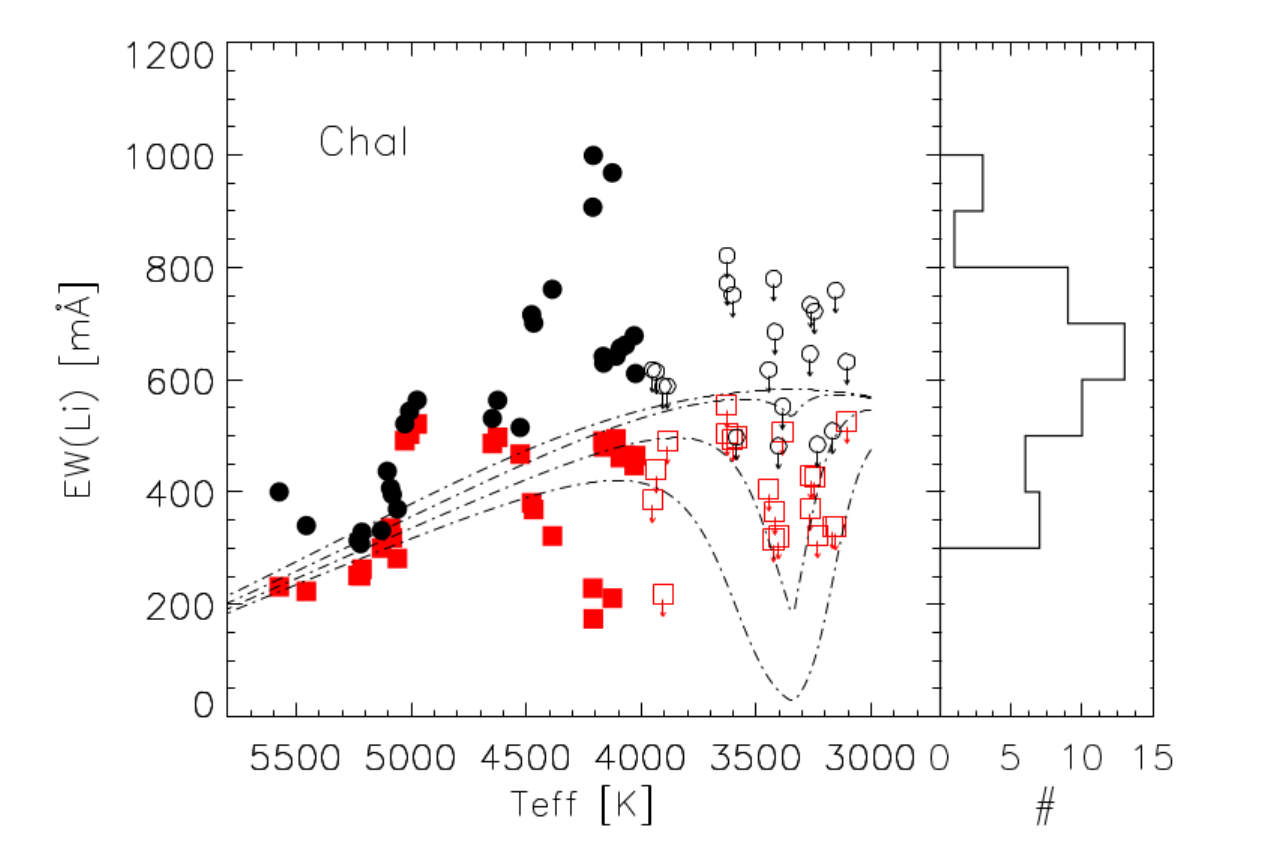}
%        \par\small (a) sOri
%		\caption{sOri}
%		\label{fig:subfig1}
	   \end{minipage}
	   \begin{minipage}{0.45\linewidth}
		\includegraphics[width=\linewidth]{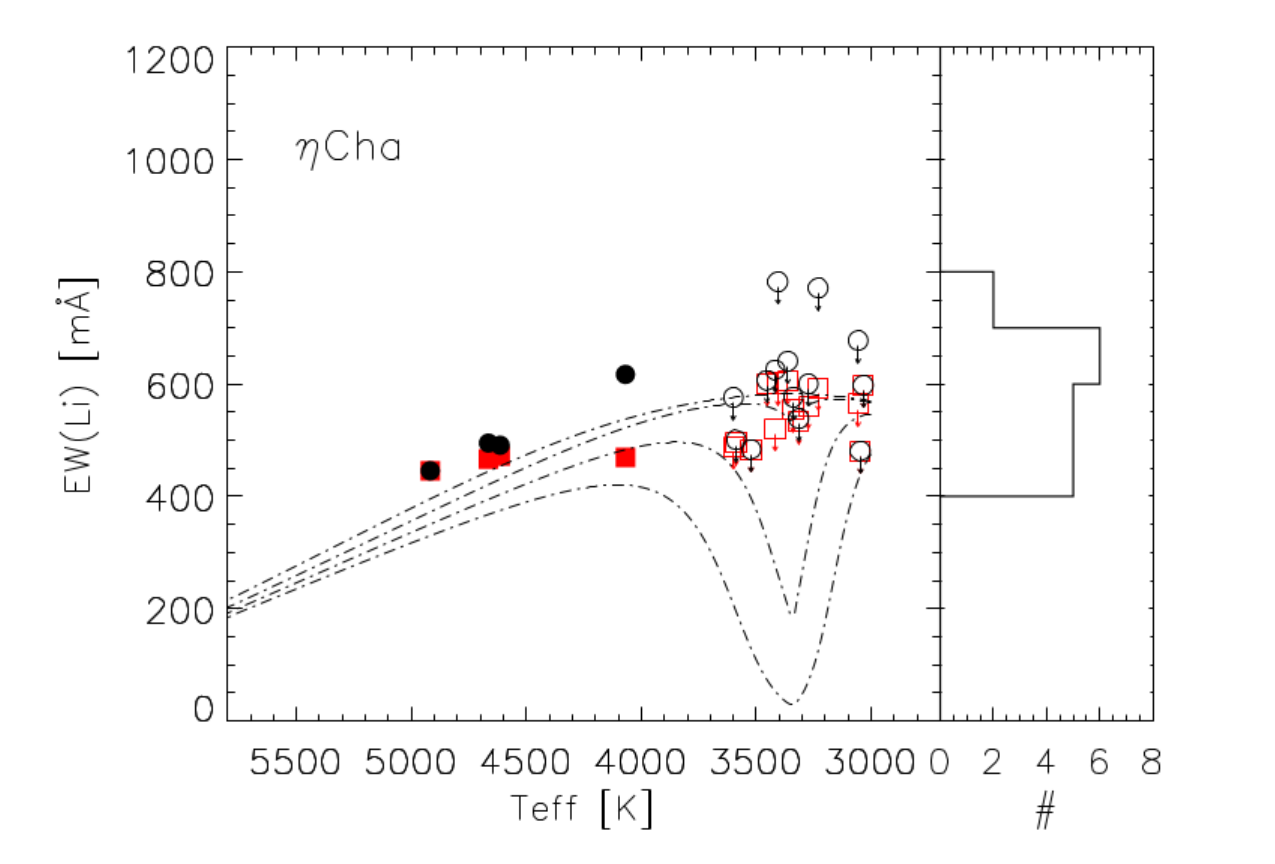}
%\par\small (b) Orion OB1
%		\caption{Orion OB1}
%		\label{fig:subfig2}
	    \end{minipage}
	\vfill
	     \begin{minipage}{0.45\linewidth}
		 \includegraphics[width=\linewidth]{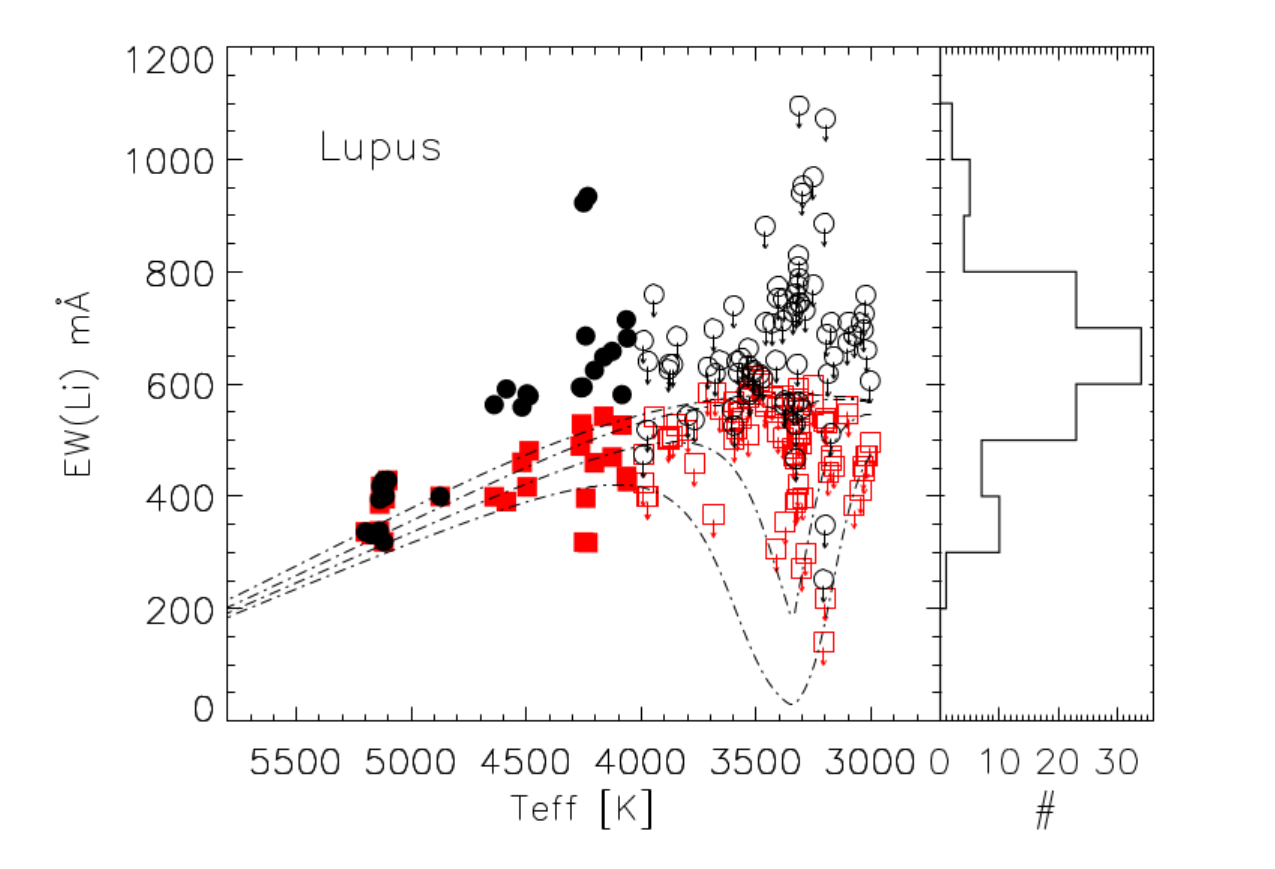}
%\par\small (c) Lupus
%		 \caption{Lupus}
%		 \label{fig:subfig3}
	      \end{minipage}
	       \begin{minipage}{0.45\linewidth}
		  \includegraphics[width=\linewidth]{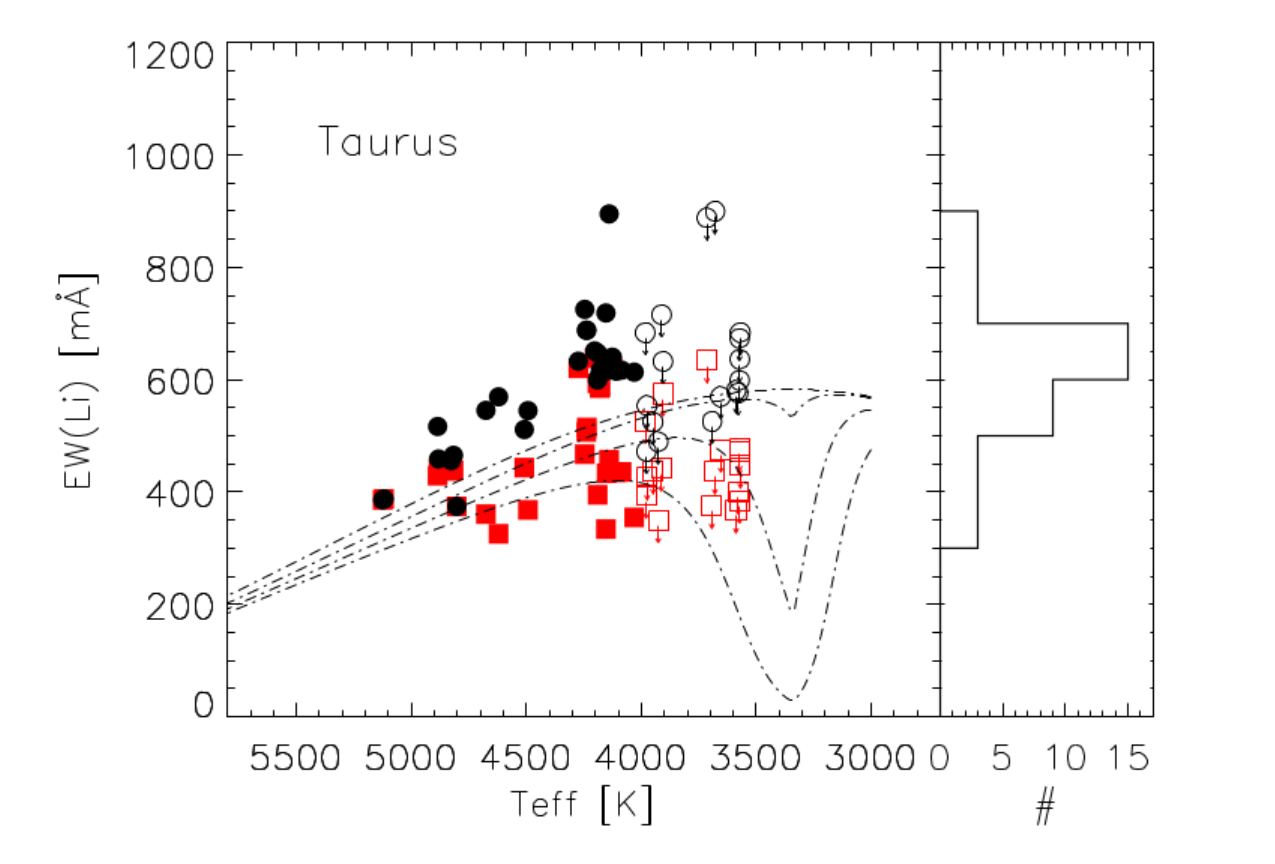}
%        \par\small (d) Taurus
%		  \caption{Taurus}
%		  \label{fig:subfig4}
	       \end{minipage}
    \vfill   
     \begin{minipage}{0.45\linewidth}
		 \includegraphics[width=\linewidth]{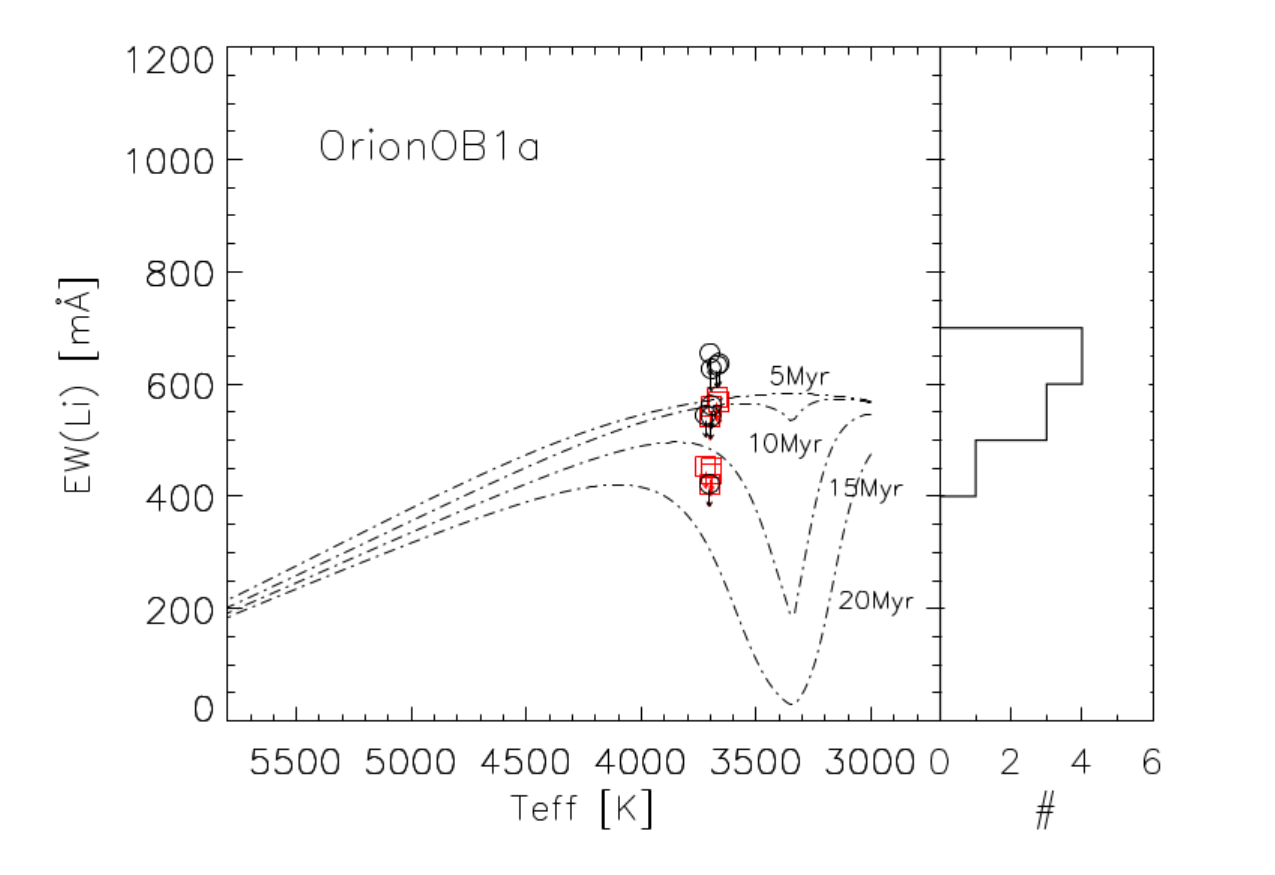}
%         \par\small (e) eCha 
%		 \caption{eCha}
%		 \label{fig:subfig3}
	      \end{minipage}
	       \begin{minipage}{0.45\linewidth}
		  \includegraphics[width=\linewidth]{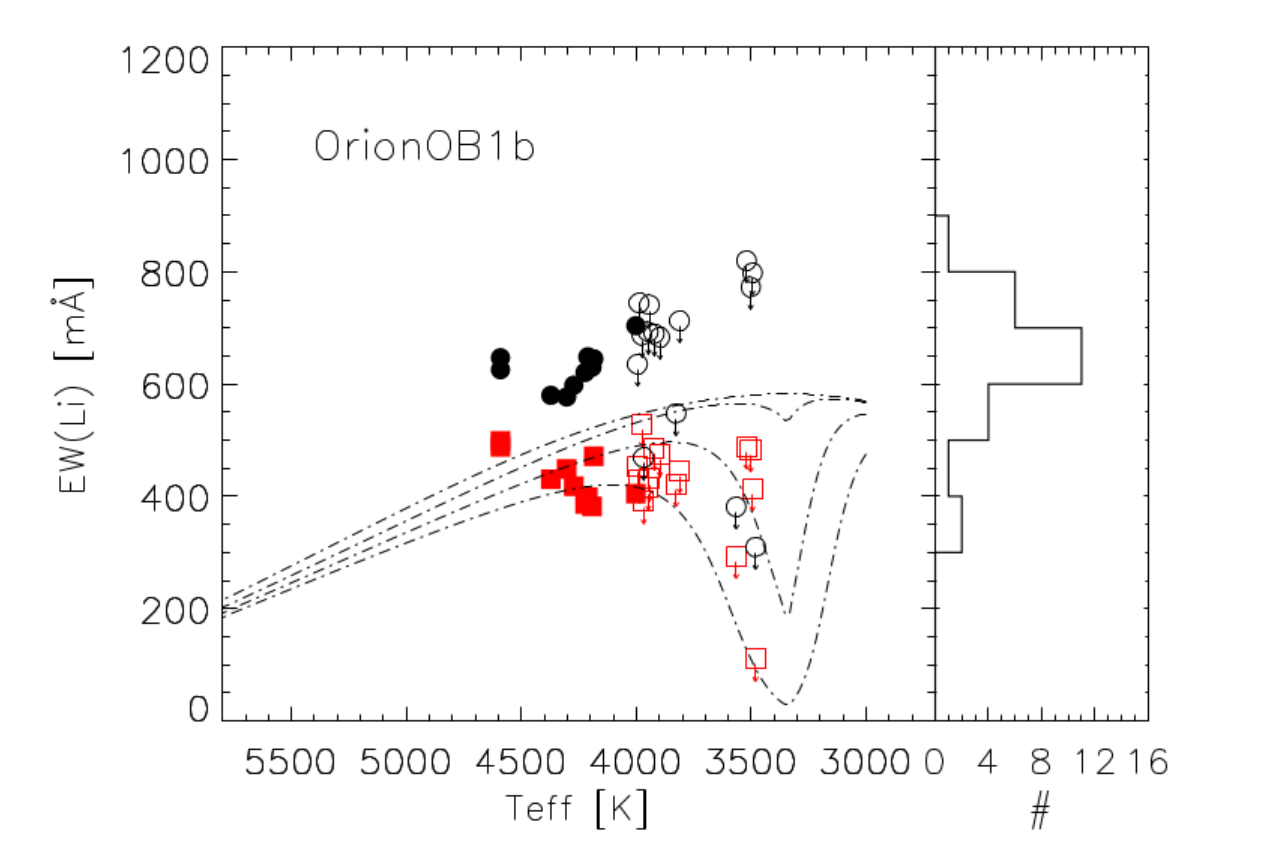}
%\par\small (f) ChaI
%		  \caption{ChaI}
%		  \label{fig:subfig4}
	       \end{minipage}
                \begin{minipage}{0.45\linewidth}
		 \includegraphics[width=\linewidth]{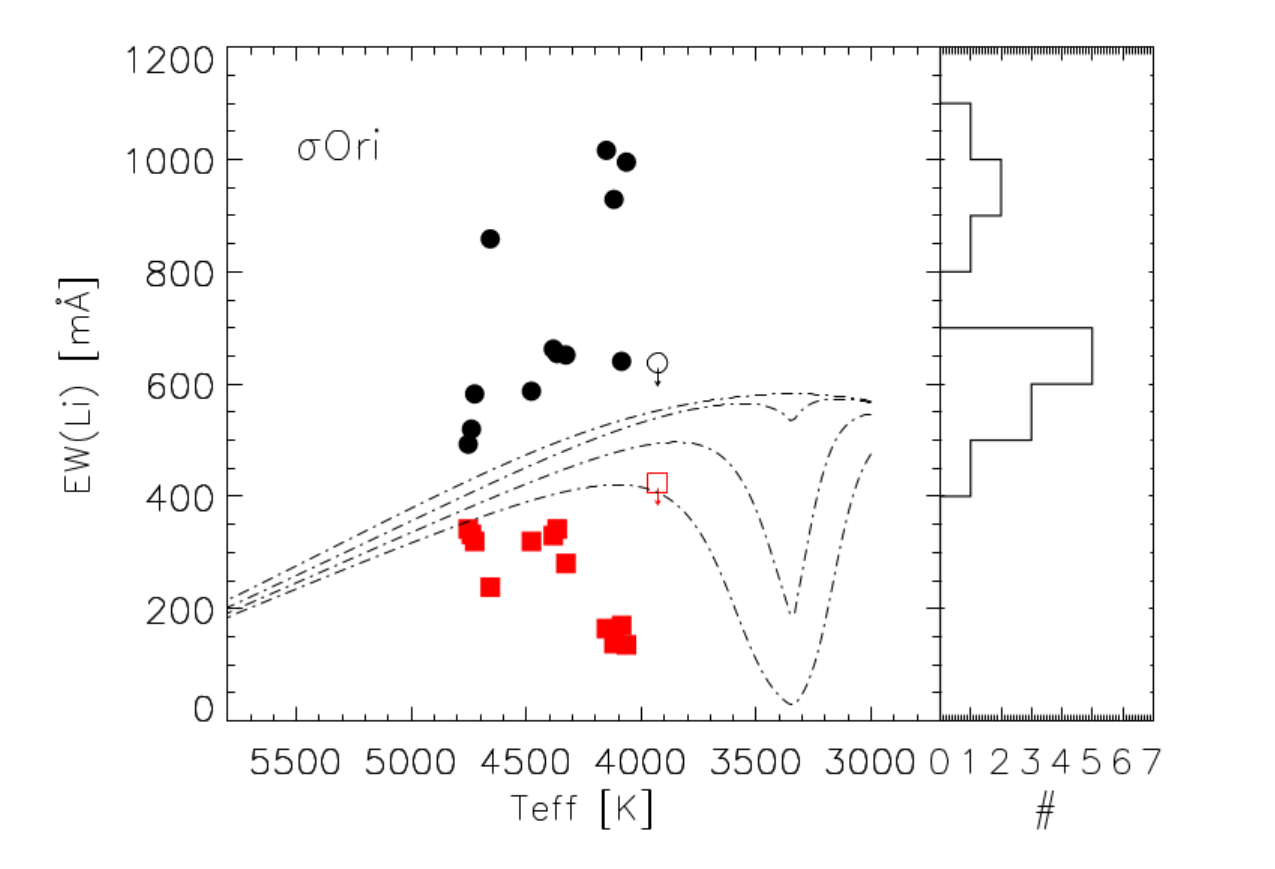}
%         \par\small (e) eCha 
%		 \caption{eCha}
%		 \label{fig:subfig3}
	      \end{minipage}
	       \begin{minipage}{0.45\linewidth}
		  \includegraphics[width=\linewidth]{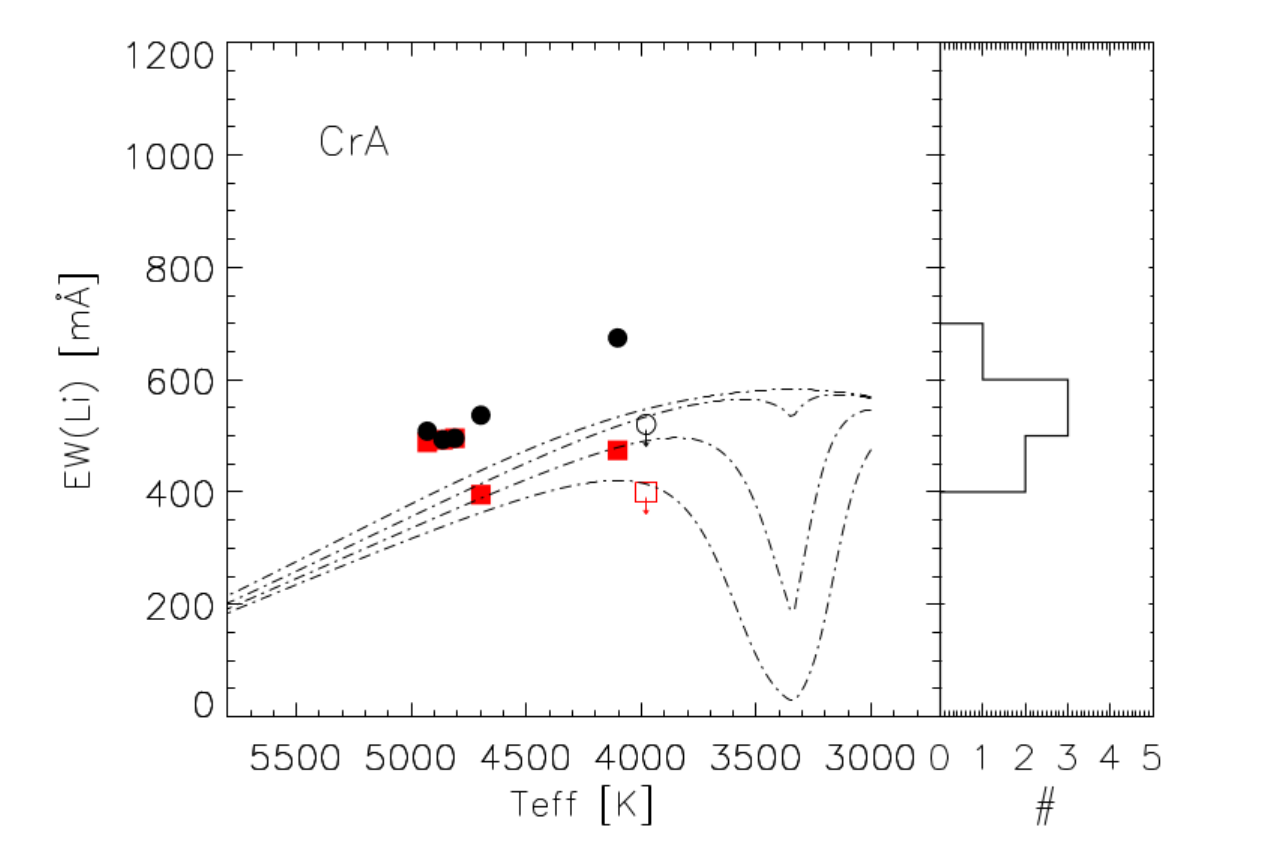}
%\par\small (f) ChaI
%		  \caption{ChaI}
%		  \label{fig:subfig4}
	       \end{minipage}
	\caption{Multi-panel overview of the lithium equivalent widths for the eight SFRs (Cha I, $\eta$ Cha, Lupus, Taurus, Orion OB1a, Orion OB1b, $\sigma$\,Ori and CrA) in our sample. For each region, the left sub-panel shows lithium equivalent width versus $T_{\rm eff}$. 
   The  red squares  represent  the EW  corrected for blending with the iron line ($EW_{\rm Li}^{\rm Fe}$),  the  black dots represent the equivalent width after further correction for spectral veiling ($EW_{\rm Li}^{\rm veil+Fe}$). K-type ( T $\geq$ 4000 K) and  M-type (T < 4000 K) stars  are denoted  by filled and open symbols, respectively. Arrows indicate upper limits due to the unresolved contribution of the FeI $\lambda$6707.4 line. The empirical model isochrones by \cite{jef23} at 5, 10, 15, 20 Myr are over plotted. The right sub-panels display the corresponding $EW_{\rm Li}^{\rm veil+Fe}$ distribution.}
	\label{ewli}
\end{figure*}

 Individual measurements for $EW_{\rm Li}^{\rm raw}$ and $EW_{\rm Li}^{\rm veil+Fe}$ (corrected for both veiling and iron blending for K-type stars), or $EW_{\rm Li}^{\rm veil}$ ( corrected only for veiling for M-type stars), along with the corresponding veiling coefficients are provided in the Appendix (\ref{tabewli1}, \ref{tabewli2}, \ref{tabewli3}). The results for the eight star-forming regions are displayed in Fig. \ref{ewli}.  Each panel compares the lithium equivalent widths corrected for both veiling and iron blending ($EW_{\rm Li}^{\rm veil+Fe}$, black dots) with those corrected only for iron blending ($EW_{\rm Li}^{\rm Fe}$, red squares) as a function of $T_{\rm eff}$ (left sub-panels).
 The corresponding $EW_{\rm Li}^{\rm veil+Fe}$ distributions are provided in the right sub-panels.
 We denoted with filled and empty symbols the K-type and M-type stars, respectively.Dash-dotted lines  represent a set of model isochrones in the 5-20 Myr range, as derived by \cite{jef23} through the fitting of the Gaia-ESO Survey training data. A higher $r$ value leads to a larger difference between $EW_{\rm Li}^{\rm veil+Fe}$ and $EW_{\rm Li}^{\rm Fe}$, and in some cases the difference is $\sim$ 800 m\AA\, as for SO\,1153  and VZ\,Cha.
Differences between the $EW_{\rm Li}$ values  obtained from spectra of the same target at different epochs are discussed in Sec. \ref{variation}.
Since M-type stars cannot be corrected for the iron blending, their  EWs represent  upper limits, as they also include  the contribution of the Fe line.
Most of the corrected equivalent width values range between 400 and 800 m\AA, in some case rising up to $\sim$ 1000 m\AA.
Some targets in our sample are spectroscopic binaries (SBs). For the single-lined systems (SB1), the effect of binarity on equivalent width of lithium  is negligible. For double-lined spectroscopic binaries (SB2), this effect is within the measurement uncertainties. Therefore, the presence of these binary stars does not impact our final results \citep{frasca18}.
 Below, we provide specific comments on the $EW_{\rm Li}$ for each star-forming region:

 %\begin{figure*}
 %  \centering
 %  \includegraphics[width=\textwidth]{EWLi_Teff_arrow.pdf}   
 %  \caption{Lithium equivalent width versus effective temperature for the studied star-forming regions (s-Ori, Lupus, $\eta$ Cha, Orion OB1, Taurus, Cha I and CrA).
 %  The  red squares  represent  the EW  corrected for blending with the iron line ($EW_{\rm Li}^{\rm Fe}$).  The  black dots represent the equivalent width after further correction for spectral veiling ($EW_{\rm Li}^{\rm veil+Fe}$). K-type ( T $\geq$ 4000 K) and  M-type (T < 4000 K) stars  are denoted  by filled and open symbols, respectively. Arrows indicate upper limits due to the unresolved contribution of the FeI $\lambda$6707.4 line. }
  %  \label{ewli}
 %   \end{figure*}

\begin{itemize}    
\item \textit{Chamaelon I}: We  measured $EW_{\rm Li}$ in 49 spectra of 15 targets. 
 The corrected $EW_{\rm Li}$ values range from approximately 300 to 1000 m\AA, in agreement, within the uncertainties, with \cite{gutierrez24}. 
The peak of the distribution is around 650 m\AA.
The highest values correspond to spectra of VZ\,Cha, taken with UVES across three different epochs.
Conversely, the target with the lowest values is Sz\,19 observed with ESPRESSO over three different epochs.

\item \textit{$\eta$ Chamaeleon}:
We  measured $EW_{\rm Li}$ in 18 spectra of 7 targets.
The raw values of $EW_{\rm Li}$ are in agreement within the uncertainties with \cite{mentuch08}.
The trend of the $EW_{\rm Li}$ with the temperature is similar to that observed in Cha I, though without the extreme low and high values.
Specifically, the measurements range from approximately  445 to 780 m\AA.
The distribution peaks at about 650 m\AA.

\item \textit{Lupus}:
The sample is composed by 108 spectra of 30 targets. The trend of the $EW_{\rm Li}$ with the temperature is consistent with what we observed in Cha I and $\eta$ Cha. However, the targets with the highest $EW_{\rm Li}$ in Lupus  are at lower temperature compared to those in Cha I.
These targets include Sz\,84, Sz\,72 and Sz\,104.

Our values are slightly higher than those reported by \cite{biazzo17}; our distribution peaks at $\sim$ 650 m\AA\ compared to their  value of 560 m\AA. This discrepancy stems primarily from the different composition of the two samples. Specifically, the sample analyzed by  \citealt{biazzo17} is richer of stars in the Li-depletion region ($T_{\rm eff}$ < 4000 K). An additional factor is the  difference in the  veiling values adopted in the two works. Despite these discrepancies, the overall trends remain consistent,  and  the $EW_{\rm Li}^{\rm veil+Fe}$  values for the targets in common  agree within the uncertainties.
%We did not find any lithium-depleted members in our sample. The lowest values of $EW_{Li}$ were measured for  Sz69 (in agreement with \cite{biazzo17}) and SS161344, both at  low temperature with spectra taken by X-Shooter.  RY Lupus also showed low $EW_{Li}$ but at high temperature, based on  five spectra taken by ESPRESSO.

\item \textit{Taurus}:
We analyzed  31 spectra of 10 targets.
The trend of the $EW_{\rm Li}$ with the temperature is consistent with what observed in other clusters, with the distribution peaking at $\sim$ 650 m\AA.
The corrected $EW_{\rm Li}$  values range from 375 to 900 mÅ. The targets exhibiting the highest $EW_{\rm Li}$ values are AAtau, DKTauB, and LkCa4.

\item \textit{$\sigma$-Orionis}:
The sample is composed by 13 spectra of 3 targets.
The general trend observed in the other star-forming regions is maintained here, with $EW_{\rm Li}^{\rm veil+Fe}$  values ranging  from 490 to 1020 m\AA. The distribution's peak  is $\sim$ 650 m\AA.
 The highest values correspond to SO\,1153, observed by ESPRESSO, mainly due to the very high veiling ($\sim$ 5.0). The raw $EW_{\rm Li}$ is  only about 150-170 m\AA.

\item \textit{Orion OB1a}:
The sample is composed by 8 spectra of 2 targets. These are M-type stars of very similar $T_{\rm eff}$. $EW_{\rm Li}^{\rm veil+Fe}$  values range  from 400 and 700 m\AA, with a peak of distribution at $\sim$ 650 m\AA.

\item \textit{Orion OB1b}:
The sample comprises  25 spectra of 7 targets.
This region shows the same general trend as the other clusters, but it lacks stars with temperatures higher than $\sim$ 4500 K.
The  corrected $EW_{\rm Li}$  values in the sample range between 310 and 820 m\AA, in agreement with \cite{piscarreta25}.
CVSO-90 observed with X-Shooter has the lowest $EW_{\rm Li}$ value.
Conversely, the three spectra of CVSO-176, acquired with UVES, show the highest value.
The peak of the distribution is about 650 m\AA, similar to what is observed for Orion OB1a.

\item \textit{Corona Australis}: We measured only only 6 spectra of 2 targets. The $EW_{\rm Li}^{\rm veil+Fe}$ values range between about 490 m\AA\, and 679 m\AA, consistently with the other SFRs analyzed.

\end{itemize}

To quantify the impact of the veiling correction for an accurate measurement of the lithium equivalent width,  Figure 2 shows the normalized difference  ($EW_{\rm Li}^{veil}$-$EW_{\rm Li}^{\rm raw}$)/$EW_{\rm Li}^{\rm veil}$, as a function of $T_{\rm eff}$. To also investigate a possible dependence on instrumental resolution, we distinguish between high-resolution (ESPRESSO + UVES) and medium-resolution (XS)  data, represented by solid black circles and filled red triangles, respectively. 
We find no significant trend between the normalized difference and the resolution. As expected,  the influence of veiling decreases for $T_{\rm eff}$ > 5000 K,  because, as the temperature of the accretion spots and the stellar temperatures become more similar, the line veiling becomes more negligible \citep{2004ApJ...617..406M}.
Overall, the normalized differences are almost uniformly distributed across the diagram.
 The average variation  in  $EW_{\rm Li}$ due to the veiling is about 30-40\%, reaching  up to  80\% in specific cases (e.g. SO\,1153 and  VZ\,Cha).
 
 These results emphasize that neglecting the veiling correction in young, active stars leads to a substantial underestimation of lithium abundances, which could result in an incorrect interpretation of stellar ages and lithium depletion history.

 \begin{figure}
  \centering
   \includegraphics[width=\hsize]{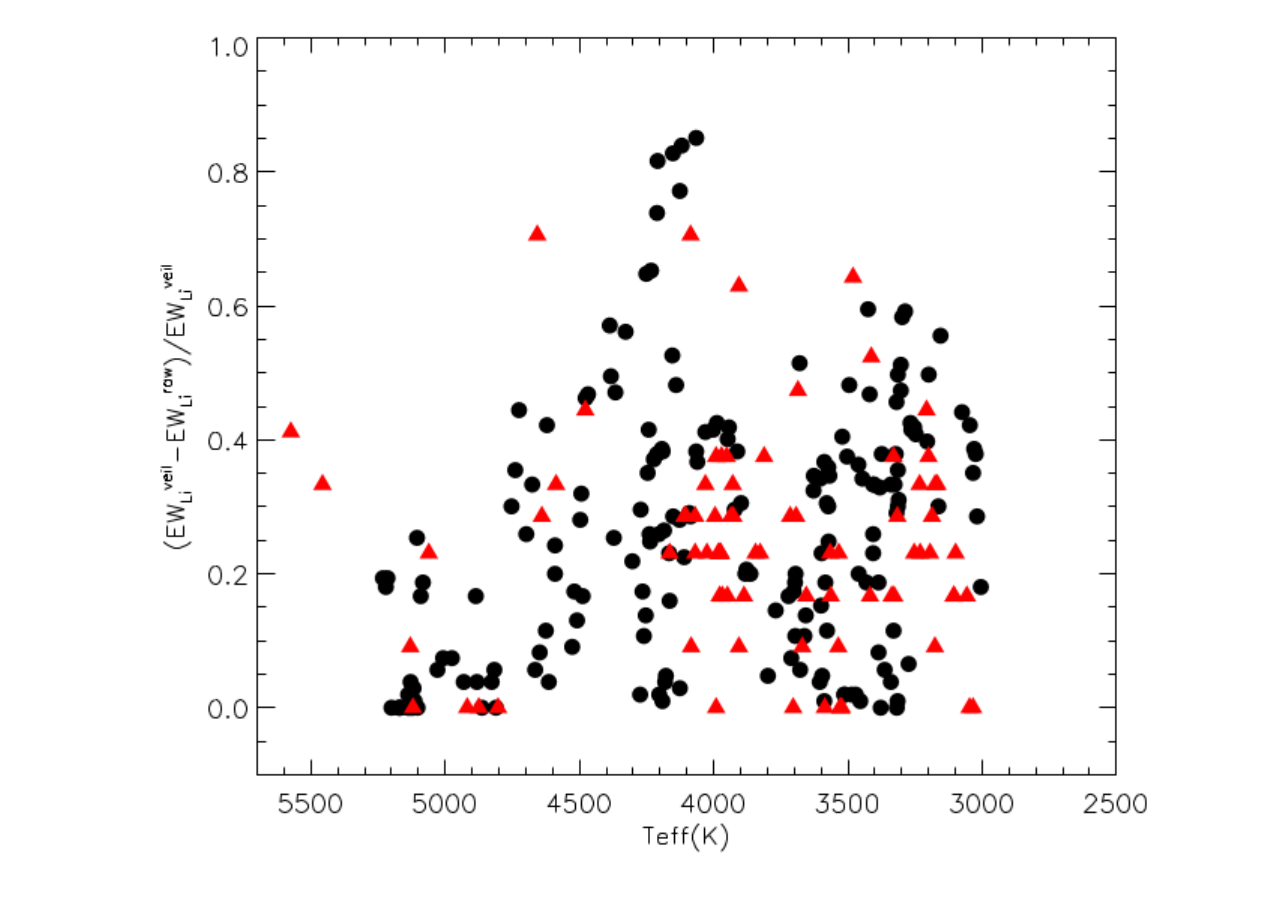}   
   \vspace{-1cm}
  \caption{($EW_{\rm Li}^{veil}$-$EW_{\rm Li}^{\rm raw}$)/$EW_{\rm Li}^{\rm veil}$ as a function of $T_{\rm eff}$. The filled black dots and the  filled red triangles represent high (ESPRESSO + UVES) and medium (X-Shooter) resolution data, respectively.}
\label{rel_err}
\end{figure}

\subsection{Lithium variation}
\label{variation}

Chromospherich activity affects the equivalent width of absorption lines in stellar spectra (\citealt{spina20} and reference within).
Magnetic fields impact spectral lines both directly, through the Zeeman effect, and indirectly, by altering the atmospheric thermodynamic structure (e.g. \citealt{borrero08,moore15,sh16}).
During a star's activity cycle, the intensity of the magnetic fields in the stellar atmosphere  and the fraction of the stellar surface covered by cool spots vary \citep{babcock1959, schwabe}.

Similarly, the accretion process is  intrinsically highly variable \citep{joy45,hartigan91}, on timescales ranging from minutes to years (\citealt{costigan14, Nguyen09}). 
Short-term variability may result from the deformation of magnetic field lines due to differential rotation (e.g. \citealt{goodson97}).
Recent studies have also demonstrated  veiling variability associated with changes in the accretion rate in low-mass PMS stars (e.g. \citealt{bouvier03,costigan14,manara21}). Consequently, both chromospheric activity and accretion variability, may play a role in  the observed variations of  the equivalent width of the absorption lines.
%Recently, \cite{spina20}, analyzing about 21000 spectra of 211 Sun-like stars, found that the EWs of the spectroscopic lines increase with the chromospheric activity along the stellar cycle.
 %Moreover they observed a decrease of $T_{\rm eff}$  with chromospheric activity, while no significant variation in $\log g$.

Our sample  which consists of multi-epoch observations, provide a unique opportunity to investigate a  potential variation  in Li abundance across the epochs. The time intervals between observations are generally a few days, except for cases where observations were repeated after longer periods (see  Tables \ref{tabewli1}, \ref{tabewli2}, \ref{tabewli3}).

\begin{figure}
   \centering
   \includegraphics[width=\hsize]{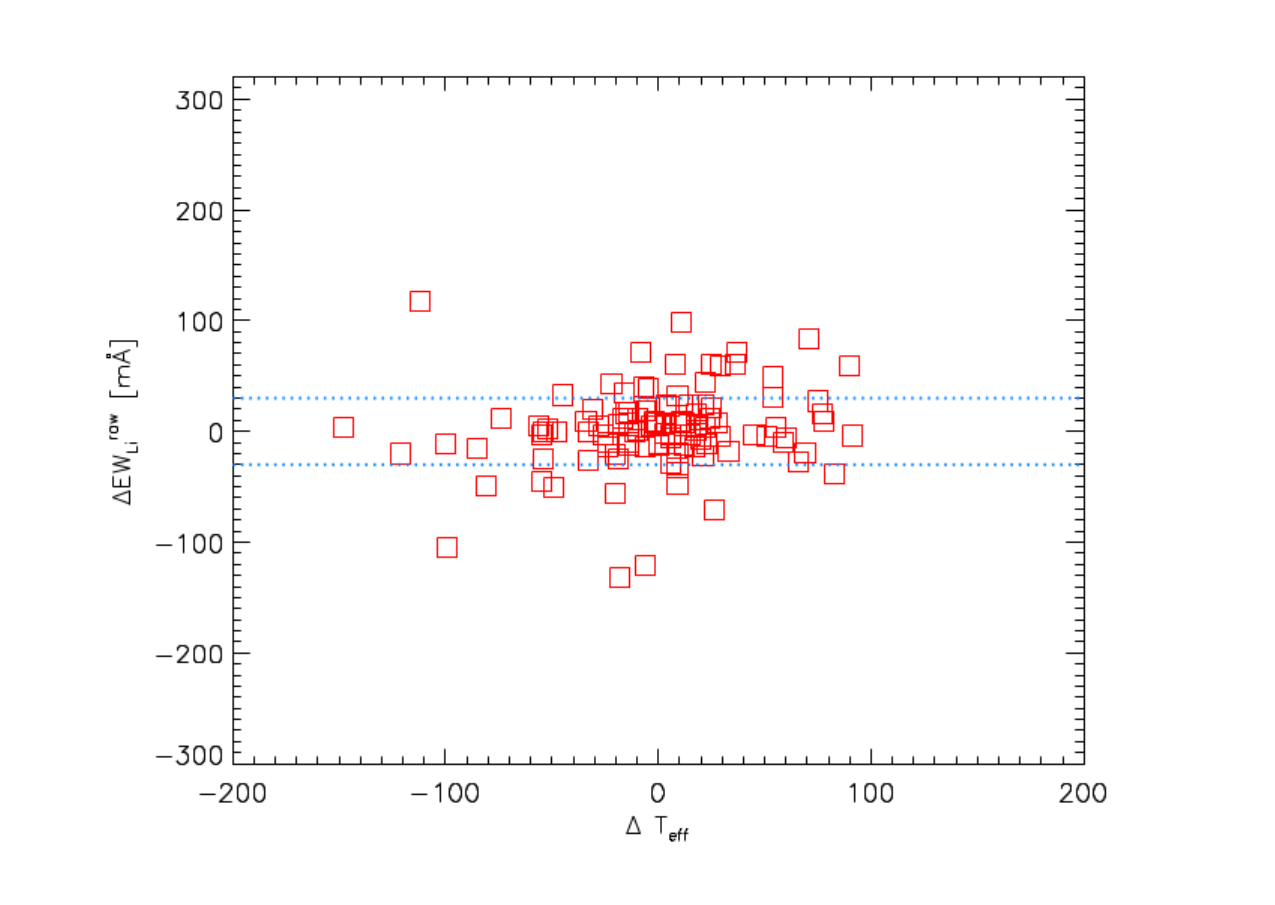} 
   \vspace{-1cm}
   \caption{Variation of $EW_{\rm Li}^{\rm raw}$ across the observing epochs as a function of $T_{\rm eff}$  variations, for ESPRESSO and UVES data. The blue dotted lines represent 3 $\sigma$ values.}
    \label{var2}
    \end{figure}

\begin{figure}
   \centering
   \includegraphics[width=\hsize]{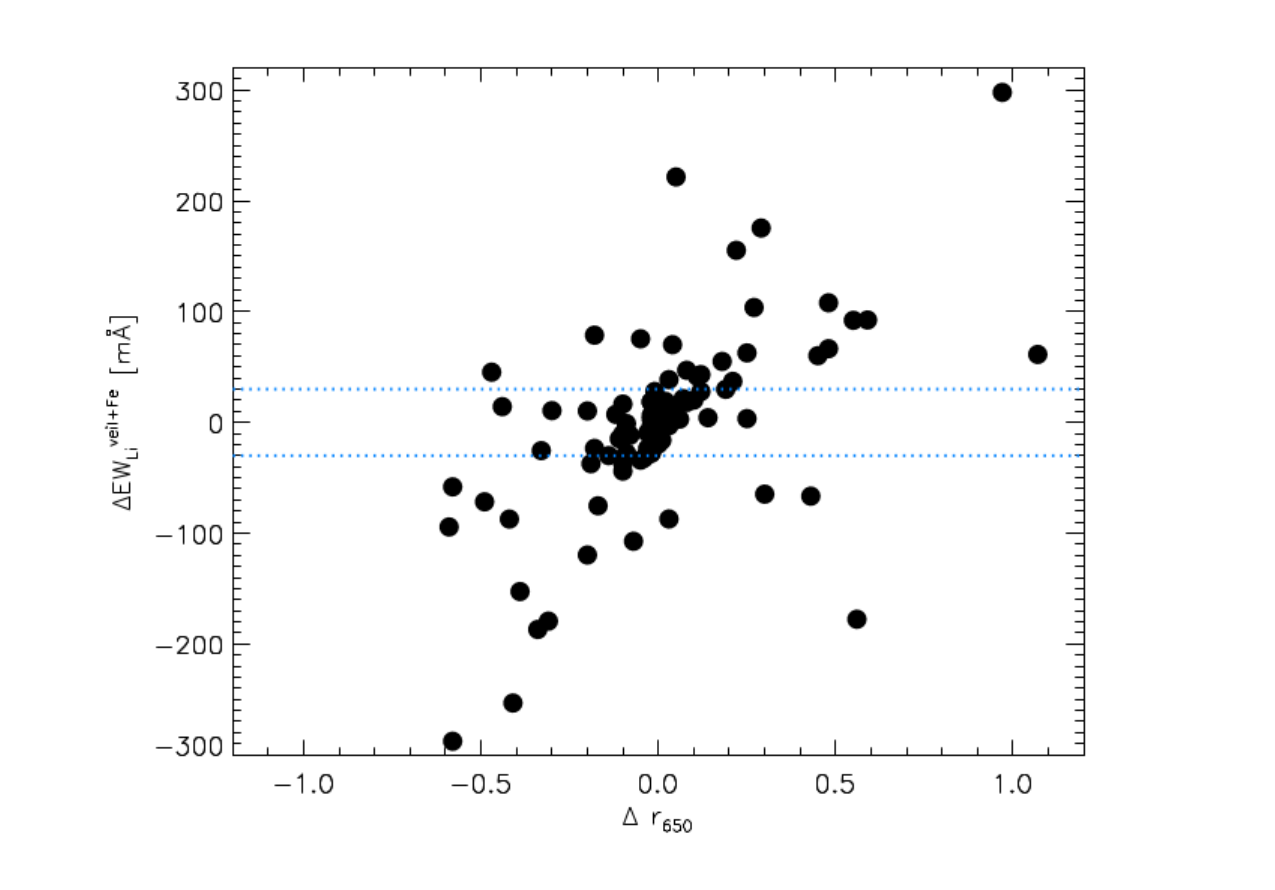}   
   \vspace{-1cm}
   \caption{Variation of $EW_{\rm Li}^{\rm veil+Fe}$ across the observing epochs as a function of veiling variations, for ESPRESSO and UVES data.The blue dotted lines represent 3 $\sigma$ values}
    \label{var1}
    \end{figure}

As first step, we analyzed the variations in the raw lithium equivalent width to investigate the variability associated with chromospheric activity.
We found that 26  targets  exhibited $EW_{\rm Li}^{raw}$ changes more than 3$\sigma$ at least once in a given time interval, of which 10 in both intervals (i.e., between the first and second, and between the second and third epochs).
The mean value of $\Delta EW_{\rm Li}^{\rm raw}$ of these 26 targets is about 62 $\pm$ 28 m\AA.
As shown in Fig. \ref{var2},  these variations are not driven by changes in $T_{\rm eff}$. 
The observed temperature variations are small (less than 150 K) and remain within the uncertainties of the $T_{\rm eff}$ determination. Subsequently, we examined the potential variations  in   $EW_{\rm Li}$ linked to the accretion process.
Fig. \ref{var1} shows the variation of $EW_{\rm Li}^{\rm veil+Fe}$ across the observing epochs as a function of the veiling difference  $\Delta r_{650}$ for each target.

We identify 30 sources showing $\Delta EW_{\rm Li}^{\rm veil+Fe}$ greater than 3$\sigma$,  15 of which in both temporal intervals.
For these specific targets,  the mean  $\Delta EW_{\rm Li}^{\rm veil+Fe}$  is significantly higher than  the values derived from raw measurements, i.e. 92.2 $\pm$ 65.9 m\AA. Interestingly, only 9 of these 30 sources overlap with the "raw" sample; this occurs because veiling variations occasionally mask the intrinsic $\Delta EW_{\rm Li}^{raw}$, while in others instances, they amply it. 
Moreover, Fig. \ref{var1} shows  a clear positive correlation  between $\Delta EW_{\rm Li}^{\rm veil+Fe}$ and $\Delta r_{650}$: as the variation in veiling increases, the variation in equivalent width increases accordingly.
This result is in agreement with recent works, such as \cite{stout00}, that suggest that higher accretion rates, and thus higher veiling, produce larger Li abundances because fresh material, with primordial levels of Li, is incorporated onto the star's surface.

In any case, this analysis reinforces the critical necessity of accounting for veiling to achieve accurate lithium equivalent width determinations.

%The same tables show that the variation in temperature between the epochs is within the margin of error, while the contribution of veiling can vary considerably.
%We therefore analyzed the variation in the equivalent width of lithium and veiling between the epochs.

 \subsection{Abundance of $^{7}\rm Li$}
We estimated lithium abundances ($A{\rm (Li)}$)\footnote{In the usual notation $A{\rm (Li)}$ = $\log{N{\rm (Li)}}/N{\rm (H)}$ +12} from the measured equivalent widths, using the atmospheric parameters ($T_{\rm eff}$, $log g$, $vsin i$)  cited above,assuming a typical  microturbulent velocity of 1.0 km/s. Since it was not possible  to determine [Fe/H] individually for each region (see Sec. \ref{feba}) with our data, we adopted a solar iron abundance  for all targets in order to ensure a consistent methodology across the entire dataset.
 This assumption is appropriate for nearby star-forming regions \citep{randich22}. Moreover, a variation in [Fe/H] of $\pm$ 0.1 dex corresponds to an uncertainty of around $\pm$ 0.01 dex in $A{\rm (Li)}$. This contribution is negligible compared to the total uncertainty in the $A(\text{Li})$ determination and, consequently, does not affect the results. Under these assumptions, we applied the LTE curves of growth (COGs)  developed by \cite{franciosini},  which are differentiated for K and M stars.
The valid  range for $A{\rm (Li)}$  is between [-1.0, 4.0] dex, values falling outside this range are determined through extrapolations.
The NLTE (Non-Local Thermodynamic Equilibrium)  effects to the $A{\rm (Li)}$ were considered, using the correction values from   \cite{lind09} available  for K stars in the range [-0.30,4.20] dex.
For stars whose  final extrapolated $A{\rm (Li)}$ value exceeds 4.0 dex, the lithium abundance was set to 4.0 dex, which represents a conservative lower limit for our dataset.

  The uncertainties  in the stellar parameters and to the measurement of $EW_{\rm Li}$ represent the main  
   sources of error in $A{\rm (Li)}$.
 The  total uncertainty for $A{\rm (Li)}$ was estimated
 taking into account every source of uncertainty ($T_{\rm eff}$, $\log g$, $EW(\rm Li)$, $\xi$, $vsin i$, [Fe/H]) and by combining them in quadrature.
 The overall uncertainties typically fall within the range of 0.1-0.2 dex, with the uncertainty in effective temperature being the main contributor.
Additionally, an uncertainty of $\sim$ 0.1 in the veiling factor introduces  an error of about 0.2 dex in $A{\rm (Li)}$.

% \begin{figure*}
%   \centering
%   \includegraphics[width=\textwidth]{Abu_NLTE_tot.pdf}
   %%%\includegraphics{empty.eps}
   %%%\includegraphics{empty.eps}
%   \caption{
%   Abundance of lithium corrected for the NLTE effect, versus effective temperature.  The open red squares and the the filled black dots  represent  the $A{\rm (Li)}$ derived from $EW_{\rm Li}^{\rm Fe}$   and $EW_{\rm Li}^{\rm veil+Fe}$, respectively. The lithium isochrones by \citealt{baraffe17} in the 2-20 Myr range are overlaid with dot-dashed lines. Arrows refer to lower or upper limits.}
%              \label{nlte}
%    \end{figure*}

\begin{figure*}
      \centering
	   \begin{minipage}{0.45\linewidth}
		\includegraphics[width=\linewidth]{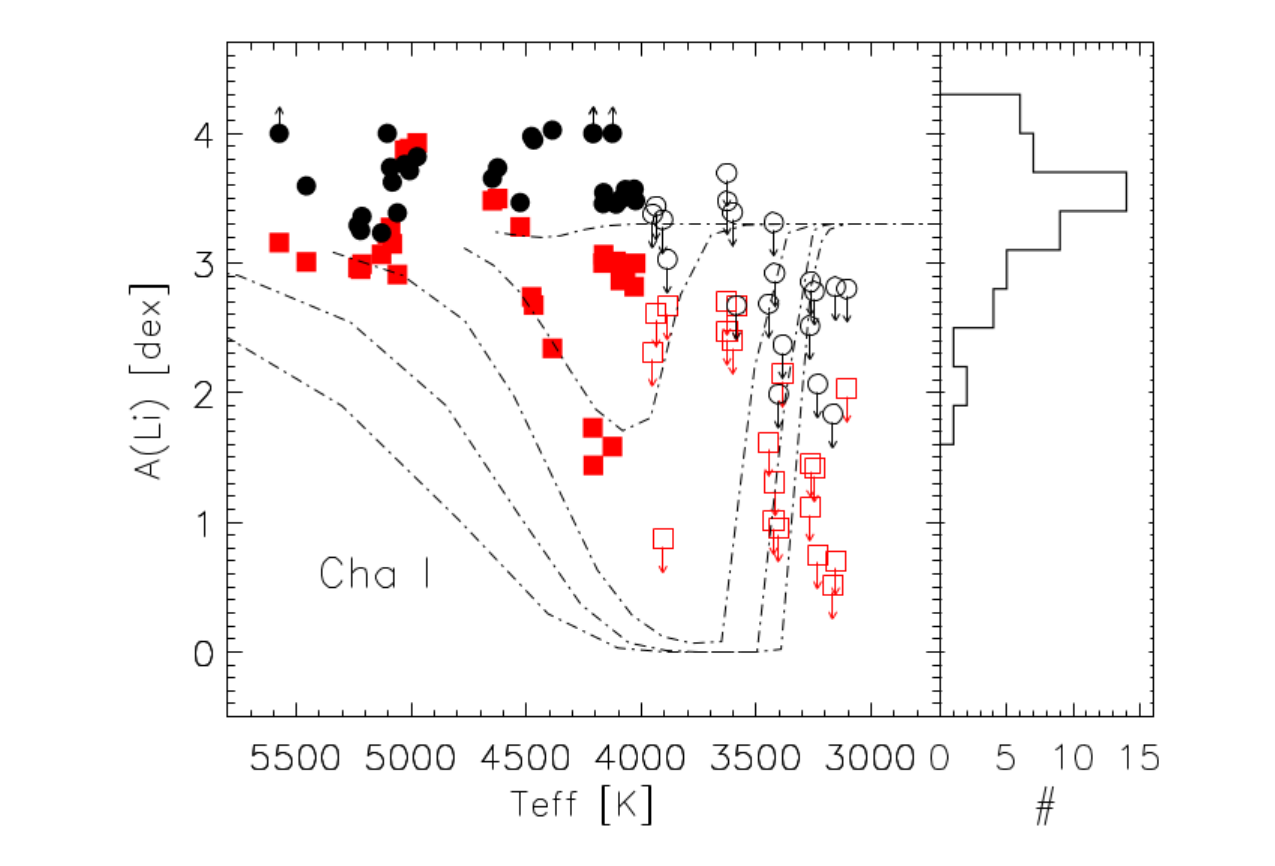}
%        \par\small (a) sOri
%		\caption{sOri}
%		\label{fig:subfig1}
	   \end{minipage}
	   \begin{minipage}{0.45\linewidth}
		\includegraphics[width=\linewidth]{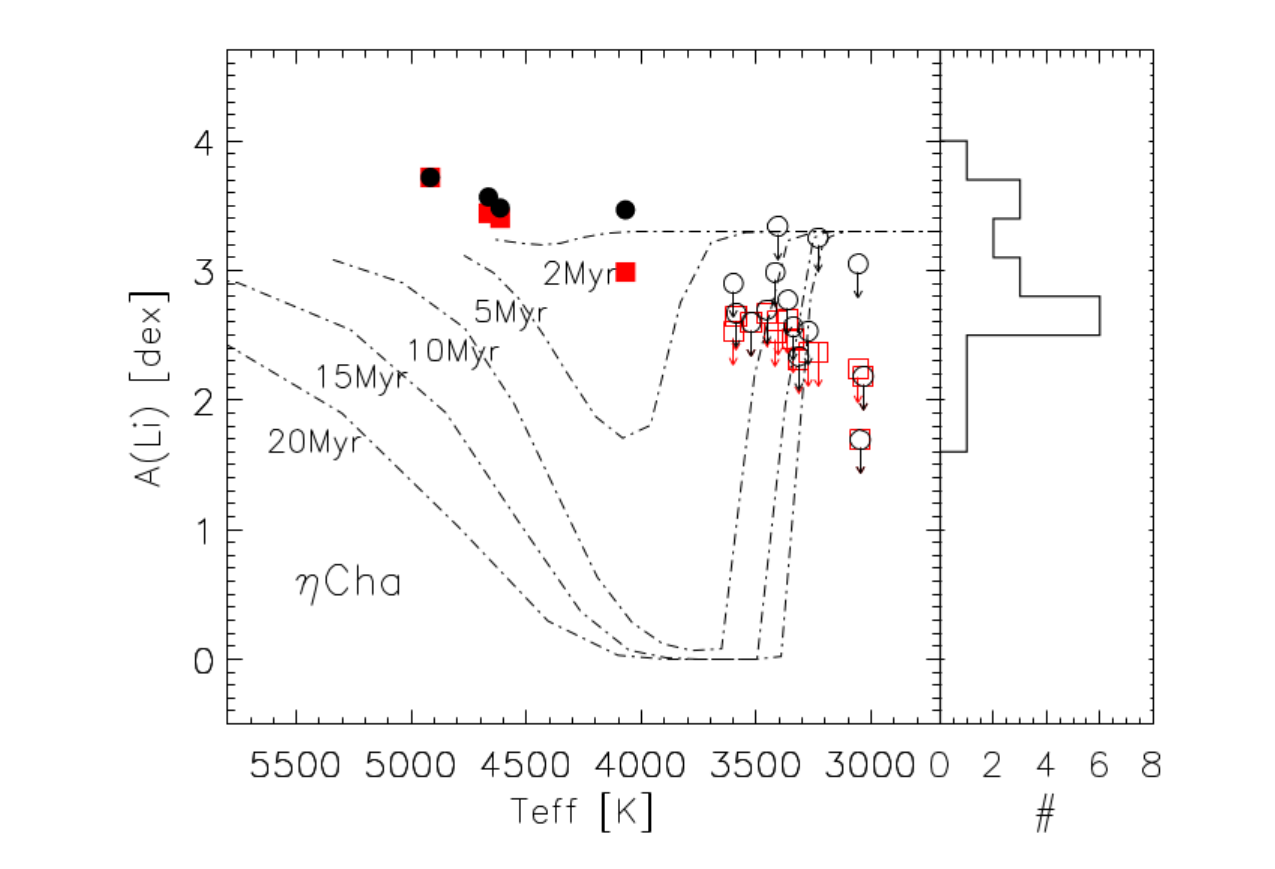}
%\par\small (b) Orion OB1
%		\caption{Orion OB1}
%		\label{fig:subfig2}
	    \end{minipage}
	\vfill
	     \begin{minipage}{0.45\linewidth}
		 \includegraphics[width=\linewidth]{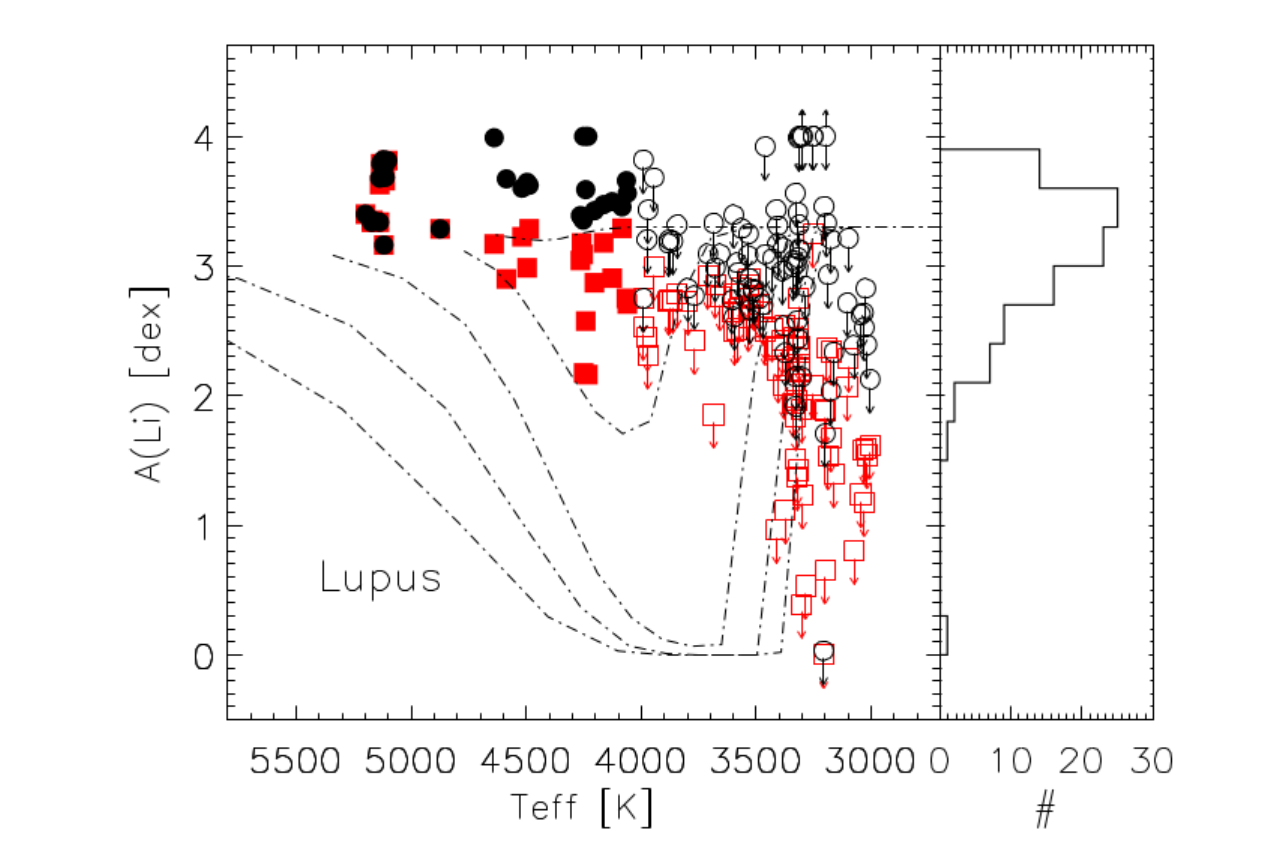}
%\par\small (c) Lupus
%		 \caption{Lupus}
%		 \label{fig:subfig3}
	      \end{minipage}
	       \begin{minipage}{0.45\linewidth}
		  \includegraphics[width=\linewidth]{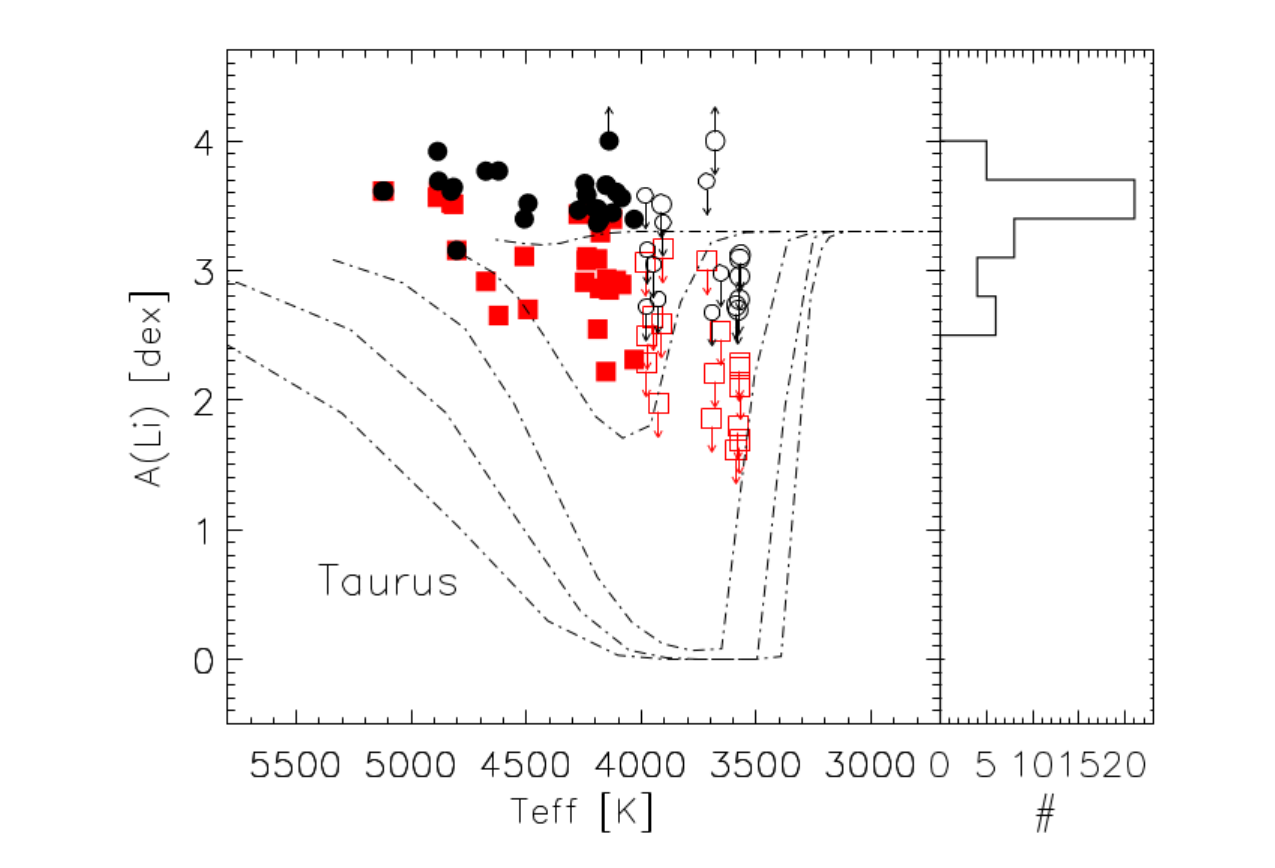}
%        \par\small (d) Taurus
%		  \caption{Taurus}
%		  \label{fig:subfig4}
	       \end{minipage}
    \vfill   
	       \begin{minipage}{0.45\linewidth}
		  \includegraphics[width=\linewidth]{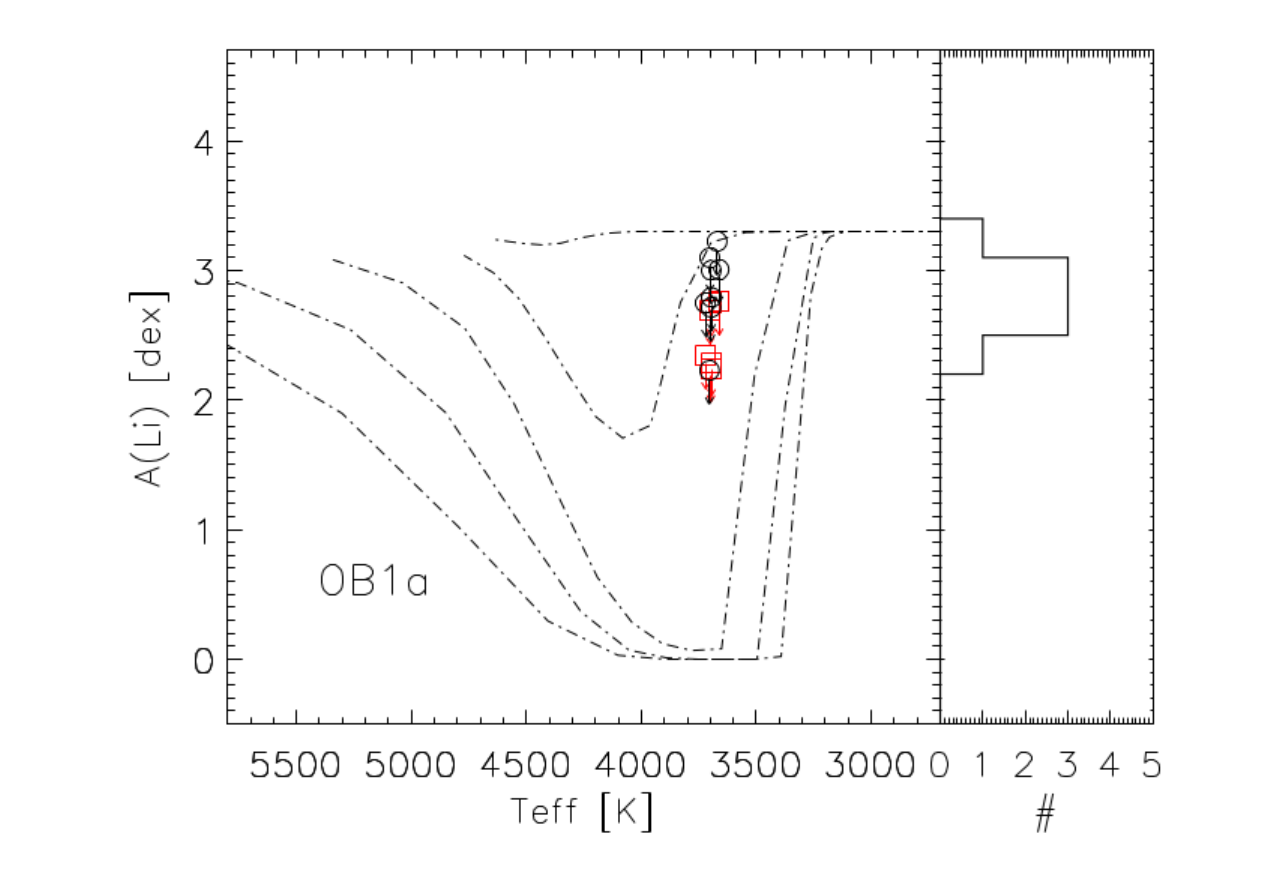}
%\par\small (f) ChaI
%		  \caption{ChaI}
%		  \label{fig:subfig4}
	       \end{minipage}          
        \begin{minipage}{0.45\linewidth}
		  \includegraphics[width=\linewidth]{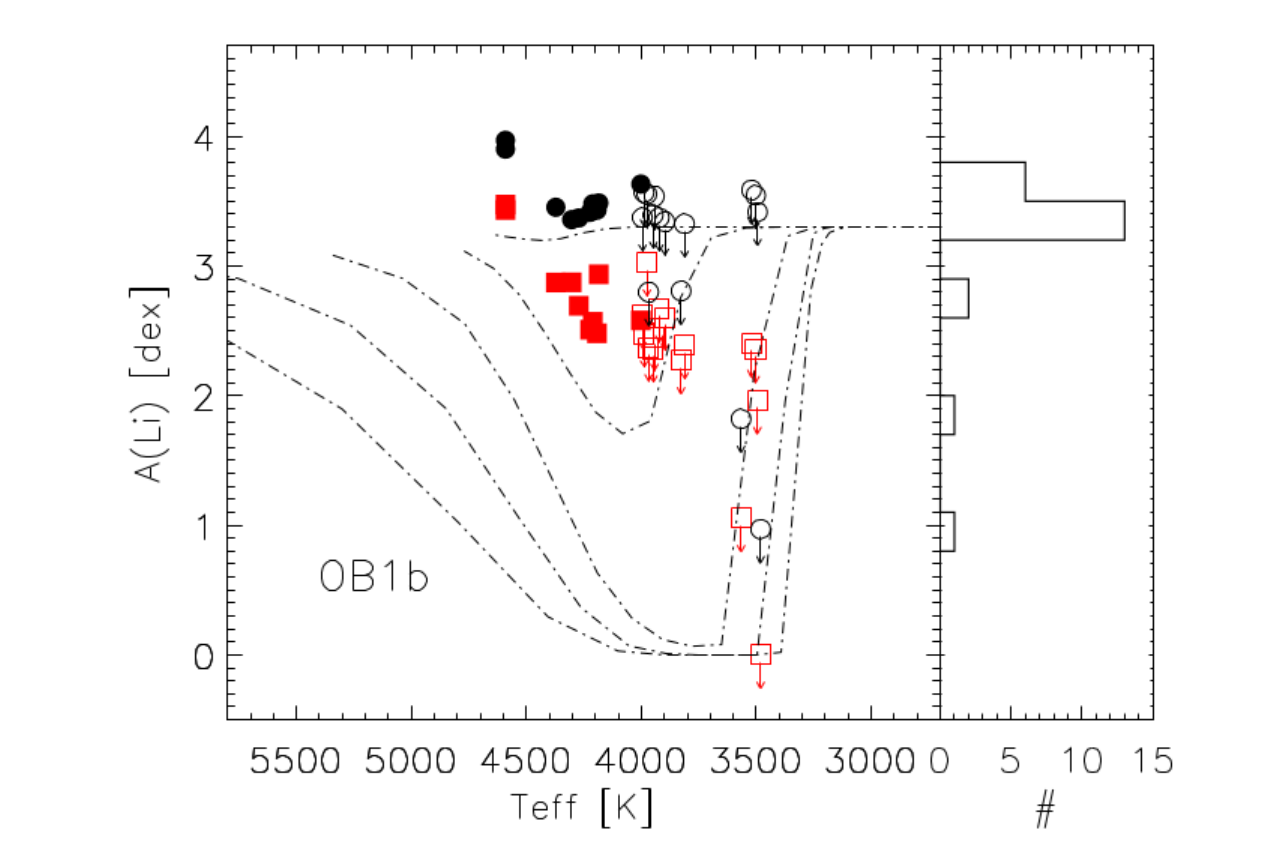}
          \end{minipage}
           \vfill
               \begin{minipage}{0.45\linewidth}
		 \includegraphics[width=\linewidth]{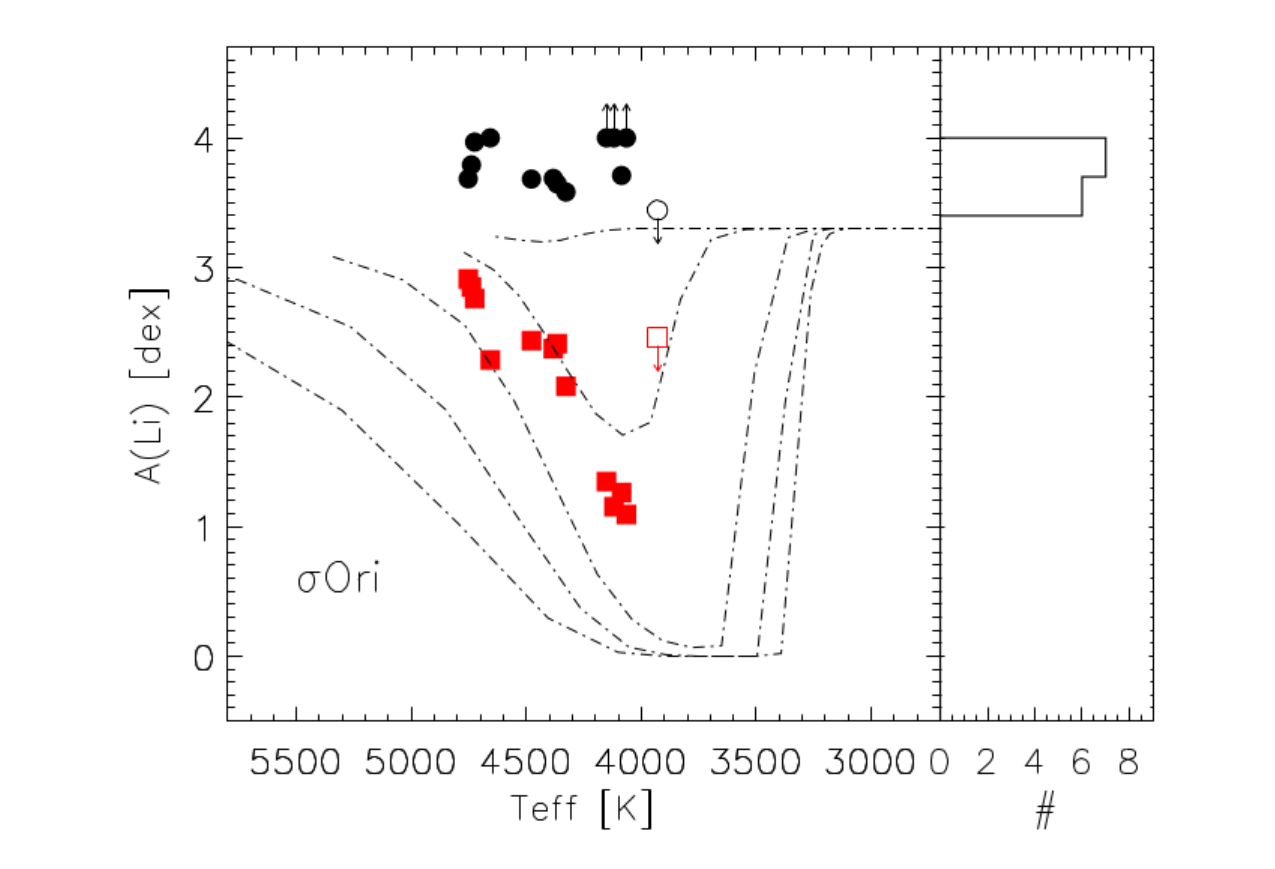}
%         \par\small (e) eCha 
%		 \caption{eCha}
%		 \label{fig:subfig3}
	      \end{minipage}
    \begin{minipage}{0.45\linewidth}
		  \includegraphics[width=\linewidth]{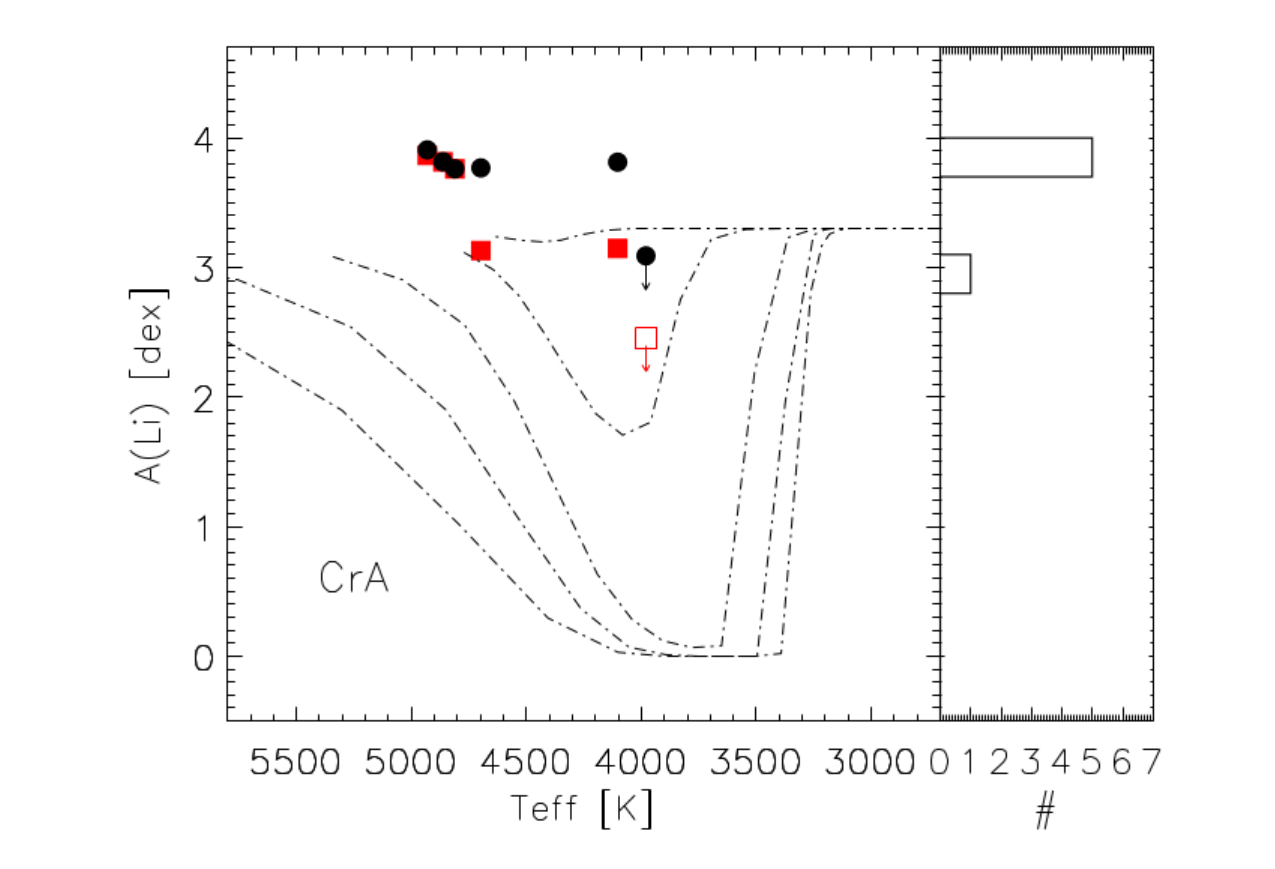}
%\par\small (f) ChaI
%		  \caption{ChaI}
%		  \label{fig:subfig4}
	       \end{minipage}
	\caption{
     Multi-panel overview of the lithium abundance for the eight SFRs (Cha I, $\eta$ Cha, Lupus, Taurus, Orion OB1a, Orion OB1b, $\sigma$\,Ori and CrA). For each region, the left sub-panel shows NLTE-corrected lithium abundance as a function of effective temperature. Red squares and black dots represent the $A{\rm (Li)}$ values derived from $EW_{\rm Li}^{\rm Fe}$ and $EW_{\rm Li}^{\rm veil+Fe}$, respectively. K-type and M-type stars are plotted with filled and open symbols, respectively. The lithium isochrones by \cite{baraffe15} in the 2-20 Myr range are overlaid with dot-dashed lines. Arrows refer to lower (stars whose
final $A{\rm (Li)}$ value exceeds 4.0 dex, see text) or upper (due to the unresolved contribution of the FeI $\lambda$ 6707.4  line) limits. Each right sub-panel displays the histogram of the corresponding $A{\rm (Li)}$ (corrected for veiling) distribution.} 
	\label{nlte}
\end{figure*}

The results or the eight SFRs are shown in Fig. \ref{nlte}. Each panel provides a comparison between  the $A{\rm (Li)}$, corrected for NLTE effect, determined from with $EW_{\rm Li}^{\rm veil+Fe}$ (black dots)  and $EW_{\rm Li}^{\rm Fe}$ (red squares) values as a function of $T_{\rm eff}$ (left), and the $A{\rm (Li)}$ (corrected for veiling)   distribution (right).
%Figure\,\ref{nlte} displays the $A{\rm (Li)}$ values, corrected for NLTE effect, as a function of the effective temperature, for each stellar association. 
 M-type stars ($T_{\rm eff}$ $\lesssim$ 4000 K) were not corrected for NLTE effect, thus the $A{\rm (Li)}$ values shown in the plot represent upper limits, as NLTE corrections tend to decrease $A{\rm (Li)}$. %Black dots and red  squares represent the $A{\rm (Li)}$ determined from $EW_{\rm Li}^{\rm veil+Fe}$ and $EW_{\rm Li}^{\rm Fe}$, respectively.
 K-type ( T $\geq$ 4000 K) and M-type (T $<$ 4000 K) stars are denoted by filled and open symbols, respectively.
Targets where the $EW_{\rm Li}$ exceeded the valid range of the COGs (as a function of $T_{\rm eff}$ and $\log{g}$), are indicated with  lower limits.
Isochrones from \cite{baraffe15} in the 2-20 Myr range are over plotted as dot-dashed lines.

Similar to its effect on $EW_{\rm Li}$,  veiling can  drastically alter the $A{\rm (Li)}$ of the targets and, consequently, the age estimate of the cluster.
Indeed, as shown in Fig. \ref{nlte}, when veiling is neglected, most sources in each cluster would result to have  $A{\rm (Li)}$ values less then 3 dex,  placing the cluster ages between  5 - 20 Myr.
However, the veiling corrected  $EW_{\rm Li}$ measurements yield  $A{\rm (Li)}$ values between 3 and 4 dex. 
Consequently, the resulting cluster ages, estimated by comparing the data to the overlaid isochrones, are constrained to be less than 5 Myr. These younger age limits are well in line with previously published values (e.g \citealt{spina14} and reference within).
The effect of veiling on age determination will be explored in depth in the next session.
The peak of the $A{\rm (Li)}$ distribution  is about 3.5-3.6 dex for Cha I, Lupus and Taurus regions, the regions with targets spanning  the wider range in temperature. This value is slightly higher than the standard expected initial abundance of $\sim$ 3.3 dex, but remains consistent with the expected value when considering our average uncertainty of $\sim 0.3$ dex.
Furthermore, our results align with recent studies that have identified a population of Li-rich  stars with abundances exceeding the meteoritic limit, ranging from 3.5 to 4.5 dex (e.g.\citealt{deli02,yan22}). Recently, \cite{zhou25}, reported NLTE $A{\rm (Li)}$ values between 3.3 and 4.6 dex for a sample of  62 unevolved stars, further supporting the consistency of our findings with the current literature.\\
\hspace*{0.5cm}$\eta$\,Cha and Orion Ob1a exhibit  peaks at lower values, at 2.6 dex and 3.0 dex, respectively; this might be because the sample is biased towards cool stars.
Conversely, CrA and $\sigma$\,Ori show the highest peak at $\sim$ 3.8 dex likely due to a sample bias toward warmer stars.
The association Orion OB1b has the peak of the distribution at $\sim$ 3.3 dex.\\
\hspace*{0.5cm}A notable spread in $A{\rm (Li)}$ is observed in each cluster for stars cooler than 3500 K, being particularly evident in the Cha I, Lupus and Orion OB1b associations. This region in $T_{\rm eff}$ is populated by  fully convective low-mass stars, which are depleting $^{7}\rm Li$ at the base of convective zone.
Adopting an $A{\rm (Li)}$ threshold of  2.0 dex to define $^{7}\rm Li$-depleted targets, our analysis identifies seven  sources falling below this limit: Sz\,10 (Cha I), Sz\,104, Sz\,69, SS61344.1-373646 (Lupus), CVSO-176, CVSO-90 (Orion OB1b) and ECHAJ0844.2-7833 ($\eta$\,Cha), all observed with X-Shooter. 
The $A{\rm (Li)}$ values measured for Sz104, Sz10 and CVSO-176 are only marginally below the 2.0 dex depletion threshold. Their abundance uncertainties are large enough to potentially place these three targets within  the non-depleted regime of the plot.
This ambiguity is further supported by a comparison with higher-resolution data: Sz\,10, Sz\,104, SS61344.1-373646, CVSO-176, exhibit a $^{7}\rm Li$ abundance  higher than 2.0 dex. While  the $A{\rm (Li)}$ values for  Sz\,10 and SS61344.1-373646 remain relatively low (2.8 dex and 2.2 dex respectively), the other two sources show the mean $A{\rm (Li)}$ $\sim$ 3.5 dex.
This discrepancy between datasets is due primarily  to the veiling factor adopted in the analysis; as explained in Sec. \ref{variation}, slight variations in the veiling correction significantly impact the measured EW of the $^{7}\rm Li$ line. %Indeed, the $A{\rm (Li)}$ values  for Sz\,104 show measurable epoch-to-epoch variations across the three observed epochs (see Tables\, \ref{tabewli1}, \ref{tabewli3}), strongly supporting the role of stellar variability as a contributing factor to the observed differences in lithium abundance.
ECHAJ0844.2-7833 has been observed only with X-shooter and therefore its lithium abundance cannot be independently cross-validated.
For the sources Sz\,69 and CVSO-90, the absence  of $^{7}\rm Li$ line at in the spectra obtained by the other instruments provides strong independent evidence supporting their classification as $^{7}\rm Li$-depleted targets (for the sake of brevity,  non-detections are not reported in the Appendix). It is worth noting that  the depleted $^{7}\rm Li$ abundance derived for Sz\,69 is consistent with previous finding in the literature \citep{biazzo17}.

\subsection{Influence of veiling on age estimate based on Li diagnostics}

One of our main goals of this work is to investigate what is the effect on the age estimates, based on the Li line intensity, when the line equivalent width is not corrected for veiling. For this purpose we use the software \textit{EAGLES} \citep{jef23}.
%To determine the age of each stellar association   we used the software  \textit{EAGLES} \citep{jef23}, %developed by fitting to a training dataset consisting of around 6200 stars in 52 open clusters,  observed as part of the Gaia-ESO survey \citep{gilmore22,randich22,jack22}.
This  code allows us to obtain age estimates and age probability distributions  from measurements of the $^{7}\rm Li$ I 6708 \AA\, equivalent width and $T_{\rm eff}$ for individual PMS stars, or associated group of coeval stars, with $3000 < T_{\rm eff} < 6500$ K, $-0.3 < [Fe/H]  < 0.2$, and 200 $\lesssim$ $EW_{\rm Li}$ $\lesssim$ 800 m\AA.
The code produces estimates of the most probable age, uncertainty and the median age of the stellar cluster.
For stars aged less than 10 Myr and more than 1 Gyr, the code provides only upper and lower limits on the age.
For intermediate values, the age is estimated with a precision that will depend on the number of stars and their $T_{\rm eff}$-$EW_{\rm Li}$ distribution (see \cite{jef23} for more details).

To determine the age of each association, we have considered the ESPRESSO, UVES, and X-Shooter data together; % corrected for the veiling contribution and the Fe at $\lambda6704\,\AA$ blending. 
in the case of multiple epoch observations, we took the average value of $EW_{\rm Li}$ and $T_{\rm eff}$ from the different epochs for one single star.
To evaluate the impact of veiling on age determination, we ran the code twice: first using  $EW_{\rm Li}^{veil+Fe}$ as input,  and then using $EW_{\rm Li}^{Fe}$, the results for both cases  are shown in Table \ref{agecl}.

\begin{table}
\caption{Ages estimated with the EAGLES code, from $EW_{\rm Li}^{\rm veil+Fe}$ (second column) and $EW_{\rm Li}^{\rm Fe}$ (third column).}             
\label{agecl}     
\centering               
\begin{tabular}{c c c }        
\hline\hline               
Name & age (Myr) & age (Myr) \\    % table heading 
Cluster & with $r$ contribution & without $r$ contribution               \\
\hline                        
Cha I &  < 5.2 & $16.1_{-0.7}^{+0.8}$  \\
$\eta$\,Cha & < 7.5 &  <10.4 \\
Lupus & < 4.9  &$12.8_{-0.6}^{+0.8}$  \\
Taurus& < 5 & $13.7_{-11.3}^{+1.0}$ \\
Orion OB1a & < 12.2 & $13.6_{-10.8}^{+1.8}$ \\
Orion OB1b &< 5.5  & $16.7_{-1.1}^{+1.4}$ \\
$\sigma$\,Orionis & < 6  &$29.7_{-2.9}^{+3.2}$\\
CrA & < 7.0 & < 17.3\\
\hline                                   %inserts single line
\end{tabular}
\end{table}

For simplicity, only the Lupus case is shown here as a representative example; the results for the remaining regions are provided in the Appendix (\ref{age1}, \ref{age2}). Fig.\,\ref{agelup} displays the $EW_{\rm Li}^{\rm veil+Fe}$ (left panel) and $EW_{\rm Li}^{\rm Fe}$ (right panel) as a function of $T_{\rm eff}$ with the error bars (blue dots)  the best-fitting empirical isochrone (black solid line) and its associated dispersion (gray region). 

 Consistent with the results shown in Fig. \ref{nlte}, using the  \textit{EAGLES} code the ages derived when accounting for veiling are significantly younger than those obtained without this correction. For the Lupus SFR, the age difference between the two cases is about 7 Myr.  The age differences found for the remaining SFRs are from around few Myr up to around 25 Myr, with the latter value obtained for the $\sigma$ Ori association.
This means that the veiling correction is crucial for an appropriate estimation of the age of YOCs based on the lithium diagnostics.
The upper age limits  obtained for each SFR considering the veiling contribution  are  $\sim$ 5-7 Myr, consistent with those reported in the literature (e.g., for  Cha\,I, see  \citealt{lum07}, \citealt{manara16}, \citealt{randich22}, \citealt{chen25}; for Lupus, \citealt{biazzo17}; for Taurus, \citealt{simon93} and \citealt{lum23}, for Orion OB1a and Orion OB1b \citealt{bice19}, and for $\sigma$\,Ori, \citealt{caballero18}).
When considering only the hotter stars  ($T_{\rm eff}$ $\ge$ 4000 K), the estimated upper age limits  increase by around 2 Myr.

An interesting aspect of the upper age limit for Orion OB1b is that an age below 5 Myr would be consistent with the high accretion rates reported by \cite{pittman22}. These rates are typically incompatible with a 5 Myr-old PMS stars, pointing instead toward a significantly younger age. Our results are further supported by the recent work of \cite{piscarreta25},  who highlighted the impact of accretion,  including veiling, on age determination and photospheric properties. They reported  that properly accounting for veiling consistently leads to younger age estimates.

\begin{figure*}
      \centering
	   \begin{minipage}{0.45\linewidth}
\includegraphics[width=\linewidth]{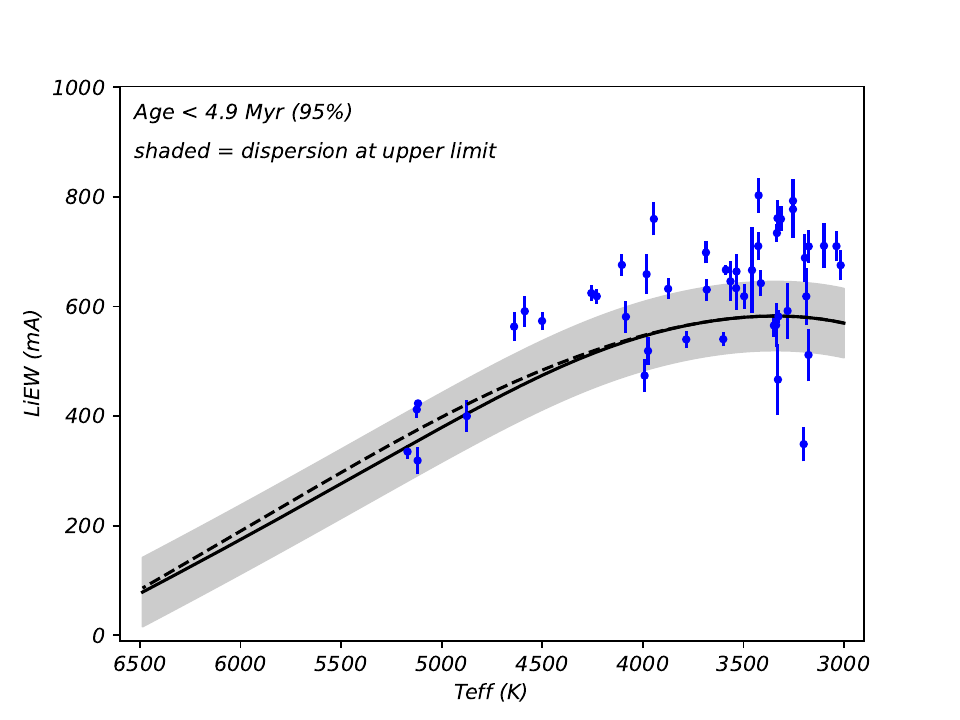}
	      \end{minipage}
	       \begin{minipage}{0.45\linewidth}
\includegraphics[width=\linewidth]{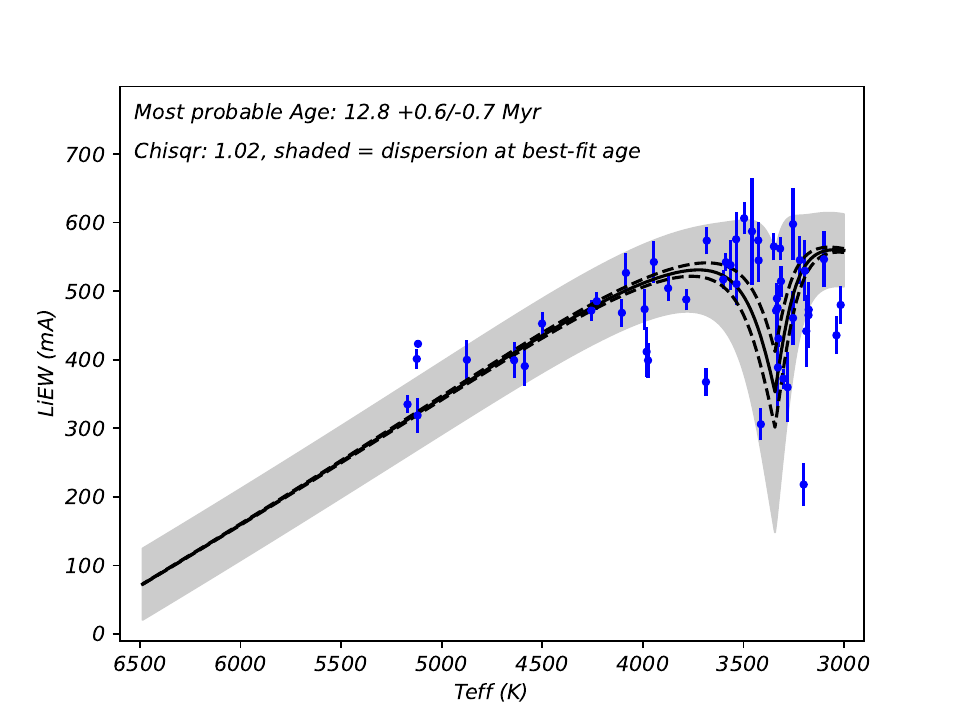}
		   \end{minipage}
           \caption{Example of lithium pattern fitting: the case of the Lupus SFR.
           The left panel shows the case in which the age was determined using the $EW_{\rm Li}^{\rm veil+Fe}$, while the right panel shows the case in which the $EW_{\rm Li}^{\rm Fe}$ have been used. The solid black line represent the best-fit isochrone in the $EW_{\rm Li}$ vs $T_{\rm eff}$ plane. The shaded region illustrates the model intrinsic dispersion at the best-fit age or its upper limit. The black dashed lines represent 95\% upper and lower limits where no clear peak is observed. 
     The blue dots show EW(Li) as a function of $T_{\rm eff}$  with the uncertainties on EW(Li) measurements. 
    The text in the  top-left corner on the plot shows maximum likelihood age.}
	\label{agelup}
\end{figure*}

\section{Iron and Barium abundance}
\label{feba}

 We focused the analysis on the iron ([Fe/H]) and barium ([Ba/H]) abundances on PMS stars with $T_{\rm eff}$ greater than 4400 K. This threshold helps to avoid the strong contribution of molecular bands \citep{biazzo17}. Additionally, we selected stars with    veiling lower than $\sim$ 0.2 to minimize uncertainties caused by the veiling contribution. Unfortunately, this initial selection criteria,  prevented us from having targets in every analyzed cluster. Our sample is therefore composed by:
\begin{itemize}
\item 6 targets observed with X-shooter: MY\,Lup, RECX\,11, RX\,J0438.6+1546, RY\,Lup, SSTc2dJ160830.7-382827, Sz\,68;
\item 2 targets  observed with UVES: CS\,Cha and CV\,Cha;
\item 8 targets observed  with ESPRESSO: CHX\,18N, LkCa\,15, MY\,Lup, RECX\,11, RX\,J0438.6+1546, RY\,Lup, and SSTc2dJ160830.7-382827.
\end{itemize}

We derived [Fe/H] and [Ba/H] through the spectral synthesis method \citep{biazzo17}.
 The iron abundances were derived using the open-source spectral analysis framework iSPEC \citep{blanco14,blanco19}, in conjunction with the radiative transfer code MOOG \citep{sneden12}. Synthetic spectra were generated using  the \cite{kur05} set of model atmospheres. We adopted the \cite{asplund09} solar abundances  and the GES line list with hyperfine structure and isotopes \citep{ges}. For this analysis, we chose  the wavelength window between 5520 \AA\ and  6800 \AA. 

For the barium abundance, we employed  spectral synthesis using  the  MOOG code \citep{sneden12} and  \cite{asplund09} model atmosphere. We considered the  spectral synthesis of the Ba II line at $\lambda$ = 5853.7 $\AA$, which is known to be strong, isolated, and not affected by Non-Local Thermodynamic Equilibrium (NLTE) effects (e.g \citealt{masho07}). To achieve the best possible result, we included the hyperfine structure and isotopic shift provided by \cite{mcwill98} in our analysis. We adopted the isotopic solar mixture by \cite{anders89} and, as done for the iron,  we considered the solar barium abundance by \cite{asplund09}.

The limb-darkening coefficients were taken from \cite{claret12}. We estimated the microturbulence $\xi$ and macroturbulence $v_{mac}$ using the relations of \cite{dutra16} and \cite{brewer16}, respectively.
The values of $\xi$ and $v_{mac}$ for the selected targets are shown in Table\,\ref{micmac}.

Table\, \ref{tabfeba}  presents the results of our [Fe/H] and [Ba/H] analysis. For the ESPRESSO and UVES data, the table reports the mean results across the multi-epoch values, obtained from the individual spectra. It also includes uncertainties related to the best-fit model ($\sigma$1) and to the errors in the stellar parameters ($\sigma$2). For more details, see Sec. \ref{secerror}.
 In Table\, \ref{cluster}, we show the mean [Fe/H] and [Ba/H]  values  along with their standard deviation for the  respective clusters. For these calculations, we assigned one value per target. For objects with multi-instrument observations, we considered the average of the two independent measurements.
 %These values represent the average abundances  of the targets within each respective cluster.

 \subsection{Error estimate}
 \label{secerror}

There are two sources of uncertainty in the abundances derived from spectral synthesis:
(i) errors associated with the fitting procedure, and
(ii) uncertainties arising from the choice of the atmospheric parameters.

In the case of  iron abundance, the first source of uncertainty, which also includes errors in the continuum placement, is about 0.05 dex for ESPRESSO, and, 0.1-0.2 dex for UVES and X-Shooter.
In the case of Ba, the uncertainties are $\sim$ 0.1 dex for ESPRESSO and UVES, $\sim$ 0.15 dex  for X-Shooter. To quantify the impact of stellar parameters ($T_{\rm eff}$, $log g$, $\xi$ and $v \sin i$) on the abundance measurement, we changed each quantity separately and evaluated the corresponding change in the derived abundance.
Specifically, a change of $\pm$ 60 K in $T_{\rm eff}$ for ESPRESSO data, and $\pm$ 100 K for UVES and X-Shooter spectra, resulted in Fe variations of 0.04, 0.03, and 0.05 dex across the three instruments. For Ba, the corresponding errors ranged from 0.05 to 0.06 dex. 
Varying $log g$ by $\pm$0.15 dex led to Fe abundance variations of 0.02-0.03 dex, and Ba variations of 0.02-0.08 dex.
Finally, a $\pm$2 km/s change in $v \sin i$ contributed as 0.01-0.04 dex in Fe uncertainties and 0.02-0.04 dex in Ba uncertainties.
%Microturbulence has the strongest impact on the abundance measurements. 
Considering $\xi$ = 0 km/s instead 2 km/s, we obtained an error on [Fe/H] of 0.03-0.07 dex, and on [Ba/H] of about 0.04-0.13 dex.
The cumulative uncertainties can be obtained by summing in quadrature the different contributions (see Table\,\ref{tabfeba}).
 
%It is evident that in the case of Ba, the uncertainties dominated is due to the best-fit procedure.

\subsection{Iron abundance in the context of nearby SFRs}
Metallicity plays a crucial role in shaping stellar evolution and possible Galactic chemical enrichment within SFRs.
Recent studies  have  shown that Fe abundance of nearby (< 500 pc) YOCs ranges between approximately $-$0.2 to 0.3 dex.
The youngest associations  ($\lesssim$ 100 Myr) are generally clustered to the lowest values \citep{biazzo11, spina14, spina17}.
%Given their recent formation, these regions haven't yet undergone substantial migration across the galactic disk. Consequently, their metallicity should closely represent the present-day chemical composition of the surrounding interstellar gas from which they formed, exhibiting negligible signs of chemical evolution (\citealt{spina14} and references therein).

In this work, we find slightly subsolar iron abundance for our SFRs (Table\,\ref{cluster}), with a value  in line with the recent studies cited above, although our dispersion is somewhat high.
 In particular, for Cha I, we find [Fe/H] =$-$0.08\,dex, which is consistent with the value reported by \cite{spina14, spina17}. For the Taurus association, we find [Fe/H] = $-$0.07\,dex, again in agreement within the uncertainties with \cite{dorazi11}.
 For Lupus we find [Fe/H] =$-$0.14\,dex, which  agrees within the errors with \cite{biazzo17} and \cite{santos08}.
 Moreover, we present the first metallicity estimate for the $\eta$\,Cha SFR, finding a value of $-$0.08\,dex, consistent with that of Cha I.
 
\begin{figure}[!h]
   \centering
   \includegraphics[width=\hsize]{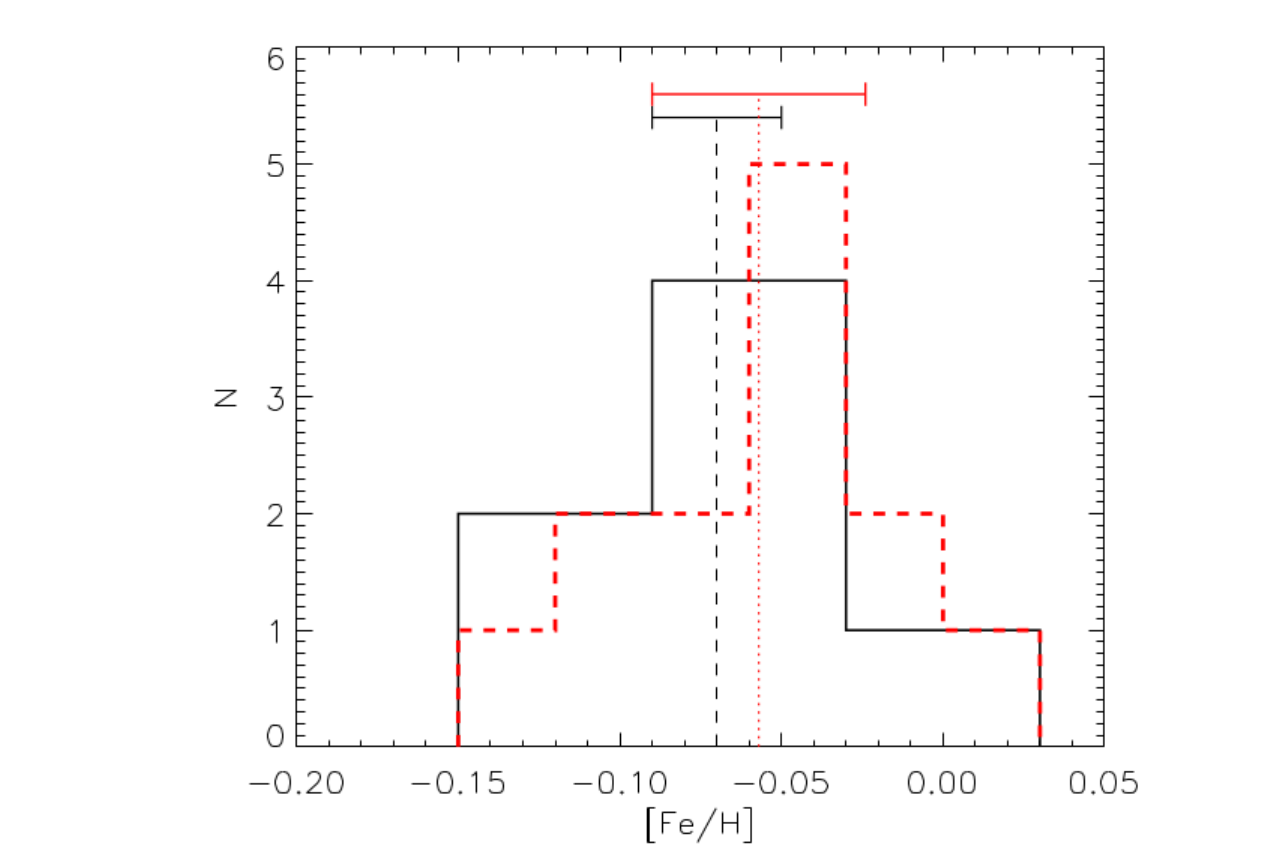}
   \vspace{-0.7cm}
   \caption{[Fe/H] distribution of the open clusters and SFRs in the solar neighborhood within a distance of 500 pc and age lower than 10 Myr. The histogram bin size is 0.03. The red  dashed line  represents data from \cite{spina14}, while the black solid line shows the same dataset after replacing the values for the clusters in common with our own measurements. The vertical lines and an error bars indicate the median values and the corresponding median absolute deviations.}
    \label{isto}
    \end{figure}

 Fig. \ref{isto} displays the [Fe/H] distribution of  young open clusters and SFRs in the solar neighborhood within a distance of 500 pc and age smaller than 10 Myr. The black line represents the distribution  based on the data from \cite{spina14}, where our measurements have replaced those for the clusters in common (i.e. Cha I, Lupus, Taurus, see Tab. \ref{cluster}). For comparison, the original distribution from \cite{spina14} is over plotted as a red dashed line.
 Our results are consistent with their estimates, yielding  a median [Fe/H] = -0.06   $\pm$ 0.03 dex  for our combined sample (indicated by a dashed black vertical line and an error bar) compared to -0.057  $\pm$ 0.03 dex for the original \cite{spina14} dataset (red dotted vertical line). The histograms reveal that the majority of the observed young sources exhibit sub-solar metallicities, with  both distributions showing a  prominent peak around [Fe/H]=$-$0.05 dex.

The common metal-poor composition of these young environments, not characteristic of the local ISM, may be the result of a complex interplay 
of chemical processes involving a wide area of the Galactic disk \citep{spina17}.

 \subsection{The barium abundance conundrum}
 
Previous works showed that the Ba abundance in star-forming regions and young associations increases  with decreasing age, reaching values up $\sim$0.6-0.7 dex (e.g. \citealt{dorazi09,biazzo17,baratella21}). This remarkably high enhancement cannot be explained by standard nucleosynthesis and Galactic Chemical Evolution (GCE), nor by NLTE effects.
 
In this work we have homogeneously measured the [Ba/H] abundance in four very young stellar associations ( < 10 Myr).
As in the previous studies, we find an overabundance of Ba (Table\,\ref{cluster}).
Specifically, for Lupus  [Ba/H]=+0.69 dex, which is in  perfect agreement, within the error, with the value of $\sim$0.7 dex reported by \cite{biazzo07}.
To our knowledge, no previous studies focusing on  barium abundance have been published to date for the SFRs Taurus, Cha I and $\eta$\,Cha.
Here, we determined the mean [Ba/H] in these regions finding 0.73 dex, 0.75 dex and 0.64 dex respectively.
However, it should be noted  that the value of 0.64 dex is based on observations of a single star (RECX\,11) using two different instruments; therefore, it may not be representative of the entire region.
 
\begin{figure}
   \centering
   \includegraphics[width=\hsize]{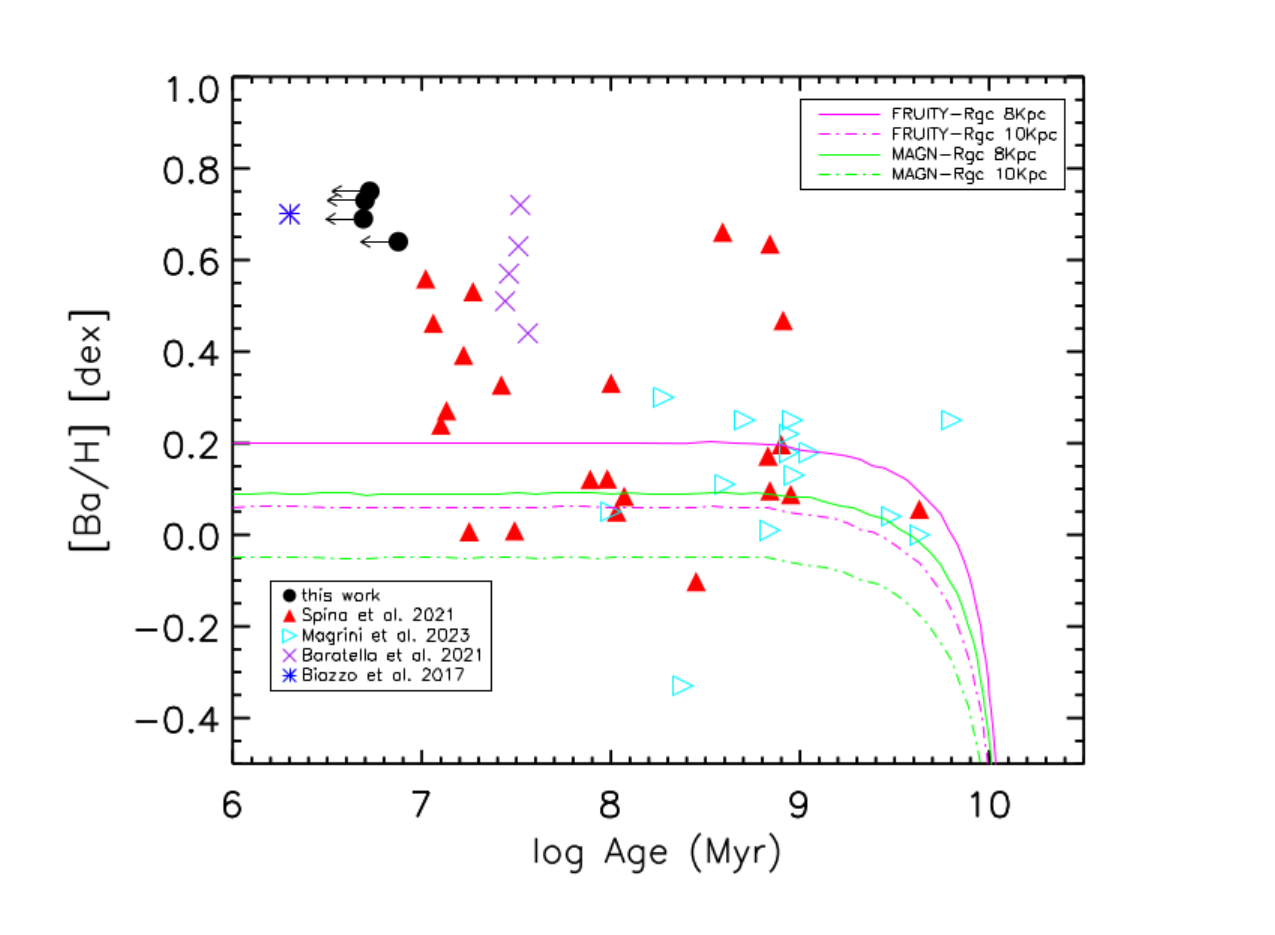}
   \vspace{-1cm}
   \caption{[Ba/H] as a function of the age of Galactic open clusters and associations. The black dots represent the estimates derived in this work. Arrows indicate upper limits in age. The other symbols represent estimates from the literature, as highlighted in the figure. When the same cluster was analyzed by more than one author, we considered the values obtained by  \cite{magrini23}. The GCE models with yields from the FRUITY (\citealt{cristallo09}, magenta lines) and MAGN models (\citealt{visconti20}, green lines) are also overplotted.}
    \label{bage}
    \end{figure}

In Fig. \ref{bage} we plot our mean cluster [Ba/H] values as function of age (black dots),  together with the results obtained by other authors: \citealt{biazzo17} (blue asterisk), \citealt{spina21} (red triangles), \citealt{baratella21} (purple crosses), and \citealt{magrini23} (cyan triangles).
We selected SFRs and clusters located at Galactocentric distances between 7.5 and of 9 pc. 
Moreover, we displayed the [Ba/H] in SFRs and stellar clusters with an age from few Myr up to 10 Gyr.
We also compare the observations with the prediction of the GCE of \cite{magrini21} at different $R_{\rm GC}$ (8 and 10 kpc).
   The GCE models used in this plot incorporate s-process yields from the FRUITY models, which are based on an exponentially decreasing convective velocity profile at the inner border of the convective envelope \citep{cristallo09}, as well as from the updated MAGN models \citep{visconti20}. The latter include the effects of magnetic-field-induced mixing.
The GCE models can  reproduce  data of clusters  older than $\sim$  100 Myr quite well.  However, for  younger clusters  and associations,  the high [Ba/H] values observed,  are sistematically underpredicted by models.

A promising mechanism of production of heavy elements is the \textit{i}-process, proposed by various authors (e.g. \citealt{mishe13,dorazi17}).
This process is characterized by neutron density intermediate between those of \textit{s}- and \textit{r}-processes. Rich \textit{i}-process nucleosynthesis can occurs during the early AGB phase of low metallicy low-mass stars \citep{choplin21}, although other types of stars (e.g super AGB, rapidly-accreting white dwarfs, massive stars) have also been proposed as possible \textit{i}-process hosts (\citealt{baratella21} and reference therein).
However further theoretical models are needed.

 \cite{baratella21} investigated whether  stellar activity,  strong magnetic field or the First Ionization Potential effect could  explain the high peculiar  Ba abundance. They concluded that these factors  play a role, but there is still no convincing evidence that  any of them provide a definitive solution.
Recently, \cite{sheminova24}, analyzing 13 solar-type F, G and K-type stars in the thin disk of the Galaxy, with ages from 2 Gyr to 14 Gyr, and confirmed the increase in the barium abundance with increasing chromospheric activity.
This suggests that it is crucial to adopt a more complex atmosphere model that includes the magnetic structure in order to obtain more reliable Ba abundances. 
In any case, at present, the high [Ba/H] values  in the SFRs still remains  a conundrum.

 \begin{table}
\caption{Iron and barium abundances measured through spectral synthesis. The uncertainties include contributions from the fitting process and the propagation of errors in the atmospheric parameters.}       
\label{tabfeba}      
\centering  
\resizebox{8.0cm}{!}{  
\begin{tabular}{c c| cc} 
\hline \hline
     Name  &  SFR &[Fe/H] & [Ba/H]  \\
    Target   &   &  (dex) &   (dex)    \\
\hline 
\noalign{\smallskip}
\multicolumn{4}{c}{ESPRESSO} \\
\noalign{\smallskip}
\hline 
CHX\,18N &  ChaI&   0.07  $\pm$ 0.05  $\pm$ 0.10  &     ... $\pm$...$\pm$ ...\\
LkCa\,15 &  Taurus &  $-$0.09 $\pm$ 0.05  $\pm$ 0.10 &     0.66 $\pm$ 0.07 $\pm$ 0.07\\
MY\,Lup  &  Lupus & 0.07  $\pm$ 0.06 $\pm$ 0.01 &      0.68 $\pm$ 0.09 $\pm$ 0.07\\
RECX\,11 & $\eta$\,Cha &  $-$0.15  $\pm$ 0.05 $\pm$ 0.10 &    0.60 $\pm$ 0.07 $\pm$ 0.07\\
RX\,J0438.6+1546 &  Taurus &  0.06 $\pm$ 0.06 $\pm$ 0.10 &  0.81 $\pm$ 0.09 $\pm$ 0.07\\
RY\,Lup   &  Lupus & $-$0.07  $\pm$ 0.07 $\pm$ 0.10 &  0.66 $\pm$ 0.07 $\pm$ 0.07\\
SSTc2dJ160830.7-382827 &  Lupus  & 0.05 $\pm$ 0.06  $\pm$ 0.1 & 0.80 $\pm$ 0.07 $\pm$ 0.07\\
Sz\,75  & Lupus & $-$0.19 $\pm$ 0.05 $\pm$ 0.10 & 0.71 $\pm$ 0.07 $\pm$ 0.07\\
\hline
\noalign{\smallskip}
\multicolumn{4}{c}{ UVES} \\
\noalign{\smallskip}
\hline
CS\,CHA & ChaI& $-$0.08 $\pm$ 0.12 $\pm$ 0.05&   0.72 $\pm$ 0.07 $\pm$ 0.07\\
CV\,CHA& ChaI& $-$0.22 $\pm$ 0.16 $\pm$ 0.05& 0.78$\pm$ 0.09 $\pm$ 0.07\\
\hline
\noalign{\smallskip}
\multicolumn{4}{c}{ XS} \\
\noalign{\smallskip}
\hline
    MY\,Lup  &  Lupus & $-$0.20 $\pm$ 0.20 $\pm$ 0.06 &0.60 $\pm$ 0.17 $\pm$ 0.12\\
   RECX\,11  & $\eta$\,Cha & $-$0.02 $\pm$ 0.10 $\pm$ 0.06&0.68 $\pm$ 0.13 $\pm$ 0.12\\
 RX\,J0438.6+1546 & Taurus &$-$0.14   $\pm$ 0.20$\pm$ 0.06 &0.76 $\pm$ 0.17 $\pm$ 0.12\\
    RY\,Lup  & Lupus &  $-$0.15 $\pm$ 0.10 $\pm$ 0.06&0.75 $\pm$ 0.17 $\pm$ 0.12\\
SSTc2dJ160830.7-382827  & \textbf{Lupus} & $-$0.10  $\pm$ 0.10 $\pm$ 0.06&0.65 $\pm$ 0.13$\pm$ 0.12 \\
     Sz\,68  & Lupus &  $-$0.30 $\pm$ 0.10 $\pm$ 0.06&0.65 $\pm$ 0.17 $\pm$ 0.12\\
\hline
\end{tabular}}
\end{table}

\begin{table}
\caption{Mean Iron and Barium abundance of 4 SFRs, including  their respective  standard deviation.}             
\label{cluster}     
\centering               
\begin{tabular}{c c c }        
\hline\hline               
Name & [Fe/H] & [Ba/H]  \\    % table heading 
     & (dex)     & (dex)               \\
\hline                        
ChaI  &    $-$0.08  $\pm$ 0.15  &  0.75    $\pm$  0.04    \\
$\eta$\,Cha  &    $-$0.08  $\pm$ 0.09  &  0.64  $\pm$  0.06    \\
Lupus &    $-$0.14 $\pm$ 0.11  &  0.69  $\pm$  0.04  \\
Taurus &  $-$0.07  $\pm$ 0.04  &  0.73 $\pm$  0.09    \\
\hline                                   %inserts single line
\end{tabular}
\end{table}

%-----------------------------------------------------------------

\section{Conclusions}

We  present the results of a study on elemental abundances in  several nearby star-forming regions, namely Cha I, $\eta$\,Cha, Lupus, Taurus, Orion OB1a, Orion OB1b, $\sigma$\,Ori, and CrA. We used spectroscopic data  gathered as part of the PENELLOPE program,   obtained using the instruments ESPRESSO, UVES and X-Shooter, all mounted on the VLT.

Our main results can be summarized as follows:
\begin{itemize}
\item We measured the equivalent width of the lithium line at $\lambda$= 6707.8 \AA. For all 75 targets in our sample, we corrected the measurements for the contribution of veiling, obtaining an average $EW_{\rm Li}^{\rm veil+Fe}$ value of 170 m\AA. 

\item Analysis of ESPRESSO and UVES multi-epoch spectra reveals significant $EW_{\rm Li}$ variability. We identified 26 targets with raw variations ($\Delta EW_{\rm Li}^{\rm raw} = 62 \pm 28$ m\AA), which could be linked to chromospheric activity. Additionally, in a subsample of 30 sources, the veiling-corrected variations ($\Delta EW_{\rm Li}^{\rm veil+Fe} = 92.2 \pm 65.9$ m\AA) appear to be more pronounced. The correlation between $\Delta EW_{\rm Li}^{\rm veil+Fe}$ and $\Delta r_{670}$ suggests that variations in the accretion process may play a significant role in driving the observed $EW_{\rm Li}$ changes.

\item We estimated  the abundance of $^{7}\rm Li$ from the corrected equivalent widths, for the  targets in the sample with $T_{\rm eff}$ higher than 3000 K.
For the stars with temperature ranging between 3000 K and 4000 K  we measured upper limits in $A{\rm (Li)}$.
We also emphasized  the crucial role of the veiling  contribution in the  determination of $A{\rm (Li)}$, which leads to an  average correction of  $\sim$ 0.74 dex.

\item We identified 7 possible Li-depleted sources: Sz\,10  in Cha I, Sz\,104, Sz\,69, SS61344.1-373646 in Lupus, CVSO-176,  CVSO-90 in Orion OB1b and ECHAJ0844.2-7833 in $\eta$\,Cha.

\item Using the \textit{EAGLES} code, we attempted to estimate the ages of all SFRs, based on  their lithium equivalent widths, both including and neglecting the contribution of  veiling.  For all the young regions, we found differences of several Myr, reaching up to 25 Myr, between the two cases. This result underscores the crucial importance of accounting for veiling in age determinations.

\item We determined the mean iron and barium abundance of the SFRs Lupus, Taurus, Cha I and $\eta$\,Cha. We find slightly sub-solar iron abundance values. This  result confirms the recent studies in which the youngest ($\lesssim$ 100 Myr) and nearby (< 500 pc) stellar associations generally  cluster around sub-solar iron values.
We found overabundance of the mean Ba in these SFRs, up to $\sim$ 0.75 dex, which still remains a conundrum, as no recent theory is able to predict such a high value at young ages.
 
\end{itemize}

The results presented of this work demonstrate that veiling significantly impacts both $A{\rm (Li)}$ and age determinations, while also inducing notable epoch-to-epoch variations in the lithium equivalent width. These findings emphasize the necessity for multi-epochs observations of PMS stars and more rigorous investigations into veiling-induced systematic effects. Furthermore, our discovery of barium overabundances in three additional young regions, extending beyond previously documented cases, strengthens the empirical evidence for this enhancement. This highlights the need for expanded theoretical and observational studies of star forming regions and young clusters (age < 100 Myr) to elucidate the physical origin of the so-called "barium puzzle". Finally, this work provides a high-resolution fundamental benchmark for future large-scale surveys. Our results will be essential for the interpretation of upcoming studies of young clusters conducted with the new 4MOST facility and the forthcoming MOONS spectrograph, both of which operate at lower spectral resolutions.

\begin{acknowledgements}
 This work has been financially supported by the grants INAF 2022 TRAME@JWST (TRacing the Accretion Metallicity rElationship with NIRSpec@JWST; PI: K. Biazzo), Can AGB stellar winds unveil the origin of the unidentified infrared emission bands? (PI: R. Carini), YSOs Outflows Disks and Accretion (YODA; PI: B. Nisini), by the European Union (ERC, WANDA, 101039452), and by and NextGenerationEU, M4C2 1.2 CUP C83C25000450006 within the project Tracing the staR and plAnet formation in different Circumstellar Environments (TRACE; PI: K. Biazzo). Views and opinions expressed are however those of the author(s) only and do not necessarily reflect those of the European Union or the European Research Council Executive Agency. Neither the European Union nor the granting authority can be held responsible for them. 
 his work was partly funded by the Deutsche Forschungsgemeinschaft (DFG, German Research Foundation) in the framework of the YTTHACA Project 469334657 under the project code MA 8447/1-1.
 This work was also supported by the NKFIH NKKP grant ADVANCED 149943 and the NKFIH excellence grant TKP2021-NKTA-64. Project no.149943 has been implemented with the support provided by the Ministry of Culture and Innovation of Hungary from the National Research, Development and Innovation Fund, financed under the NKKP ADVANCED funding scheme.
 This work has been also supported by Large Gran INAF-2024 "Spectral Key features of Young stellar objects: Wind-Accretion LinKs Explored in the infraRed (SKYWALKER)".
 I.M. is funded by grant PID2022-138366NA-I00, by the Spanish Ministry of Science and Innovation/State Agency of Research MCIN/AEI/10.13039/501100011033 and by the European Union. JFG was supported by Funda\c{c}\~{a}o para a Ci\^{e}ncia e Tecnologia (FCT) through the research grants UID/04434/2025
 This work benefited from discussions with the ODYSSEUS team (HST AR-16129), \url{https://sites.bu.edu/odysseus/}.

\end{acknowledgements}

   \bibliographystyle{aa} % style aa.bst
   \bibliography{biblio} % your references Yourfile.bib

\onecolumn
\par % O \mbox{} o \vspace*{1em}

\begin{appendix}

    \section{ Tables of effective temperature, veiling, Equivalent Widths, and Lithium Abundances values with fit errors.}

\begin{center}
\scriptsize
\begin{longtable}{|c|c c| c c |c c |c|}  
\caption{Results from the UVES spectra.\\
{\scriptsize \textit{Notes}: Targets marked with an asterisk (*) are M stars; their $EW_{\rm Li}$ values are corrected only for veiling. NLTE corrections are not available, so the lithium abundances listed are LTE values.\\
Stars marked with ($^{**}$) have $A({\rm Li})$ values higher than 4.0 dex. We fixed the lithium abundance of these stars at 4.0 dex.}}
\label{tabewli1}  \\
\hline 
\textbf{Name}&\textbf{epoch} & \textbf{Observation date}& \bm{$T_{\rm eff}$} & \bm{$r_{650}$} & \bm{$EW_{\rm Li} raw$} & \bm{$EW_{\rm Li}^{\rm veil+Fe}$}& \bm{$A{\rm (Li)}NLTE$}\\ 
  &  & [K] & & m\AA & m\AA & dex\\ 
\hline 
\endfirsthead
\caption{ \scriptsize continued from previous page} \\
\hline
\textbf{Name}&\textbf{epoch} & \textbf{Observation date}& \bm{$T_{\rm eff}$} & \bm{$r_{650}$} & \bm{$EW_{\rm Li} raw$} & \bm{$EW_{\rm Li}^{\rm veil+Fe}$}& \bm{$A{\rm (Li)}NLTE$}\\ 
 & & [K] & & m\AA& m\AA& dex\\ 
\hline 
\endhead
\caption{\scriptsize Continued on next page}\\
\hline\hline
\endfoot
\hline \hline
\endlastfoot
\hline
\hline 
\noalign{\smallskip}
\multicolumn{7}{c}{ Orion OB1} \\
\noalign{\smallskip}
\hline
     CVSO-17  & 1  & 2020-12-04& 3721 $\pm$ 72 & 0.20 $\pm$ 0.10 &  454.2 $\pm$  13.8&  545.0 &   2.8 \\
     CVSO-17$^*$  & 2   &2020-12-05& 3697 $\pm$ 88 &0.23 $\pm$ 0.10& 440.8 $\pm$  13.3 & 542.2 &  2.7\\
     CVSO-17$^*$  & 3 & 2020-12-06& 3695 $\pm$ 88 &0.25 $\pm$ 0.09   &  449.7 $\pm$  15.6&  562.1 &  2.8 \\
\hline
     CVSO-36$^*$  & 1 &2020-12-02& 3702 $\pm$ 88 & 0.21 $\pm$ 0.03 &     541.3 $\pm$ 17.5 & 655.0&   3.1 \\
     CVSO-36$^*$  & 2  &2020-12-03& 3696 $\pm$ 91 & 0.12 $\pm$ 0.04 &  559.9 $\pm$ 19.3& 627.1 &  3.0 \\
     CVSO-36$^*$  & 3  &2020-12-04& 3662 $\pm$ 69 & 0.12 $\pm$ 0.04   & 568.4 $\pm$ 18.4& 636.6 &  3.0\\
\hline
     CVSO-58  & 1  &2020-11-30& 4193 $\pm$ 103 & 0.63 $\pm$ 0.12 & 393.6 $\pm$10.5&    629.9 &  3.4  \\
     CVSO-58  & 2  &2020-12-01& 4211 $\pm$ 110 & 0.61 $\pm$ 0.08 & 409.7$\pm$ 10.3&    648.5&  3.5   \\
     CVSO-58  & 3  &2020-12-02& 4223 $\pm$ 105 & 0.59 $\pm$ 0.10 & 397.2$\pm$ 11.4&    620.7 & 3.4\\
\hline
    CVSO-107$^*$  & 1 & 2020-12-03& 3988 $\pm$ 118 & 0.74 $\pm$ 0.18 & 427.8 $\pm$ 9.8& 744.4  &  3.6\\
    CVSO-107$^*$  & 2 & 2020-12-04&  3943 $\pm$ 93 & 0.72 $\pm$ 0.18 & 430.8 $\pm$ 9.6& 741.0 & 3.5\\
    CVSO-107    & 3  & 2020-12-05&  4002 $\pm$ 119 & 0.71 $\pm$ 0.14 & 421.4 $\pm$  9.6 &704.0 &   3.6\\
\hline
    CVSO-109$^*$  & 1 & 2020-11-26&  3898 $\pm$ 112& 0.44 $\pm$ 0.08 & 474.5 $\pm$ 12.4&  683.3 &   3.3 \\
    CVSO-109$^*$  & 2 & 2020-11-27 & 3922 $\pm$ 106 & 0.42 $\pm$ 0.12 &  485.7 $\pm$ 12.5& 689.7  & 3.4 \\
    CVSO-109$^*$  & 3   &2020-11-28&  3948 $\pm$ 91 &0.67 $\pm$ 0.14 & 415.0 $\pm$ 10.1& 693.1  & 3.4 \\
\hline
    CVSO-176$^*$  & 1   & 2020-11-28&  3495 $\pm$ 85 & 0.93 $\pm$ 0.60 & 413.6 $\pm$ 11.3& 798.3 &   3.4\\
    CVSO-176$^*$  & 2 & 2020-11-29 & 3503 $\pm$ 82 & 0.60 $\pm$ 0.38 & 483.0 $\pm$ 10.3&  772.8  & 3.5  \\
    CVSO-176$^*$  & 3 & 2020-11-30& 3521 $\pm$ 77& 0.68 $\pm$ 0.48 &  488.0 $\pm$ 11.7&  819.8 &  3.6  \\
\hline
\noalign{\smallskip}
\multicolumn{7}{c}{ $\sigma$\,Orionis} \\
\noalign{\smallskip}
\hline
       SO\,518 & 1   & 2020-11-29&  4328 $\pm$ 168 & 1.28$\pm$ 0.13&  290.1 $\pm$6.4 &  651.7  &  3.6 \\
       SO\,518 & 2    & 2020-11-30&  4383 $\pm$ 141& 0.98 $\pm$ 0.04&  339.1 $\pm$10.2 &   662.3 &  3.7 \\
       SO\,518 & 3    &2020-12-01&  4366 $\pm$ 150 & 0.89 $\pm$ 0.08 & 351.0 $\pm$10.1 &  654.2  &  3.7 \\
\hline
       SO\,583 & 1   &2020-11-29& 4753 $\pm$ 119 &0.43 $\pm$ 0.08 & 351.3 $\pm$ 8.5&   492.9  &3.7\\
       SO\,583 & 2  &2020-11-30& 4739 $\pm$ 118 & 0.55 $\pm$ 0.11&  341.0  $\pm$ 8.0 & 519.7  &  3.8  \\
       SO\,583 & 3    & 2020-12-01 &4725 $\pm$ 117 &0.80 $\pm$ 0.07 & 328.5$\pm$  7.0 &   582.4 &  4.0 \\
\hline
\noalign{\smallskip}
\multicolumn{7}{c}{Cha I} \\
\noalign{\smallskip}
\hline
CS\,Cha  &   1 & 2022-05-11  & 4625 $\pm$ 169 & 0.13 $\pm$ 0.09 &  506.3 $\pm$  4.2  & 563.2 &   3.7   \\
CS\,Cha  &   2 & 2022-05-12&  4648$\pm$ 153&0.09 $\pm$ 0.08& 495.5 $\pm$  9.5&    531.0   &3.7\\
CS\,Cha   &  3 & 2022-05-16& 4527 $\pm$ 183 &0.10 $\pm$ 0.01& 475.4 $\pm$ 9.9  & 514.9     &  3.5  \\
\hline
CV\,Cha  &  1 &2022-05-11& 5083 $\pm$ 71 &0.23 $\pm$ 0.06 & 334.3 $\pm$ 5.0 &     395.4  &   3.6\\
CV\,Cha   &  2 & 2022-05-13&5105 $\pm$ 73 & 0.34 $\pm$0.05 & 337.4 $\pm$ 5.3 &  436.9    &   4.0$^{**}$\\ 
CV\,Cha   &  3 &2022-05-16& 5091 $\pm$ 79& 0.20 $\pm$0.08& 354.4 $\pm$  5.0 &   406.9    & 3.7  \\
\hline
Hn\,5$^*$ & 3 &2021-06-03& 3446 $\pm$ 118 &0.52 $\pm$ 0.47 & 406.3 $\pm$  18.0 & 617.6 &  2.7\\
\hline
IN\,Cha$^*$ & 1& 2021-06-03& 3386 $\pm$ 125 &0.09 $\pm$ 0.05 & 506.5  $\pm$ 17.4&  552.1 & 2.4\\ 
\hline
VW\,Cha &1 & 2022-05-11& 4468 $\pm$  176 &0.88 $\pm$ 0.20 & 377.1 $\pm$  8.8 & 700.7 & 4.0 \\
VW\,Cha &2 & 2022-05-12 &4387 $\pm$ 139 & 1.33 $\pm$ 0.16 & 330.3 $\pm$ 6.6 &  760.9  & 4.0\\
VW\,Cha & 3 & 2022-05-16& 4477 $\pm$ 167 & 0.86 $\pm$ 0.15 & 389.4 $\pm$ 9.5 &  715.6 & 4.0 \\
\hline
VZ\,Cha  &1 &2022-05-04& 4211 $\pm$ 111 & 2.83 $\pm$ 0.39 & 239.7 $\pm$  6.1 &  907.0  & 4.0 \\
VZ\,Cha &2 & 2022-05-07&4126 $\pm$ 146 & 3.38 $\pm$ 0.45 & 224.1  $\pm$  5.7  &  968.3   & 4.0$^{**}$\\
VZ\,Cha &3 & 2022-05-11& 4209 $\pm$ 143 & 4.45 $\pm$ 0.68 & 185.4 $\pm$ 4.5 &   999.1   &4.0$^{**}$\\
\hline
WZ\,Cha$^*$&1bis&2022-06-23&  3419 $\pm$ 118 & 0.88 $\pm$ 0.55 &364.4 $\pm$ 20.3 & 685.1 & 2.9\\
WZ\,Cha$^*$&2& 2022-05-07& 3403 $\pm$ 66 & 0.50 $\pm$ 0.16 & 321.1 $\pm$  13.6 & 481.7    &2.0\\
WZ\,Cha$^*$&3& 2022-05-11&  3425 $\pm$ 110& 1.47 $\pm$ 0.24 & 315.6 $\pm$ 15.0  & 779.5   &3.3 \\
\hline
XX\,Cha$^*$&1 &2021-06-03& 3627 $\pm$ 78 &0.53 $\pm$ 0.16 & 504.3 $\pm$ 12.8  & 771.6  &3.5 \\
XX\,Cha$^*$ &2&2021-06-04&  3603 $\pm$ 64& 0.52 $\pm$ 0.18 & 494.0 $\pm$ 12.4 & 750.9   &3.4\\
XX\,Cha$^*$&3&2021-06-06&  3628 $\pm$ 77& 0.48 $\pm$ 0.32 & 554.7 $\pm$  13.5  &  821.0  &3.7 \\
\hline
\noalign{\smallskip}
\multicolumn{7}{c}{Lupus} \\
\hline
SSTc2dJ160000.6-422158$^*$ &1&  2021-07-21&  3318 $\pm$  104 & 0.00 $\pm$  0.00 &567.9  $\pm$ 13.1 & 567.9 & 2.4\\   
SSTc2dJ160000.6-422158$^*$&2& 2021-07-22& 3378$\pm$  130 & 0.00 $\pm$  0.00 &562.6 $\pm$ 15.6&  562.6 & 2.5\\ 
\hline
SSTc2dJ161243.8-381503$^*$&1& 2022-04-27 &3878 $\pm$  103 & 0.26 $\pm$  0.11 &504.2 $\pm$ 11.5&  635.3 & 3.2\\
SSTc2dJ161243.8-381503$^*$&2bis&2022-05-04&3863 $\pm$  114 & 0.25 $\pm$  0.11 &507.9 $\pm$ 8.6&  634.9 & 3.2\\
SSTc2dJ161243.8-381503$^*$&3& 2022-05-02&  3882 $\pm$  104 & 0.25 $\pm$  0.14 & 501.1 $\pm$ 12.3&  626.4 & 3.2\\
\hline
SSTc2dJ161344.1-373646$^*$&1&2022-05-02& 3303 $\pm$  115 &1.05 $\pm$  0.25 &271.2 $\pm$18.0  &  556.0& 2.1\\
SSTc2dJ161344.1-373646$^*$&2&2022-05-04&3374 $\pm$  160 &0.61 $\pm$  0.30 & 354.2 $\pm$  28.9&  570.3 & 2.3\\
SSTc2dJ161344.1-373646$^*$&3&2022-05-07& 3163 $\pm$  167 & 0.43 $\pm$  0.43 &454.2 $\pm$ 38.2&  649.5 & 2.3\\
\hline
Sz\,84$^*$&1 &2022-05-10&  3253 $\pm$  128 & 0.72 $\pm$  0.26 & 563.4 $\pm$ 22.1  &  969.1 & 4.0 \\ 
Sz\,84$^*$&2&2022-05-12& 3199 $\pm$  117 &0.99 $\pm$  0.48 &539.1 $\pm$ 20.9  & 1072.8 & 4.0$^{**}$\\
Sz\,84$^*$&3& 2022-05-15&  3205 $\pm$  117 & 0.66 $\pm$  0.21 &  533.7 $\pm$  18.5&  885.9 & 3.5\\
\hline
Sz\,97$^*$& 1 &2022-05-11& 3314 $\pm$  107 &0.43 $\pm$  0.39& 521.3 $\pm$ 14.0 & 745.5 & 3.1\\    
Sz\,97$^*$& 2 &2022-05-12& 3314 $\pm$  107 &0.55 $\pm$  0.60 & 508.8 $\pm$13.5 & 788.6 & 3.3\\
Sz\,97$^*$& 3 &2022-05-14&  3311 $\pm$  109 & 0.45 $\pm$  0.55 &513.5 $\pm$ 11.9 & 744.6 & 3.1\\ 
\hline
Sz\,98& 1 &2022-05-03& 4265 $\pm$ 123 &0.21 $\pm$  0.13 &498.8 $\pm$ 8.9 &  594.3    &3.4\\
Sz\,98& 2 &2022-05-06& 4260 $\pm$  115 &0.12 $\pm$  0.06 &538.0 $\pm$  9.3 &   593.2   &3.4\\
Sz\,98& 3 &2022-05-10& 4242 $\pm$  108 & 0.71 $\pm$  0.13 &406.6 $\pm$ 7.3 &  685.6    & 3.6\\
\hline
Sz\,100$^*$ & 1&2022-06-17&  3024 $\pm$  139 &0.61 $\pm$  0.26 &471.2 $\pm$ 18.6 & 758.6& 2.8\\
Sz\,100$^*$ & 2& 2022-06-30& 3005 $\pm$  133 &0.22 $\pm$  0.36 &496.6 $\pm$ 13.7  & 605.9 & 2.1\\    
Sz\,100$^*$ & 3&2022-07-04& 3019 $\pm$  138 &0.40 $\pm$  0.40 &472.1 $\pm$15.1  & 660.9&2.4 \\
\hline
Sz\,103$^*$ & 1&2022-04-28&3046 $\pm$  143&0.73 $\pm$  0.39&409.7 $\pm$ 18.0 & 708.9 & 2.6\\   
Sz\,103$^*$ & 2&2022-05-01&3030 $\pm$  141& 0.63 $\pm$  0.19 &445.0 $\pm$  13.2 & 725.4 &2.5 \\ 
Sz\,103$^*$ & 3&2022-05-04& 3034 $\pm$  140& 0.54 $\pm$  0.38 & 452.4 $\pm$ 15.9 & 696.7 & 2.6\\ 
\hline
Sz\,104$^*$ & 1&2022-06-24& 3303 $\pm$  126&0.90 $\pm$  0.82&   494.7 $\pm$ 32.8 & 939.9 &4.0 \\
Sz\,104$^*$ & 2&2022-07-05& 3381 $\pm$  126&0.49 $\pm$  0.58&   504.3 $\pm$ 17.9 & 751.4 & 3.1 \\   
Sz\,104$^*$ & 3&2022-06-30& 3075 $\pm$  202&0.79 $\pm$  0.72& 383.6 $\pm$ 14.1 & 686.6 & 2.4\\ 
\hline
Sz\,112$^*$ & 1&2022-07-23& 3406 $\pm$  43& 0.35 $\pm$  0.57& 557.7 $\pm$  18.7 & 752.9 & 3.2\\ 
Sz\,112$^*$ & 2&2022-07-24&3461 $\pm$  84 & 0.57 $\pm$  0.63& 561.3 $\pm$ 19.5 & 881.2 & 3.9\\  
Sz\,112$^*$ & 3&2022-07-25& 3406 $\pm$  43& 0.50 $\pm$  0.66 &515.9 $\pm$ 16.9  & 773.9 & 3.3\\
\hline
Sz\,115$^*$ & 1&2022-06-03& 3298 $\pm$  104 &0.20 $\pm$  0.51& 0.0 $\pm$   0.0  & 0.0 & 0.0 \\  
Sz\,115$^*$ & 2bis&2022-06-30&3319 $\pm$  97 &0.41 $\pm$  0.47& 573.4 $\pm$ 10.9   & 808.5 & 3.4\\  
Sz\,115$^*$ & 3&2022-06-09&3314 $\pm$  103 &0.99 $\pm$  0.90 &550.9 $\pm$ 12.5 &1096.3 & 4.0$^{**}$ \\
\hline
Sz\,129 & 1&2022-05-01& 4127 $\pm$  156 & 0.39 $\pm$  0.06 &483.2 $\pm$ 10.0 & 658.3    &3.5\\
Sz\,129 & 2&2022-05-03&4164 $\pm$  142 & 0.19 $\pm$  0.07& 555.0 $\pm$  11.0   &   648.0   &3.5 \\ 
Sz\,129 & 3&2022-05-06& 4065 $\pm$  116 & 0.62 $\pm$  0.04&    450.3 $\pm$ 11.0    &   714.6  & 3.7\\
Sz\,129 & 3b &2022-05-07& 4061 $\pm$  116 & 0.58 $\pm$  0.04 &441.1 $\pm$ 9.0 &  682.0    &3.6  \\
\hline
\noalign{\smallskip}
\multicolumn{7}{c}{Taurus} \\
\hline
DK\,TauA& 1&2021-11-25& 4237 $\pm$ 104 & 0.33 $\pm$ 0.04& 525.0$\pm$ 9.5 &  688.1   &3.6 \\
DK\,TauA& 2&2021-12-01&4246 $\pm$ 104 &0.54 $\pm$ 0.05 &477.4 $\pm$ 7.9&  725.1  &3.7 \\
DK\,TauA& 3&2021-12-02& 4239 $\pm$ 104 &0.35 $\pm$ 0.05 & 517.0 $\pm$  9.5 &   687.8   &3.6\\
\hline
DK\,TauB$^*$& 3&2021-12-02& 3680 $\pm$ 98 &1.06$\pm$ 0.20& 436.7 $\pm$  14.1 & 899.6 & 4.1\\    
\hline
%\multicolumn{7}{p{\linewidth}}{\footnotesize 
%\textit{Notes}: Targets marked with an asterisk (*) are M stars; their $EW_{\rm Li}$ values are corrected only for veiling. NLTE corrections are not available, so the lithium abundances listed are LTE values.
%}\\
%\multicolumn{7}{p{\linewidth}}{\footnotesize 
%Stars marked with ($^{**}$) have $A({\rm Li})$ values higher than 4.0 dex. We fixed the lithium abundance of these stars at 4.0 dex.
%}\\
\end{longtable}
\end{center}

%\begin{threeparttable}
\begin{center}
\scriptsize
\begin{longtable}{|c|c c | c c |c c |c|} 
\caption{ \scriptsize Results from the ESPRESSO spectra.\\
{\scriptsize \textit{Notes}: Targets marked with an asterisk (*) are M stars; their $EW_{\rm Li}$ values are corrected only for veiling. NLTE corrections are not available, so the lithium abundances listed are LTE values.\\
Stars marked with ($^{**}$) have $A({\rm Li})$ values higher than 4.0 dex. We fixed the lithium abundance of these stars at 4.0 dex.}}
\label{tabewli2}  \\
\hline 
\textbf{Name}&\textbf{epoch} & \textbf{Obs. Date}& \bm{$T_{\rm eff}$} & \bm{$r_{650}$} & \bm{$EW_{\rm Li} raw$} & \bm{$EW_{\rm Li}^{\rm veil+Fe}$}& \bm{$A{\rm (Li)}NLTE$}\\ 
  &  & YYYY-MM-DD &[K] & & m\AA & m\AA & dex\\ 
\hline 
\endfirsthead
\caption{ \scriptsize continued from previous page} \\
\hline
\textbf{Name}&\textbf{epoch} &\textbf{Obs. Date}& \bm{$T_{\rm eff}$} & \bm{$r_{650}$} & \bm{$EW_{\rm Li} raw$} & \bm{$EW_{\rm Li}^{\rm veil+Fe}$}& \bm{$A{\rm (Li)}NLTE$}\\ 
 & & YYYY-MM-DD & [K] & & m\AA& m\AA& dex\\ 
\hline 
\endhead
\caption{\scriptsize Continued on next page}\\
\hline\hline
\endfoot
\hline \hline
\endlastfoot
\hline
\hline
\noalign{\smallskip}
\multicolumn{7}{c}{Orion OB1} \\
\noalign{\smallskip}
\hline
CVSO-146& 1& 2020-12-09& 4303 $\pm$ 97& 0.28 $\pm$ 0.04&458.5 $\pm$ 9.8 &   577.0    &3.4 \\
CVSO-146& 2& 2020-12-10& 4372 $\pm$ 101&0.34 $\pm$ 0.09&439.6 $\pm$   9.4 &   579.9  &3.5 \\  CVSO-146& 3& 2020-12-11& 4272 $\pm$ 113&0.42 $\pm$ 0.08&428.4 $\pm$ 9.1 &    597.9   &3.4 \\
\hline
CVSO-165 &1 &2020-12-13&4591 $\pm$ 167&0.25 $\pm$ 0.05&507.2 $\pm$ 7.3 &   625.5  & 3.9\\  
CVSO-165 &2 &2020-12-14&4591 $\pm$ 169&0.32 $\pm$ 0.04&496.5 $\pm$ 7.3 &    646.9  & 3.8\\  
CVSO-165 &3&2020-12-15&4585 $\pm$ 167 &0.36 $\pm$ 0.05&482.4 $\pm$ 6.6  &    645.0  &3.5 \\  
\hline 
\noalign{\smallskip}
\multicolumn{5}{c}{ $\sigma$\,Orionis} \\
\noalign{\smallskip}
\hline
SO\,1153 & 1 &2020-12-08& 4152 $\pm$ 158& 4.81 $\pm$ 0.62& 177.1 $\pm$ 6.8& 1016.2  & 4.0$^{**}$ \\
SO\,1153 & 2 &2020-12-09& 4119 $\pm$ 181&5.23 $\pm$ 0.56& 151.3 $\pm$ 4.5&    928.9   &4.0$^{**}$\\ 
SO\,1153 & 3 &2020-12-10 & 4065 $\pm$ 146&5.71 $\pm$ 0.80& 150.6 $\pm$ 5.6  &   995.4 &4.0$^{**}$ \\
\hline 
\noalign{\smallskip}
\multicolumn{7}{c}{Cha I} \\
\hline
CHX\,18N &1&2021-04-28 &4975 $\pm$ 93&0.08 $\pm$ 0.08&530.4 $\pm$ 4.5&   563.5     &3.8 \\
CHX\,18N &2& 2021-04-29 & 5008 $\pm$ 115& 0.08 $\pm$ 0.08& 512.1 $\pm$ 7.3  &    543.9 &3.7  \\  
CHX\,18N &3&2021-05-01 & 5029 $\pm$ 119&0.06 $\pm$ 0.09&500.0 $\pm$ 5.5 &    521.3 & 3.8\\ 
\hline
Sz\,10$^*$ & 1&2021-05-01&3264 $\pm$ 82&0.71 $\pm$ 0.33&428.8 $\pm$ 16.2&  733.25 & 2.9\\
Sz\,10$^*$ & 2&2021-05-05&3247$\pm$ 93& 0.69 $\pm$ 0.29&426.9 $\pm$ 14.6 &  721.5 & 2.8 \\   
Sz\,10$^*$ & 2b &2021-05-07& 3155$\pm$ 102& 1.25 $\pm$ 0.46& 337.2 $\pm$ 19.8&  758.7 & 2.8\\ 
Sz\,10$^*$ & 3&2021-05-03& 3267 $\pm$ 87& 0.74 $\pm$ 0.46& 371.3 $\pm$ 15.3&  646.1 &  2.5\\  \hline
Sz\,19& 1&2022-03-11&5215 $\pm$ 78&0.24 $\pm$ 0.11& 274.7 $\pm$ 4.5&   328.5  &3.4 \\ 
Sz\,19& 2&2022-03-13& 5232 $\pm$ 74&0.24 $\pm$ 0.11&260.3 $\pm$ 4.0 &    313.8   & 3.3 \\ 
Sz\,19& 3&2022-03-15& 5221 $\pm$ 76&0.22 $\pm$ 0.15& 259.6 $\pm$ 3.5 &   307.9    &3.3 \\
\hline
Sz\,45& 1&2021-05-15& 4091 $\pm$ 52&0.41 $\pm$ 0.08&475.5 $\pm$ 8.0&    656.2   & 3.5\\ 
Sz\,45& 2&2021-05-16& 4166 $\pm$94&0.30 $\pm$ 0.08&502.9$\pm$ 7.1&     641.4  &3.5 \\  
Sz\,45& 3& 2021-05-17& 4110 $\pm$ 57&0.29 $\pm$ 0.09& 508.1 $\pm$ 8.6 &    641.7    & 3.5  \\
\hline 
\noalign{\smallskip}
\multicolumn{7}{c}{ $\eta$\,Cha } \\
\hline
RECX\,5$^*$ & 1 &2022-01-28& 3363 $\pm$ 77& 0.06 $\pm$ 0.12&   604.7 $\pm$ 12.6 &   641.0 &2.8\\   
RECX\,5$^*$ & 2 &2022-01-29&  3454 $\pm$ 104&0.01 $\pm$ 0.06&   600.9 $\pm$ 12.4 &   606.9 & 2.7\\
RECX\,5$^*$ & 3 &2022-01-30& 3406 $\pm$ 111&0.30 $\pm$ 0.06&  601.8 $\pm$ 12.4  &   782.3 & 3.3 \\ 
\hline
RECX\,6$^*$ & 1 &2022-03-02& 3600 $\pm$ 69&0.18 $\pm$ 0.06& 488.2 $\pm$ 7.1 &   576.1 & 2.9\\ 
RECX\,6$^*$ & 2 &2022-03-04& 3588 $\pm$ 52&0.01 $\pm$ 0.06&   495.8 $\pm$  6.1 &   500.8 & 2.7\\ 
\hline
RECX\,9$^*$ & 1 &2022-01-26& 3274$\pm$ 59&0.07 $\pm$ 0.22  &  561.1 $\pm$ 10.2 &   600.4 &2.5  \\   
RECX\,9$^*$ & 2 &2022-01-29& 3315 $\pm$ 139&0.01 $\pm$ 0.05&    533.0 $\pm$ 8.5 &   538.3 & 2.3\\   
RECX\,9$^*$ & 3 &2022-01-28&  3340 $\pm$ 154&0.04 $\pm$ 0.15&    554.7 $\pm$ 9.3 &   576.9 & 2.6 \\ 
\hline
RECX\,11 & 1 &2022-04-10&4614 $\pm$ 91&0.04$\pm$ 0.05& 480.2$\pm$ 2.2&  491.0 & 3.5   \\
RECX\,11 & 2 &2022-04-13& 4665 $\pm$ 83&0.06 $\pm$ 0.05& 475.1$\pm$ 1.9 &   494.8 & 3.6\\  
\hline 
\noalign{\smallskip}
\multicolumn{7}{c}{ Lupus } \\
\hline
SSTc2dJ160830.7-382827 &1&2022-07-03& 5113 $\pm$ 70& 0.00 $\pm$ 0.00&425.6 $\pm$ 4.8 &   417.7    & 3.8\\
SSTc2dJ160830.7-382827 &2&2022-07-05& 5103 $\pm$ 70&0.00 $\pm$ 0.00&436.6 $\pm$  1.8 &  428.6   &3.8  \\
\hline
MY\,Lup& 2&2022-07-03& 5118 $\pm$ 76&0.03 $\pm$ 0.05& 411.1 $\pm$  6.8  &   415.5   & 3.8\\
MY\,Lup& 3&2022-07-06&5129 $\pm$ 97& 0.04 $\pm$ 0.05& 4207. $\pm$ 6.1 &   422.0      & 3.8\\
MY\,Lup& 4&2022-07-07&5138 $\pm$ 64& 0.02 $\pm$0.04&401.4  $\pm$ 6.9 &  394.4   &3.7 \\
MY\,Lup& 3bis&2022-08-21 &5114 $\pm$ 69&0.01 $\pm$0.03&411.4 $\pm$ 6.5&  399.9     & 3.7 \\
MY\,Lup& 5bis& 2022-08-25& 5121 $\pm$87& 0.03 $\pm$ 0.04&423.6 $\pm$  6.4 &   428.3   & 3.8\\
\hline
RULup&4&2022-08-16& 4233$\pm$ 62 & 1.88 $\pm$ 0.39 &328.1 $\pm$ 8.3 &934.2 & 4.0$^{**}$\\
RULup&5&2022-08-23&4251 $\pm$ 58& 1.84 $\pm$ 0.39 &328.5 $\pm$ 6.1 & 923.0 & 4.0$^{**}$ \\
\hline
RY\,Lup& 1&2022-05-27&5167 $\pm$ 60& 0.00 $\pm$ 0.00&343.9 $\pm$ 5.5&   333.9     & 3.3\\
RY\,Lup& 2&2022-05-28&5139 $\pm$ 67&0.00 $\pm$ 0.00& 348.5 $\pm$ 5.3&  338.2    & 3.3\\
RY\,Lup& 3&2022-05-30&5168 $\pm$ 75&0.00 $\pm$ 0.00&344.0 $\pm$ 5.7&  334.8    &3.4 \\
RY\,Lup& 4&2022-05-31&5174 $\pm$ 63&0.00 $\pm$ 0.00&340.6 $\pm$ 6.0&   331.6    & 3.3 \\
RY\,Lup& 5&2022-06-04& 5200 $\pm$ 73&0.00 $\pm$0.00&346.3 $\pm$ 6.2&   336.5     & 3.4  \\
\hline
Sz\,66$^*$ &1&2021-05-15&3340 $\pm$ 88& 0.50 $\pm$ 0.12&485.7 $\pm$ 10.7& 728.6 & 3.0\\
Sz\,66$^*$ &2&2021-05-16&3326 $\pm$ 82&0.50 $\pm$ 0.10&492.8 $\pm$ 11.9& 739.2 & 3.0\\
\hline
Sz\,71$^*$ &1&2021-05-05& 3598 $\pm$ 107& 0.30 $\pm$ 0.21 &569.1 $\pm$ 9.5&739.8 & 3.4   \\
Sz\,71$^*$ &2&2021-05-09&3578 $\pm$ 51&0.13 $\pm$ 0.11& 548.8 $\pm$ 6.0&  620.1 & 3.0\\
Sz\,71$^*$ &3&2021-05-12& 3584 $\pm$ 52& 0.23 $\pm$ 0.10&520.5 $\pm$ 6.2& 640.2 &3.0 \\
\hline
Sz\,72$^*$ &1& 2021-05-02&3287 $\pm$ 65& 1.45 $\pm$ 0.25&  298.7 $\pm$ 6.2& 731.8 & 2.8 \\
Sz\,72$^*$ &2& 2021-05-05& 3298 $\pm$ 63& 1.40 $\pm$ 0.34&397.2 $\pm$ 8.7& 953.3 & 4.0\\
Sz\,72$^*$ &3&2021-05-12&3319 $\pm$ 54&0.84 $\pm$ 0.17& 421.5 $\pm$ 9.7& 775.6 & 3.1\\
\hline
Sz\,75 &1&2021-05-02&4497 $\pm$ 105&0.39 $\pm$ 0.03&424.9 $\pm$ 8.8&   582.5  & 3.7  \\
Sz\,75 &2&2021-05-05&4519 $\pm$ 82&0.21 $\pm$ 0.06&468.8 $\pm$  9.6&  559.0  & 3.6\\
Sz\,75 &3&2021-05-07& 4488 $\pm$ 89& 0.20 $\pm$ 0.04&489.1 $\pm$ 10.4&  578.8  & 3.6\\
\hline
Sz\,76$^*$ &1&2021-05-09&3512 $\pm$ 77& 0.02 $\pm$ 0.07&608.7 $\pm$ 11.4& 620.9 & 2.8\\
Sz\,76$^*$ &2&2021-05-10&3486 $\pm$ 88& 0.02 $\pm$ 0.07&605.3 $\pm$ 11.6& 617.4 & 2.8 \\
Sz\,76$^*$ &3&2021-05-18& 3514 $\pm$ 81& 0.02 $\pm$ 0.07&612.7 $\pm$11.6 & 625.0 & 2.8\\
Sz\,76$^*$ &4&2021-08-07&3471 $\pm$ 80&0.02 $\pm$ 0.07&598.9 $\pm$ 11.7& 610.9& 2.7\\
\hline
Sz\,77 &2 bis &2021-05-12&4204 $\pm$ 59&0.35 $\pm$ 0.09&470.8 $\pm$ 8.1&   624.5    & 3.4\\
Sz\,77 &3&2021-05-09& 4253 $\pm$ 57&0.16 $\pm$ 0.08&521.4 $\pm$ 10.6 &  594.7    &3.4 \\ 
\hline
Sz\,110$^*$ &2&2022-05-28&3330 $\pm$ 76& 0.13 $\pm$ 0.13&467.0 $\pm$ 9.2& 527.7 & 2.2\\
Sz\,110$^*$ &3&2022-05-31&3322 $\pm$ 81& 0.61 $\pm$ 0.10&394.8 $\pm$ 7.5& 635.6 & 2.6\\
\hline
Sz\,111$^*$ &1&2021-06-14&3769 $\pm$ 46& 0.17 $\pm$ 0.11&458.3 $\pm$ 10.4 & 536.2 & 2.8\\
Sz\,111$^*$ &2 bis&2021-08-31&3798 $\pm$ 55& 0.05 $\pm$ 0.07& 517.5 $\pm$  11.8& 543.4 & 2.8\\
\hline
Sz\,114$^*$ &1&2022-05-27&3460 $\pm$ 97& 0.25 $\pm$ 0.13& 567.6 $\pm$ 15.1& 709.5 & 3.1\\
Sz\,114$^*$ &2&2022-05-29& 3386 $\pm$ 70& 0.23 $\pm$ 0.10& 579.2 $\pm$ 15.1& 712.4& 3.0\\
Sz\,114$^*$ &3&2022-05-31&3431 $\pm$ 114& 0.23 $\pm$ 0.08& 575.9  $\pm$  15.3&708.4 & 3.1\\
\hline
Sz\,117$^*$ &1&2022-05-30&3596 $\pm$ 74& 0.05 $\pm$ 0.14&501.2 $\pm$ 9.3& 526.3& 2.6\\
Sz\,117$^*$ &2&2022-06-01& 3605 $\pm$78& 0.04 $\pm$ 0.10& 532.9 $\pm$ 9.5& 554.2& 2.7\\
\hline
Sz\,130$^*$ &1&2021-06-13& 3657 $\pm$ 92& 0.16 $\pm$ 0.13&553.9 $\pm$ 11.8& 642.5 & 3.1\\
Sz\,130$^*$ &2&2021-07-21&3711 $\pm$ 65& 0.08 $\pm$ 0.14& 584.1 $\pm$  12.1 & 630.8 & 3.1\\
Sz\,130$^*$ &3&2021-07-22&3678 $\pm$ 76& 0.06 $\pm$ 0.14&583.5 $\pm$ 11.0& 618.5 & 3.0 \\
\hline 
\noalign{\smallskip}
\multicolumn{7}{c}{ Taurus } \\
\hline
AA\,Tau &2&2021-12-02&4140 $\pm$ 130&0.93 $\pm$ 0.30&470.5 $\pm$ 28.1 &    895.1    & 4.0$^{**}$\\
AA\,Tau$^*$ &3&2021-12-03&3912 $\pm$ 255& 0.62 $\pm$ 0.25&441.8 $\pm$ 35.5 & 715.7 & 3.5 \\  
\hline
BP\,Tau & 3 bis &2021-09-07&4190 $\pm$ 77&0.62 $\pm$ 0.21&406.6 $\pm$ 15.7&647.0&3.5\\
BP\,Tau & 5 &2021-09-02& 4154 $\pm$ 97&1.11 $\pm$ 0.11 &346.6$\pm$  6.3&718.7&3.7\\
\hline
DE\,Tau$^*$ &1&2021-11-23&3569 $\pm$ 57&0.53 $\pm$ 0.11&447.0 $\pm$  6.6& 683.9 & 3.1\\   
DE\,Tau$^*$ &2&2021-11-24&3573 $\pm$ 58& 0.43 $\pm$ 0.10& 471.1 $\pm$ 6.1 & 673.7 &3.1 \\    
DE\,Tau$^*$ &3&2021-11-25&3572 $\pm$57& 0.33 $\pm$ 0.09& 478.4 $\pm$ 5.9&  636.3 & 3.0\\  
\hline
DM\,Tau$^*$ &1&2021-11-27&3579 $\pm$ 48&0.44 $\pm$ 0.13&400.2 $\pm$ 12.6 & 576.3 & 2.7\\   
DM\,Tau$^*$ &2&2021-11-28&3588 $\pm$ 53&0.58 $\pm$ 0.19&367.4 $\pm$ 11.4 & 580.5 &2.7 \\   
DM\,Tau$^*$ &3&2021-11-29& 3573 $\pm$ 53&0.56 $\pm$ 0.15&383.9 $\pm$ 12.1 & 598.9 & 2.9\\   
\hline
DN\,Tau &1&2021-12-01&4178 $\pm$ 111&0.05 $\pm$ 0.07&598.8 $\pm$ 8.8  & 618.7   &  3.4 \\ 
DN\,Tau &2&2021-12-02&4181 $\pm$ 112& 0.04 $\pm$ 0.07&597.0 $\pm$  7.8  & 608.8 &  3.4\\  
DN\,Tau &3&2021-12-03&4191 $\pm$ 110& 0.01 $\pm$ 0.05& 605.0 $\pm$ 9.7  &599.3 &  3.4\\  
\hline
GMAur&1&2021-10-22&4621 $\pm$ 144&0.73 $\pm$ 0.07& 334.1 $\pm$ 7.6&569.4&3.8\\
GMAur&2&2021-12-05& 4509 $\pm$    81&0.15 $\pm$ 0.05& 452.1 $\pm$ 9.6&  511.2  & 3.4\\
GMAur&3&2021-12-06&4886  $\pm$  164&0.20 $\pm$ 0.00&438.1  $\pm$ 14.4&  516.6  &3.9\\
GMAur&4&2021-12-07&4493  $\pm$   95&0.47 $\pm$ 0.05&376.2 $\pm$7.1&   544.8  & 3.5 \\
GMAur&5&2021-12-08&4676  $\pm$  137 &0.50 $\pm$ 0.02&369.4 $\pm$ 8.2&    545.3 & 3.8 \\
\hline
LkCa\,15&1&2021-12-03&4882 $\pm$ 95& 0.04 $\pm$ 0.05&450.2 $\pm$ 9.4  & 458.7  & 3.7\\  
LkCa\,15&2&2021-12-04&4827 $\pm$ 71& 0.04 $\pm$ 0.05&447.2 $\pm$ 8.5 &  455.7 & 3.6\\   
LkCa\,15&3&2021-12-05&4817 $\pm$ 66& 0.06 $\pm$ 0.05& 447.8 $\pm$ 8.5 &465.3  & 3.6 \\  
\hline
LkCa\,4&1&2021-11-23&4274 $\pm$ 83&0.02 $\pm$ 0.06&630.4 $\pm$ 6.2 & 632.8  & 3.5 \\  
LkCa\,4&2&2021-11-24&4126 $\pm$ 196& 0.03 $\pm$ 0.08&634.5 $\pm$ 6.0 &  639.9  & 3.4\\  
LkCa\,4&3&2021-11-25&4203 $\pm$ 152& 0.02 $\pm$ 0.06& 649.7 $\pm$ 5.9 &   651.2  & 3.5  \\  
\hline
RX\,J0438.6+1546& 1&2021-12-07&5119$\pm$ 66& 0.00 $\pm$ 0.00& 403.4 $\pm$  6.7 &   387.8  & 3.6\\  
RX\,J0438.6+1546& 2&2021-12-08&5125 $\pm$ 60&0.00 $\pm$ 0.00&  400.7 $\pm$  6.8  & 385.9 & 3.6 \\  
\hline
\noalign{\smallskip}
\multicolumn{7}{c}{CrA} \\
\noalign{\smallskip}
\hline
RXJ1842.9-1546&3&2022-07-02&4698 $\pm$ 162&0.35$\pm$ 0.09&404.2$\pm$6.9&536.7&3.8\\
\hline
RXJ1852.3-3532&1&2022-07-02&4863 $\pm$ 114&0.00$\pm$ 0.00 &502.4$\pm$8.9&493.0&3.8\\
RXJ1852.3-3532&2&2022-07-03&4811 $\pm$101&0.00$\pm$ 0.00&505.0$\pm$8.2&495.7&3.8\\
RXJ1852.3-3532&3&2022-07-06&4931 $\pm$107&0.04 $\pm$ 0.05& 497.8$\pm$8.7&508.3& 3.9\\
\hline
%\multicolumn{7}{p{\linewidth}}{\footnotesize 
%\textit{Notes}: Targets marked with an asterisk (*) are M stars; their $EW_{\rm Li}$ values are corrected only for veiling. NLTE corrections are not available, so the lithium abundances listed are LTE values.
%}\\
%\multicolumn{7}{p{\linewidth}}{\footnotesize 
%Stars marked with ($^{**}$) have $A({\rm Li})$ values higher than 4.0 dex. We fixed the lithium abundance of these stars at 4.0 dex.
%} \\
\end{longtable}
%\end{threeparttable}
\end{center}

\begin{center}
\scriptsize
\begin{longtable}{|c|c c| c c |c c |c|} 
\caption{\scriptsize Results from the X-Shooter spectra.\\
{\scriptsize \textit{Notes}: Targets marked with an asterisk (*) are M stars; their $EW_{\rm Li}$ values are corrected only for veiling. NLTE corrections are not available, so the lithium abundances listed are LTE values.\\
Stars marked with ($^{**}$) have $A({\rm Li})$ values higher than 4.0 dex. We fixed the lithium abundance of these stars at 4.0 dex.
} }
\label{tabewli3}  \\
\hline 
\textbf{Name}&\textbf{epoch} & \textbf{Obs. Date}& \bm{$T_{\rm eff}$} & \bm{$r_{710}$} & \bm{$EW_{\rm Li} raw$} & \bm{$EW_{\rm Li}^{\rm veil+Fe}$}& \bm{$A{\rm (Li)}NLTE$}\\ 
  &  & YYYY-MM-DD &[K] & & m\AA & m\AA & dex\\ 
\hline 
\endfirsthead
\caption{\scriptsize  continued from previous page} \\
\hline
\textbf{Name}&\textbf{epoch} &\textbf{Obs. Date}& \bm{$T_{\rm eff}$} & \bm{$r_{710}$} & \bm{$EW_{\rm Li} raw$} & \bm{$EW_{\rm Li}^{\rm veil+Fe}$}& \bm{$A{\rm (Li)}NLTE$}\\ 
 & & YYYY-MM-DD &[K] & & m\AA& m\AA& dex\\ 
\hline 
\endhead
\caption{\scriptsize Continued on next page}\\
\hline\hline
\endfoot
\hline \hline
\endlastfoot
\hline
\hline
\noalign{\smallskip}
\multicolumn{7}{c}{ Orion OB1} \\
\noalign{\smallskip}
\hline
CVSO-17$^*$ &1&2020-12-05&3704 $\pm$ 25& 0.0&422.2 $\pm$ 39.3 &   422.2 & 2.2\\ 
CVSO-36$^*$&1&2020-12-03&3670 $\pm$ 38& 0.1&576.0 $\pm$   29.7 &   633.6& 3.2\\ 
CVSO-58$^*$&1&2020-12-02&3968 $\pm$ 36& 0.2& 390.7 $\pm$ 23.4 &   468.8 & 2.8\\ 
CVSO-90$^*$&1&2020-12-15&3481 $\pm$ 32& 1.8&110.6 $\pm$ 21.5 &   309.7 & 1.0  \\ 
CVSO-107$^*$&1&2020-12-04&3812 $\pm$ 49& 0.6&445.5 $\pm$ 31.0  &   712.8 & 3.3 \\ 
CVSO-109$^*$&1&2020-11-28&3827 $\pm$ 34& 0.6& 421.4 $\pm$ 24.2  &   547.8 & 2.8\\ 
CVSO-146$^*$&1&2020-12-09&3995 $\pm$ 62& 0.4& 453.9 $\pm$ 25.7 &   635.4 &3.4\\ 
CVSO-165$^*$&1&2020-12-14&3976 $\pm$ 52& 0.3& 528.6 $\pm$ 25.6  &   687.2 & 3.6\\ 
CVSO-176$^*$&1&2020-12-02&3566 $\pm$ 23& 0.3&293.3 $\pm$ 28.7 &   381.3 & 1.8 \\ 
\hline
\noalign{\smallskip}
\multicolumn{7}{c}{ $\sigma$\,Orionis} \\
\noalign{\smallskip}
\hline
SO\,518$^*$&1&2020-12-02&3929 $\pm$ 61& 0.5& 425.1   $\pm$ 50.5  &   637.7 & 3.4\\
SO\,583&1&2020-12-02&4478 $\pm$ 157&0.8& 334.5 $\pm$ 24.1  & 587.3   & 3.7 \\ 
SO\,1153&1&2020-12-07& 4086$\pm$ 63 &2.4& 196.1 $\pm$ 20.7 &   640.5 &3.7  \\ 
SO\,1153&2&2021-02-13& 4657 $\pm$ 304 & 2.4 & 258.5 $\pm$ 6.0 &   858.6    & 4.0$^{**}$\\  
\hline
\noalign{\smallskip}
\multicolumn{7}{c}{ Cha I} \\
\noalign{\smallskip}
\hline
CHX\,18N& 1&2021-04-28&4025 $\pm$ 42&0.3&487.5 $\pm$ 33.3  &    610.9    &3.5\\
CHX\,18N&2&2021-04-29&4164 $\pm$ 110& 0.3& 497.7 $\pm$  29.4  &   629.4    & 3.5 \\
CS\,CHa&1&2022-05-11&4069 $\pm$ 86& 0.4& 490.0 $\pm$ 34.2 &   661.5 & 3.6\\
CV\,Cha& 1&2022-05-11&5061 $\pm$ 122& 0.3&  292.9 $\pm$ 20.1 &   369.6  & 3.4 \\ 
CV\,Cha&2&2022-05-13&5130 $\pm$ 158& 0.1& 313.2 $\pm$ 21.4 &   331.6     & 3.2\\
EPCHA&1&2022-04-12&4031 $\pm$ 56&0.5&462.1 $\pm$ 29.3 &   678.2      &3.6 \\ 
INCha$^*$&1&2021-06-08&3106 $\pm$ 72& 0.5&526.6  $\pm$ 29.0 &    631.9 & 3.8\\
SYCha$^*$&1&2022-03-23&3951 $\pm$ 60& 0.6&386.2 $\pm$ 16.0 & 617.9 & 3.4\\
Sz\,10$^*$&1&2021-04-29&3167 $\pm$ 63&0.5&339.0 $\pm$ 35.0 & 508.5 & 1.8\\
Sz\,19&1&2022-03-12&5457  $\pm$  121&0.5& 234.1 $\pm$ 20.3 &   340.2     & 3.6 \\
Sz\,19&2&2022-03-12&5575  $\pm$  185& 0.7& 241.2 $\pm$ 21.8 &   400.2  & 4.0$^{**}$ \\
Sz\,45$^*$&1&2021-05-16&3887 $\pm$ 28&0.2&490.2 $\pm$ 27.2 & 588.2 & 3.0\\ 
VWCha$^*$&1&2022-05-13&3936 $\pm$ 52& 0.4&438.8 $\pm$ 26.3 & 614.3 & 3.4\\ 
VZ\,Cha$^*$&1&2022-05-07&3906 $\pm$ 61& 1.7&218.0 $\pm$ 13.0 & 588.6 & 3.3\\ 
WZCha$^*$&1&2022-05-04& 3233 $\pm$ 58&0.5& 323.1 $\pm$ 35.2 & 484.7 &  2.1 \\    
XXCha$^*$&1&2021-06-05&3568 $\pm$ 22&0.0&497.4 $\pm$ 28.2 & 497.4 & 2.7\\
\hline
\noalign{\smallskip}
\multicolumn{7}{c}{ $\eta$\,Cha} \\
\noalign{\smallskip}
\hline
ECHA\,J0843.3-7915$^*$&1&2022-04-09&3418 $\pm$ 40&0.2&521.1 $\pm$ 63.8 &  625.3 & 3.0 \\ 
ECHA\,J0844.2-7833$^*$&1&2021-04-27&3034 $\pm$ 29&0.0&598.0 $\pm$ 80 & 598.0 & 2.2\\
ECHA\,J0844.2-7833$^*$& 2&2021-05-01&3047 $\pm$ 34&0.0&480.9 $\pm$ 78.5 & 480.9 & 1.7\\
RECX-1&1&2022-04-09&4069 $\pm$ 88&0.3&492.6 $\pm$ 30.3  &   617.12  &  3.5 \\  
RECX-5$^*$&1&2022-01-28&3231 $\pm$ 56&0.3&593.1 $\pm$ 51.3 & 771.0 & 3.3\\
RECX-6$^*$&1&2022-03-02&3523 $\pm$ 15&0.0&    483.7 $\pm$ 29.1 & 483.7& 2.6\\
RECX-9$^*$&1&2022-01-26&3057 $\pm$ 41&0.2&564.6 $\pm$ 51.0 & 677.5 & 3.1\\
RECX\,11&1&2022-04-12&4918 $\pm$ 74 & 0.0 &462.0 $\pm$  29.5 & 445.5   & 3.7 \\  
\hline
\noalign{\smallskip}
\multicolumn{7}{c}{Lupus} \\
\noalign{\smallskip}
\hline
MY\,Lup &1&2022-06-30&4587 $\pm$ 165&0.5&401.1 $\pm$ 0.03 &    591.3    & 3.8\\
RX\,J1556.1-3655$^*$&1&2022-06-23&3686 $\pm$ 40&0.9&367.6 $\pm$  0.02  &  698.4   & 3.3\\
RY\,Lup&1&2022-05-28&5120 $\pm$ 81&0.0&  330.8 $\pm$ 0.03 &  318.9   & 3.2\\ 
SSTc2dJ160000.6-422158$^*$ &1&2021-07-21&3105 $\pm$ 59&0.2& 561.4 $\pm$ 37.4  & 673.7     & 2.7 \\
SSTc2dJ160830.7-382827&1&2022-07-02&4875 $\pm$ 121&0.0&412.8 $\pm$ 28.9 &   399.9   & 3.3\\
SSTc2dJ161243.8-381503$^*$&1&2022-05-01&3844 $\pm$ 48&0.3&526.8 $\pm$ 35.1 & 684.8   & 3.3\\  
SSTC2DJ161344.1-373646$^*$ &1&2022-05-03&3207 $\pm$ 46&0.8& 140.00 $\pm$ 30 & 252.0 &0.0 \\
Sz\,66$^*$&1&2021-05-16&3337 $\pm$ 43&0.2&471.7 $\pm$ 40.9 & 566.0 & 3.1 \\ 
Sz\,68&1&2022-06-30&4640 $\pm$ 178&0.4&409.9 $\pm$ 26.3  &   563.2 & 4.0\\    
Sz\,69$^*$&1&2021-05-02&3200 $\pm$ 47&0.6&218.0 $\pm$ 31.4 & 348.8 & 1.7\\ 
Sz\,69$^*$&2&2021-05-03&3255 $\pm$ 135&0.7&0.00  $\pm$   0.0   &  0.0 & 0.0\\ 
Sz\,71$^*$&1&2021-05-04&3564 $\pm$ 23&0.2&538.2 $\pm$ 36.2 & 645.8 & 3.3\\ 
Sz\,72$^*$&1&2021-05-03&3413 $\pm$ 62&1.1&305.9 $\pm$ 23.5 & 642.4 & 3.4\\ 
Sz\,75$^*$&1&2021-05-02&3971 $\pm$ 69&0.6&400.2 $\pm$ 23.5 & 640.3 &3.4 \\ 
Sz\,75$^*$&2&2021-05-03&3991 $\pm$ 87&0.6&423.3 $\pm$ 26.3 & 677.3 & 3.8\\ 
Sz\,76$^*$&1&2021-05-08&3316 $\pm$ 84&0.5&592.4 $\pm$ 41.7 & 829.4 & 4.0\\ 
Sz\,76$^*$&2&2021-08-08& 3527 $\pm$ 10&0.0&587.2 $\pm$ 49.1 & 587.2 & 2.7\\ 
Sz\,76$^*$&2b&2021-08-08&3529 $\pm$ 13&0.0& 582.3 $\pm$ 44.6  & 582.3 & 2.9\\ 
Sz\,77$^*$&1&2021-05-08&3945 $\pm$ 60&0.4& 542.6 $\pm$  30.0 & 759.6 &3.7   \\ 
Sz\,82$^*$&1&2022-06-23&3974 $\pm$ 71&0.5&399.1 $\pm$ 25.7& 518.8 &3.2 \\ 
Sz\,84$^*$&1&2022-05-11&3194 $\pm$ 87&0.3&529.7 $\pm$ 44.4  & 688.6 & 3.3 \\ 
Sz\,97$^*$&1&2022-05-13&3175 $\pm$ 60&0.5&473.0 $\pm$ 29.4 & 709.5 & 3.2\\ 
Sz\,98&1&2022-05-04&4084 $\pm$ 88&0.1&544.4 $\pm$ 29.4 &   581.2 & 3.5 \\   
Sz\,100$^*$&1&2022-06-24&3176 $\pm$ 110&0.1&465.0 $\pm$ 48.1 & 511.5 & 2.0\\
Sz\,103$^*$&1&2022-05-01&3187 $\pm$ 61&0.4&441.7 $\pm$ 52.5 & 618.4  & 2.9\\
Sz\,104$^*$&1&2022-06-24&3328 $\pm$ 55&0.2&388.8 $\pm$ 65.0  & 466.6     & 1.9 \\
Sz\,110$^*$&1&2022-05-24&3330 $\pm$ 73&0.6&475.6 $\pm$ 33.3 & 761.0 & 3.6\\ 
Sz\,114$^*$&1&2022-05-26&3099 $\pm$ 52&0.3&546.6 $\pm$ 40.9  &710.6 & 3.2\\ 
Sz\,115$^*$&1&2022-06-24&3253 $\pm$ 68&0.3&597.9 $\pm$ 52.3 & 777.3 & 4.0$^{**}$ \\ 
Sz\,117$^*$&1&2022-05-30&3534 $\pm$ 11&0.3&510.6 $\pm$ 31.4 & 663.8 & 3.3 \\ 
Sz\,129$^*$&1&2022-05-01&3991 $\pm$ 39&0.0&473.8 $\pm$ 30.0 & 473.8 & 2.8\\ 
Sz\,130$^*$&1&2021-07-20&3536 $\pm$ 21&0.1&575.7 $\pm$ 39.2 & 633.3 & 3.1\\ 
\hline
\noalign{\smallskip}
\multicolumn{7}{c}{Taurus} \\
\noalign{\smallskip}
\hline
AATau$^*$  &1& 2021-12-02&3949 $\pm$ 38&0.2& 437.1 $\pm$ 52.6 & 524.5& 3.1\\
BPTau$^*$&1&2021-08-22&3975 $\pm$ 40&0.30&426.0 $\pm$ 24.3&553.8  & 3.2\\
BPTau$^*$&2&2021-08-26&3978 $\pm$ 38&0.20&393.7$\pm$ 22.3& 472.4  &   2.7\\ 
BPTau$^*$&3&2021-09-03&3927 $\pm$ 52&0.40& 349.6 $\pm$18.7& 489.4&     2.9 \\
DETau$^*$  &1&2021-11-26& 3655 $\pm$ 22&0.2&474.7 $\pm$ 28.8 & 569.6 & 2.3\\
DKTau$^*$  &1&2021-11-26& 3983 $\pm$ 49&0.3&525.9 $\pm$ 25.5 & 683.7 & 3.6\\
DMTau$^*$  &1&2021-11-28& 3693 $\pm$ 86&0.4&375.4 $\pm$ 24.7  & 525.6 &2.7 \\
DNTau$^*$  &1&2021-12-02& 3906 $\pm$ 24&0.1&574.9 $\pm$ 30.1  & 632.4 &3.4 \\
GMAur &1 &2021-10-17&4151 $\pm$ 108&0.40&456.8 $\pm$ 30.2&    616.1   &3.6\\
GMAur &1b &2021-10-17&4086 $\pm$ 83&0.40&452.9 $\pm$ 27.7&  616.9   &3.7\\
GMAur &3 &2021-12-08&4031 $\pm$ 51&0.70&369.5 $\pm$22.0&   613.3    &3.5\\
LkCa\,15&1&2021-12-04&4109 $\pm$ 85&0.4& 449.5 $\pm$ 29.5&  614.9    & 3.6 \\
LkCa\,4$^*$ & 1 &2021-11-24& 3715 $\pm$ 71&0.4&634.3 $\pm$ 19.5&    888.0 & 3.7 \\ 
RX\,J0438.6+1546 &1& 2021-12-08& 4803 $\pm$ 124&0.0&  383.3 $\pm$ 32.4 &  374.3  & 3.2 \\  
\hline
\noalign{\smallskip}
\multicolumn{7}{c}{CrA} \\
\noalign{\smallskip}
\hline
RXJ1842.9-3532$^*$& 1&2022-06-24& 3980 $\pm$ 89 &0.30&400.3 $\pm$28.4 &  520.4 & 3.1\\
RXJ1852.3-3700&1&2022-07-02&  4103 $\pm$ 107 &0.40 &500.4 $\pm$ 37.6&674.3& 3.8 \\
\hline
%\multicolumn{7}{p{\linewidth}}{\footnotesize 
%\textit{Notes}: Targets marked with an asterisk (*) are M stars; their $EW_{\rm Li}$ values are corrected only for veiling. NLTE corrections are not available, so the lithium abundances listed are LTE values.
%}\\
%\multicolumn{7}{p{\linewidth}}{\footnotesize 
%Stars marked with ($^{**}$) have $A({\rm Li})$ values higher than 4.0 dex. We fixed the lithium abundance of these stars at 4.0 dex.
%} \\
\end{longtable}
\end{center}

\section{ Microturbulence ($\xi$) and macroturbulence ($v_{mac}$) velocities for the subsample of targets selected for [Fe/H] and [Ba/H] abundance measurements.}

%\FloatBarrier

\begin{table}[h!]
\scriptsize
\caption{}         
\label{micmac}      
\centering  
\begin{tabular}{l|c| ccc} 
\hline \hline
     Name &  Name  & ep & $\xi$ & $v_{mac}$ \\
    Target &  Cluster  &   &  km/s &   km/s    \\
\hline 
\noalign{\smallskip}
\multicolumn{5}{c}{ESPRESSO} \\
\noalign{\smallskip}
\hline  
CHX\,18N & Cha I & 1 &   0.74  &  1.78\\
CHX\,18N & Cha I & 2 &   0.75  &  1.81 \\
CHX\,18N & Cha I & 3 &  0.76   & 1.83\\
\hline
LkCa\,15 & Taurus&  1&  0.71  &  1.70\\
LkCa\,15 & Taurus&  2& 0.69   & 1.66 \\
LkCa\,15 & Taurus&  3&  0.69  &  1.66\\
\hline
MY\,Lup  & Lupus&   2&  0.79  &  1.93\\
MY\,Lup  & Lupus&   3&  0.80  &  1.94\\
MY\,Lup  & Lupus&   4&  0.80  &  1.95\\
MY\,Lup  & Lupus&   2bis & 0.79&    1.92\\
MY\,Lup  & Lupus&   5bis & 0.79&    1.93 \\
\hline
RECX\,11 &  $\eta$\,Cha& 1 &  0.63 &   1.54 \\
RECX\,11 &  $\eta$\,Cha& 2 & 0.64 &  1.57\\
\hline
RX\,J0438.6+1546 &  Taurus&  1&0.79  &  1.93\\
RX\,J0438.6+1546 &  Taurus&  2&     0.80 &   1.94\\
\hline
RY\,Lup   &  Lupus &  1& 0.87 &  1.99 \\
RY\,Lup   &  Lupus &  2& 0.86 &   1.96\\
RY\,Lup   &  Lupus &  3& 0.88 &   1.99\\
RY\,Lup   &  Lupus &  4& 0.89 &   2.00\\
RY\,Lup   &  Lupus &  5& 0.88 &   2.04 \\
\hline
SSTc2dJ160830.7-382827  &  Lupus&  2 & 0.80  &  1.95  \\
SSTc2dJ160830.7-382827  &  Lupus&  3 & 0.79  &  1.91\\
\hline
Sz\,75  & Lupus &3 & 0.59  &  1.49 \\
\hline
\hline
\noalign{\smallskip}
\multicolumn{4}{c}{ UVES} \\
\noalign{\smallskip}
\hline
CS\,Cha &Cha I & 1&   0.63  &  1.55  \\
CS\,Cha &Cha I & 2&   0.63  &  1.56 \\
CS\,Cha &Cha I & 3&    0.60 &   1.50\\
CV\,Cha & Cha I & 3&    0.87  &  1.90 \\
\hline
\hline
\noalign{\smallskip}
\multicolumn{4}{c}{ XS} \\
\noalign{\smallskip}
\hline
    MY\,Lup  & Lupus & 1&  0.61&1.53\\
   RECX\,11  & ECha &1& 0.73&1.73\\
 RX\,J0438.6+1546 & Taurus &1&0.69  &  1.65\\
    RY\,Lup  &Lupus&1 &0.78 &  1.93\\
SSTc2dJ160830.7-382827&Lupus &1&0.72 &   1.70\\
     Sz\,68  &Lupus &1& 0.72  &  1.55 \\
\hline
\end{tabular}
\end{table}

\newpage
\section{Upper age limit estimates}

\begin{figure*}[h!]
      \centering
	   \begin{minipage}{0.4\linewidth}	
       \centering
       \includegraphics[width=\linewidth]{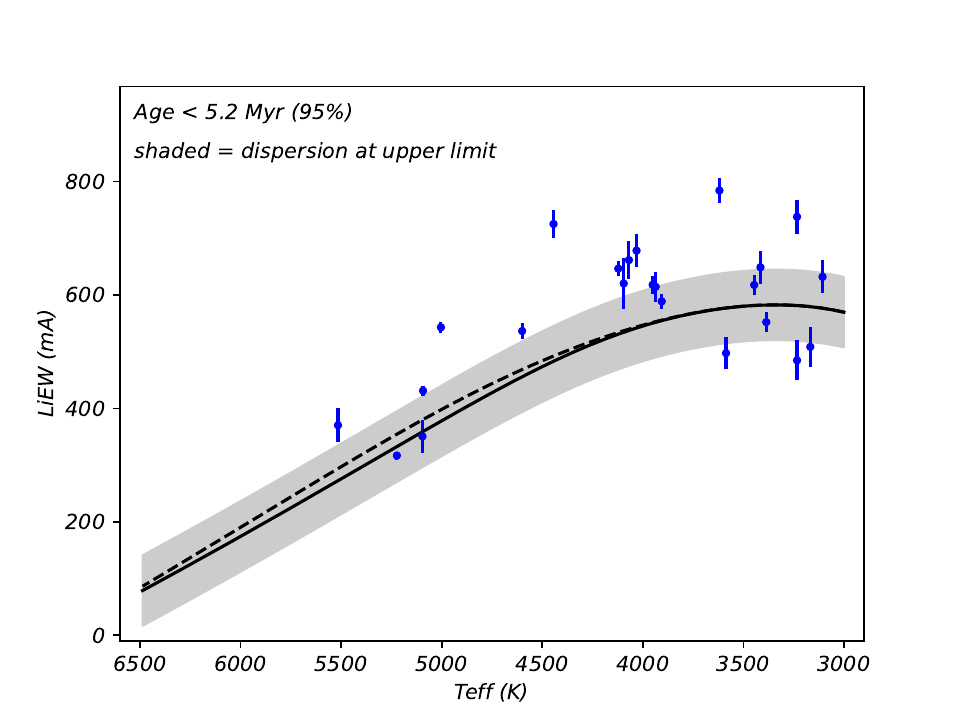}
       \par\small  Cha I
%		\caption{Cha I}
%		\label{fig:subfig1}
	   \end{minipage}
   \begin{minipage}{0.40\linewidth}
   \centering
   \includegraphics[width=\linewidth]{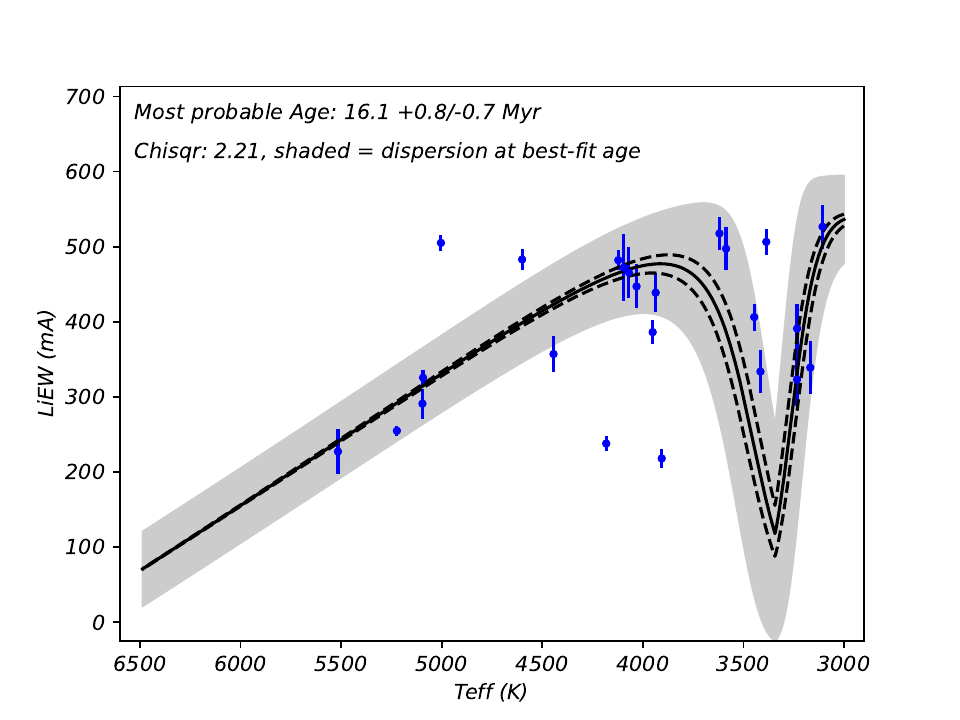}
%\par\small (b) Orion OB1
%		\caption{Orion OB1}
%		\label{fig:subfig2}
	    \end{minipage}
	\vfill
	     \begin{minipage}{0.40\linewidth}
         \centering
		 \includegraphics[width=\linewidth]{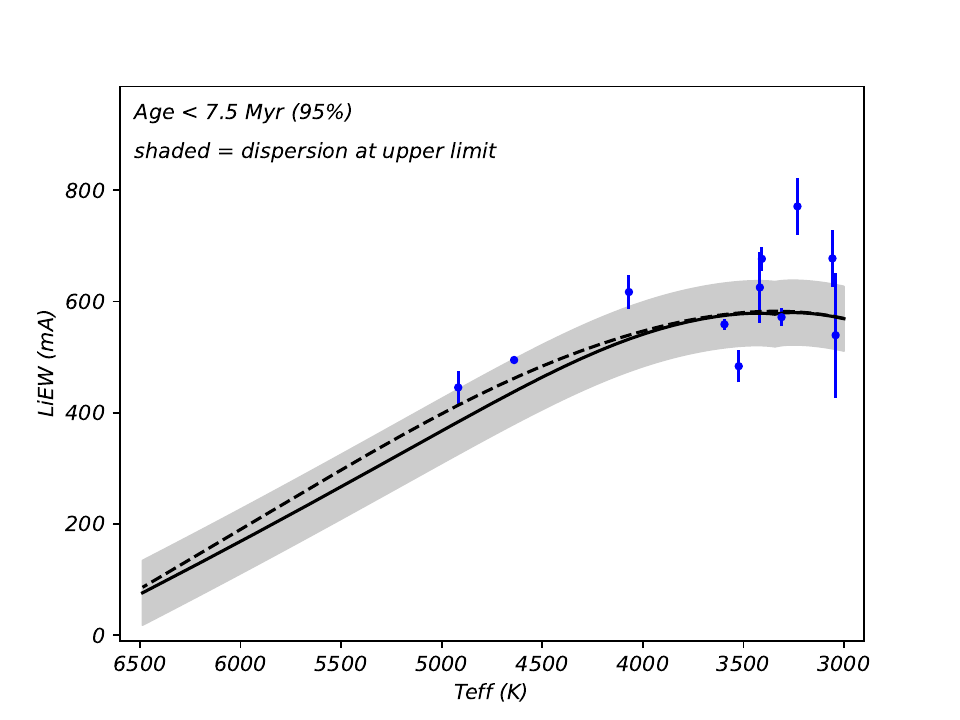}
\par\small  $\eta$ Cha
%		 \caption{Lupus}
%		 \label{fig:subfig3}
	      \end{minipage}
	       \begin{minipage}{0.4\linewidth}
           \centering
		  \includegraphics[width=\linewidth]{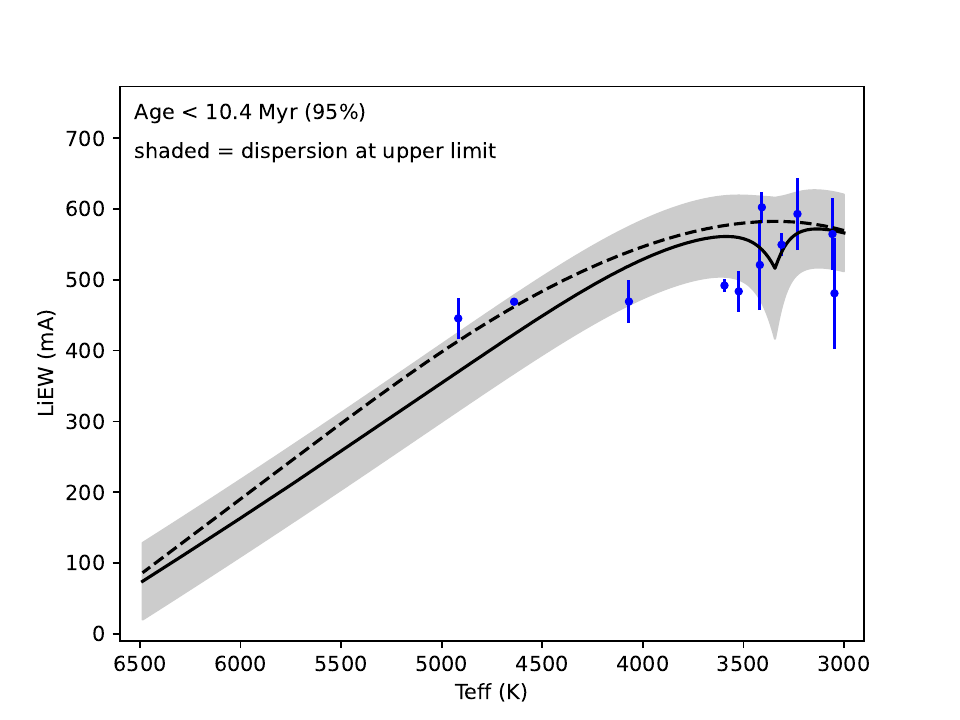}
%        \par\small (d) Taurus
%		  \caption{Taurus}
%		  \label{fig:subfig4}
	       \end{minipage}
    \vfill   
     \begin{minipage}{0.40\linewidth}
     \centering
		 \includegraphics[width=\linewidth]{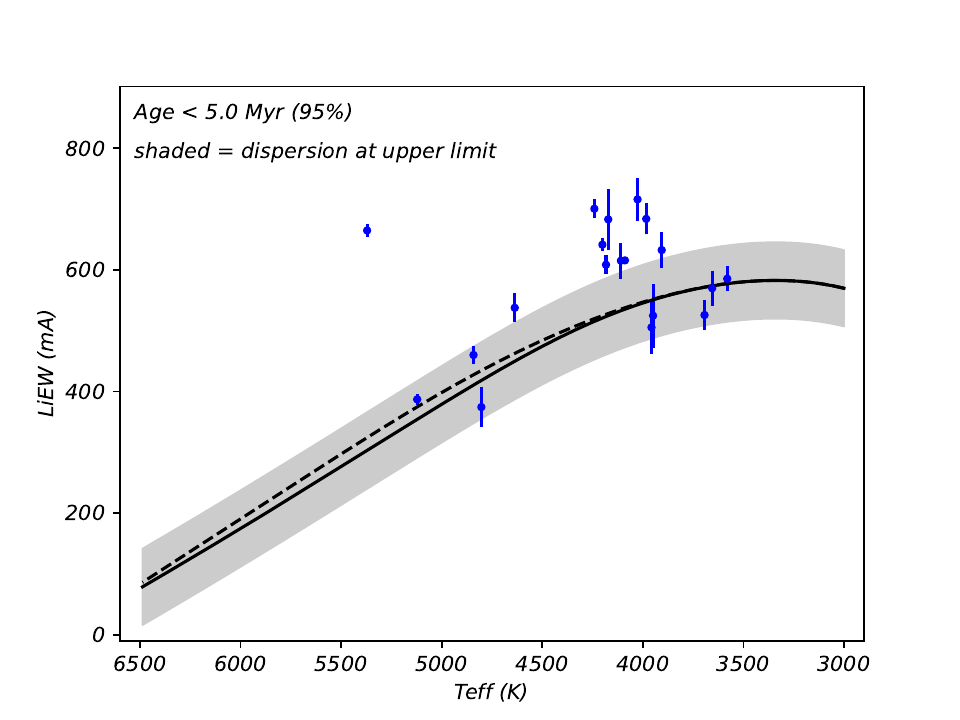}
        \par\small  Taurus
%		 \caption{eCha}
%		 \label{fig:subfig3}%
	      \end{minipage}
	       \begin{minipage}{0.40\linewidth}
           \centering
		  \includegraphics[width=\linewidth]{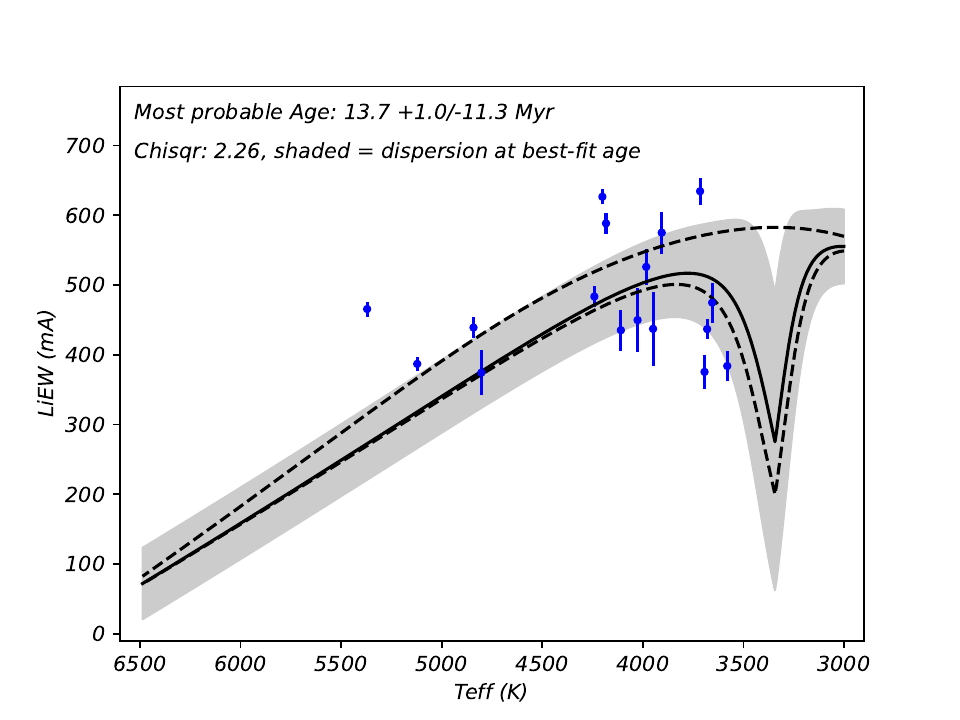}
%\par\small (f) ChaI
%		  \caption{ChaI}
%		  \label{fig:subfig4}
	       \end{minipage}
\caption{Lithium  pattern fitting  for the SFRs Cha I, $\eta$\,Cha and Taurus}
	\label{age1}
\end{figure*}

\begin{figure*}[h!]
      \centering
     \begin{minipage}{0.4\linewidth}
     \centering
		 \includegraphics[width=\linewidth]{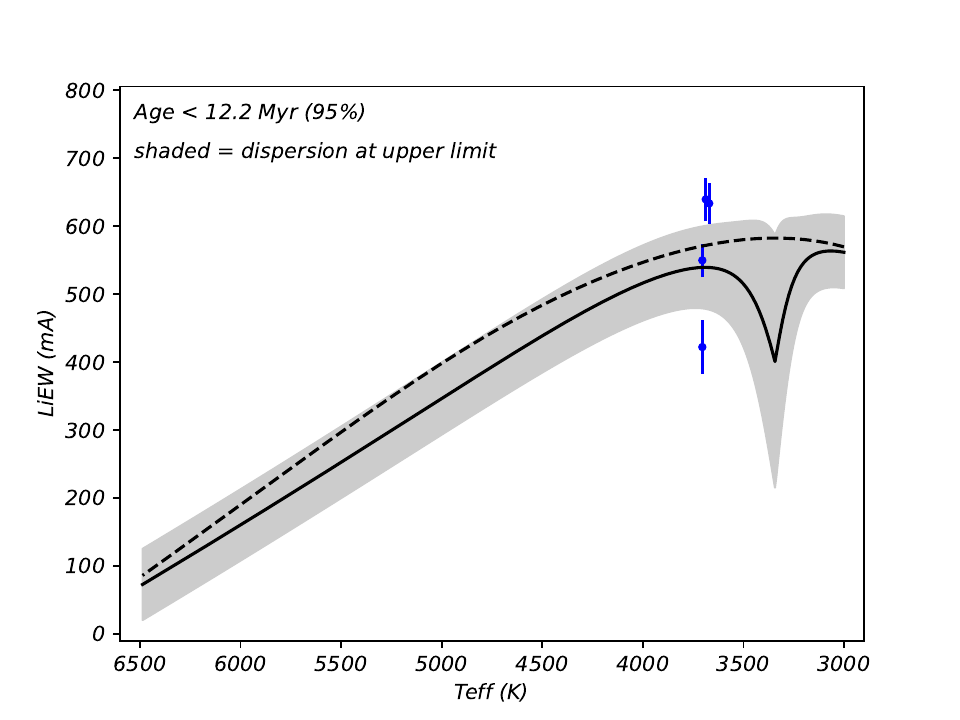}
        \par\small  Orion Ob1a
%		 \caption{eCha}
%		 \label{fig:subfig3}
	      \end{minipage}
	       \begin{minipage}{0.4\linewidth}
           \centering
		  \includegraphics[width=\linewidth]{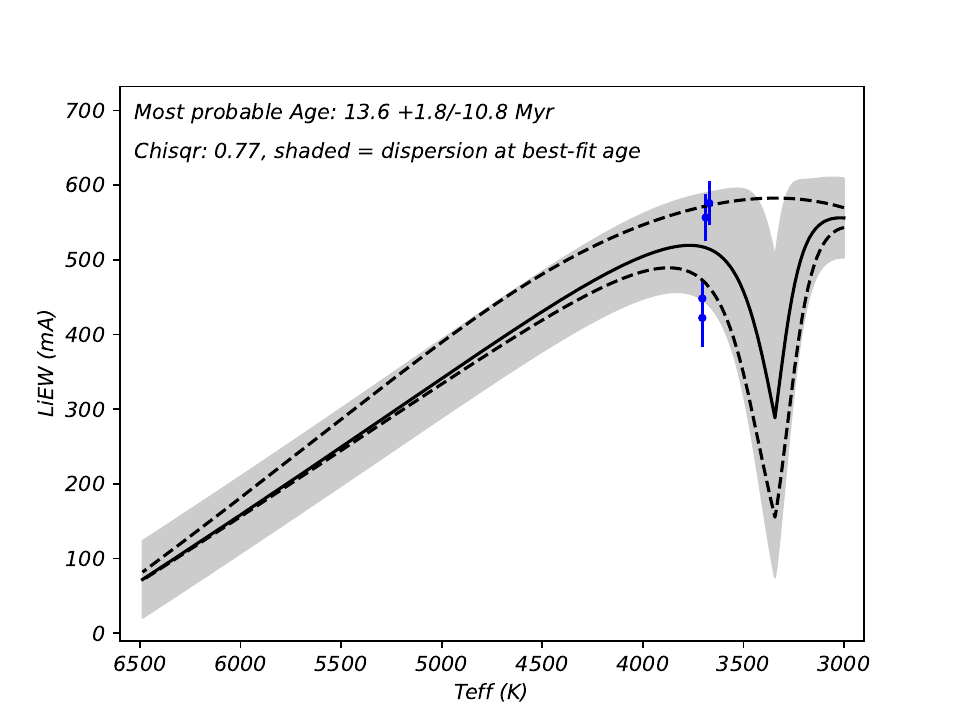}
%\par\small (f) ChaI
%		  \caption{ChaI}
%		  \label{fig:subfig4}
	       \end{minipage}
    \vfill   
     \begin{minipage}{0.4\linewidth}
     \centering
		 \includegraphics[width=\linewidth]{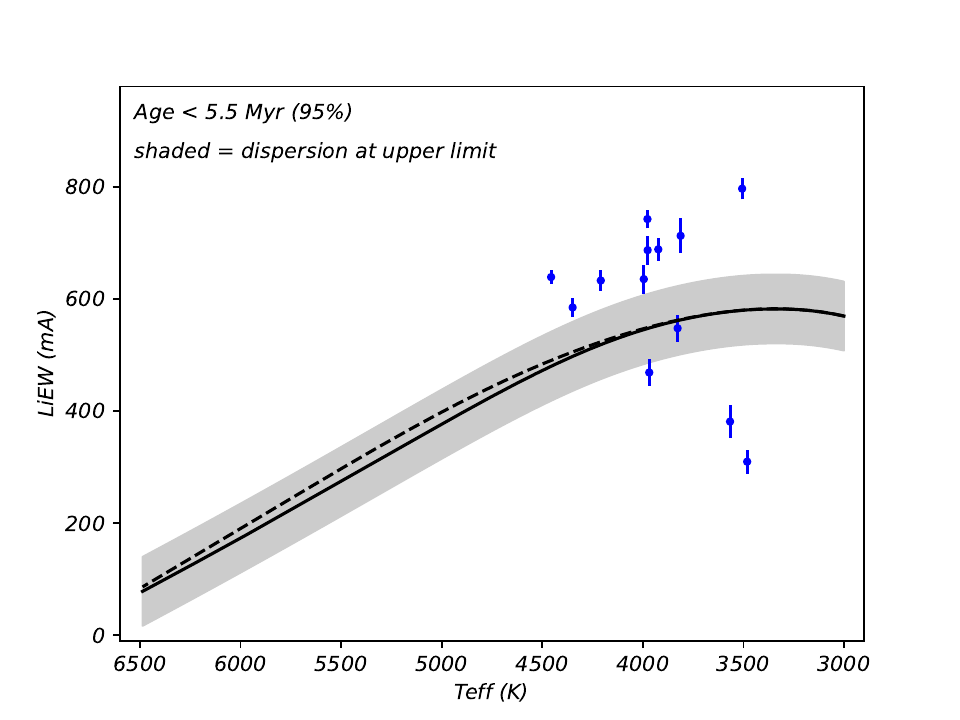}
        \par\small  Orion OB1b
%		 \caption{eCha}
%		 \label{fig:subfig3}
	      \end{minipage}
	       \begin{minipage}{0.40\linewidth}
           \centering
		  \includegraphics[width=\linewidth]{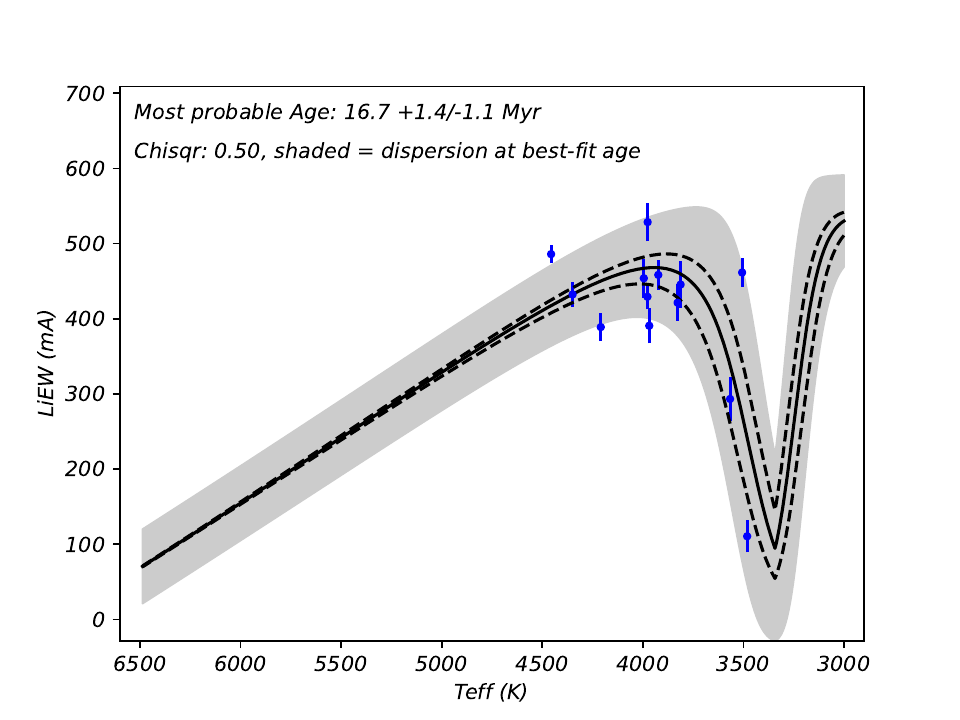}
%\par\small (f) ChaI
%		  \caption{ChaI}
%		  \label{fig:subfig4}
	       \end{minipage}
    \vfill   
     \begin{minipage}{0.4\linewidth}
     \centering
		 \includegraphics[width=\linewidth]{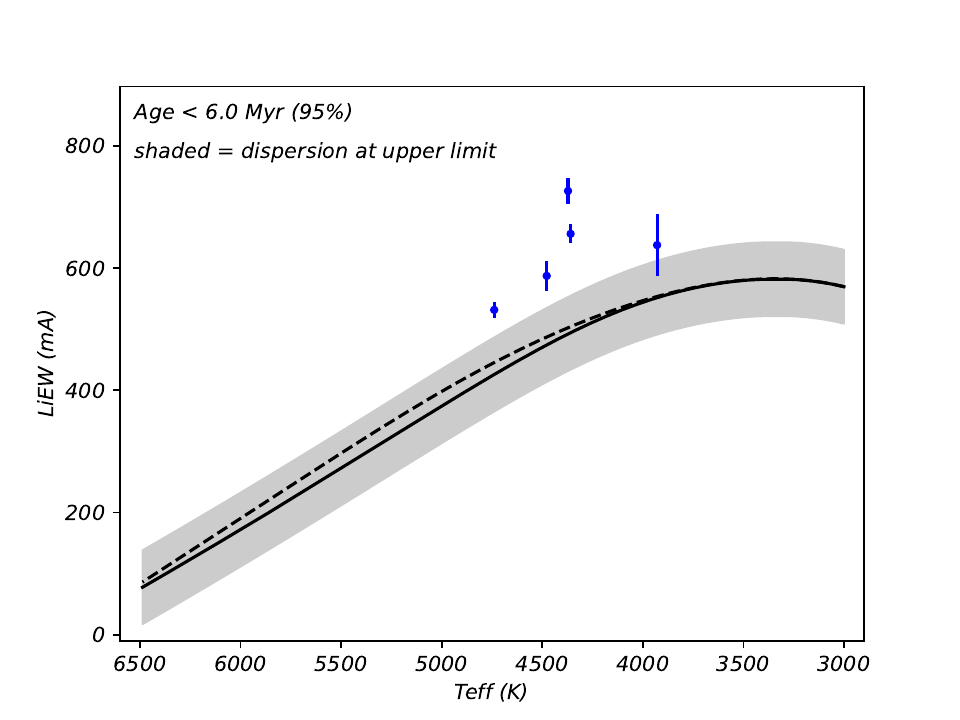}
        \par\small  $\sigma$\,Ori
%		 \caption{eCha}
%		 \label{fig:subfig3}
	      \end{minipage}
	       \begin{minipage}{0.4\linewidth}
           \centering
		  \includegraphics[width=\linewidth]{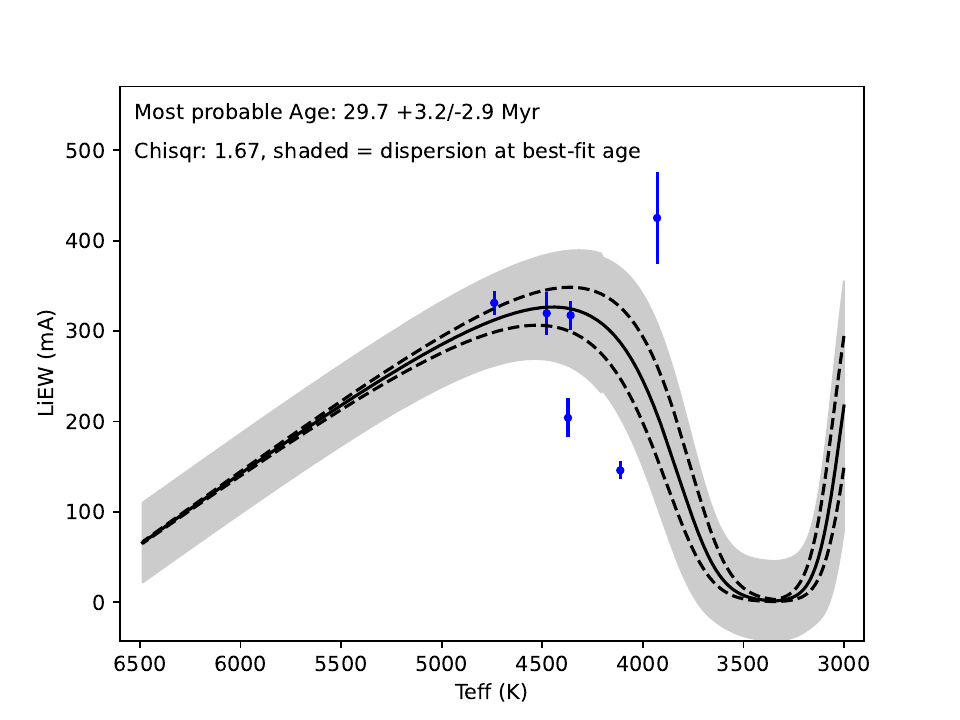}
%\par\small (f) ChaI
%		  \caption{ChaI}
%		  \label{fig:subfig4}
	       \end{minipage}
               \vfill   
     \begin{minipage}{0.40\linewidth}
     \centering
		 \includegraphics[width=\linewidth]{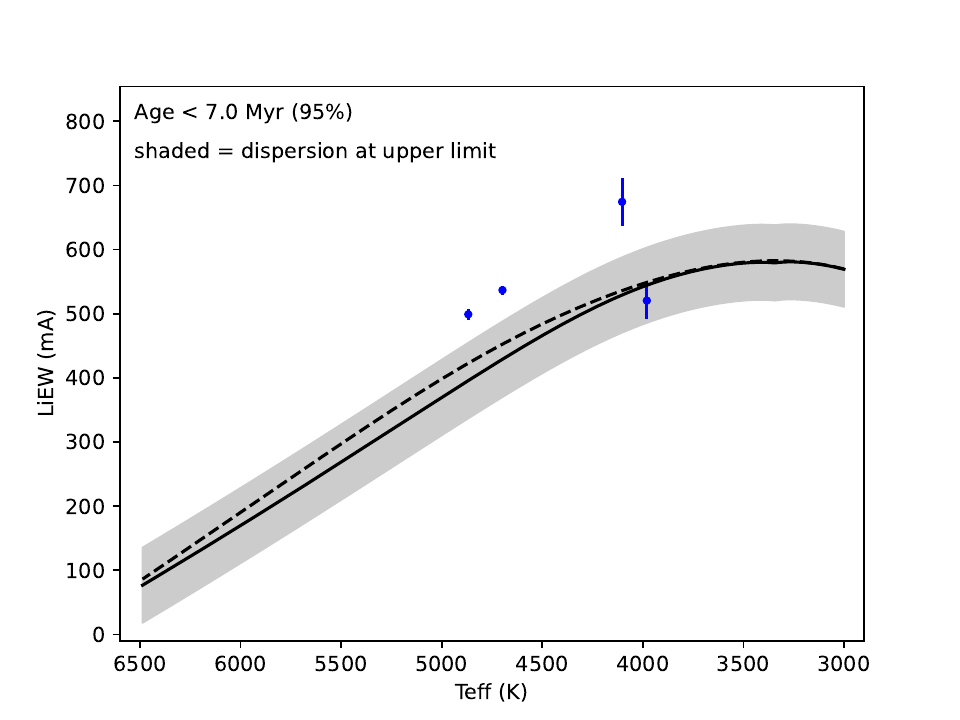}
        \par\small  CrA
%		 \caption{eCha}
%		 \label{fig:subfig3}
	      \end{minipage}
	       \begin{minipage}{0.4\linewidth}
           \centering
		  \includegraphics[width=\linewidth]{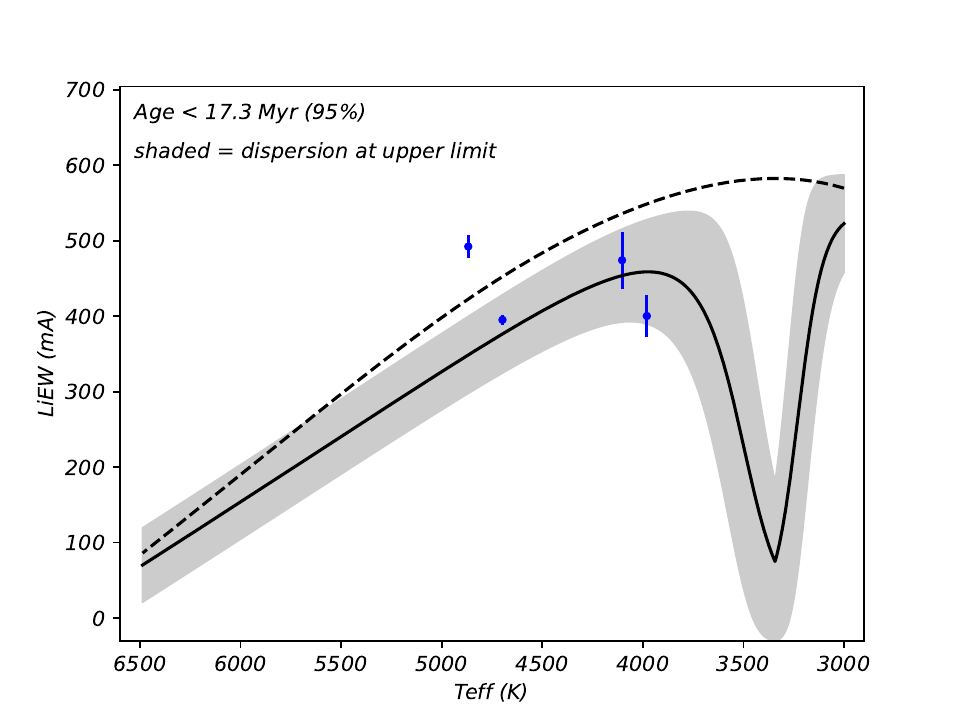}
%\par\small (f) ChaI
%		  \caption{ChaI}
%		  \label{fig:subfig4}
	       \end{minipage}
	\caption{Lithium  pattern fitting  for the SFRs   Orion OB1a, Orion OB1b, $\sigma$\,Ori and CrA. }
	\label{age2}
\end{figure*}

Lithium  pattern fitting  for the SFRs Cha I, $\eta$\,Cha, Taurus,  Orion OB1a, Orion OB1b, $\sigma$\,Ori and CrA. 
The left panels of Fig. \ref{age1} and \ref{age2}  show  the case in which the age was determined using the
$EW_{\rm Li}^{\rm veil+Fe}$ , while the right panel shows the case in which the $EW_{\rm Li}^{\rm Fe}$ have been used. The solid black line represent the best-fit isochrone in the $EW_{\rm Li}$ vs $T_{\rm eff}$ plane. The shaded region illustrates the model intrinsic dispersion at the best-fit age or its upper limit. The black dashed lines represent
95\% upper and lower limits where no clear peak is observed. The blue dots show $EW_{\rm Li}$ as a function of $T_{\rm eff}$ with the uncertainties on $EW_{\rm Li}$
measurements. The text in the top-left corner on the plot shows maximum likelihood age.


%\newpage

\end{appendix}

\end{document}